\documentclass[usenames,dvipsnames,a4paper]{article}
\usepackage{amsmath}
\usepackage{authblk}
\usepackage{amsfonts}
\usepackage{amssymb}
\usepackage{wasysym}
\usepackage{amsthm}
\usepackage{tikz-cd}
\usepackage{longtable}
\usepackage{empheq}
\usepackage{ytableau}
\usepackage{url}
\usepackage{dsfont}
\usepackage{cancel}
\usepackage{slashed}
\usepackage{scalerel}
\usepackage[pdftex]{pict2e}

\usepackage{mathtools}

\newcommand{\verteq}{\rotatebox{90}{$\,=$}}
\newcommand{\vertLra}{\rotatebox{270}{$\,\Lra$}}
\newcommand{\equalto}[2]{\underset{\scriptstyle\overset{\mkern4mu\verteq}{#2}}{#1}}

\definecolor{bright-gray}{rgb}{0.89, 0.90, 0.92}	

\newcommand\smallmath[2]{#1{\raisebox{\dimexpr \fontdimen 22 \textfont 2
      - \fontdimen 22 \scriptscriptfont 2 \relax}{$\scriptscriptstyle #2$}}}

\newcommand\smallotimes{\smallmath\mathbin\otimes}

\def\be{\begin{eqnarray}}
\def\ee{\end{eqnarray}}
\def\nn{\nonumber}

\def\ba{\begin{equation}\begin{aligned}}
\def\ea{\end{aligned}\end{equation}}

\def\horr{{{\smallsmile}\atop{\smallfrown}}}

\def\Kh{{\rm Kh}}
\def\Ker{{\rm Ker}}
\def\Im{{\rm Im}}
\def\Sh{{\rm Sh}}
\def\bphi{\bar \phi}
\def\vth{\vartheta}
\def\bvth{\bar\vartheta}
\def\hd{\hat{d}}
\def\cH{{\cal H}}

\def\Hopf{{\rm Hopf}}

\def\Lra{\Longrightarrow}

\usepackage[hidelinks,colorlinks=true,unicode]{hyperref}
\hypersetup{
	linkcolor=blue,
	citecolor=blue,
	filecolor=magenta,
	urlcolor=blue,
}
\usepackage[left=2cm,right=2cm,top=2cm,bottom=2cm,bindingoffset=0cm]{geometry}

\usepackage[natbib=true, backend = bibtex, style=numeric-comp, sorting=none,maxbibnames=99]{biblatex}
\usepackage{xcolor}

\date{}

\begin{document}

\title{\bf Khovanov--Rozansky cycle calculus for bipartite links
}

\author[2,3]{{\bf A. Anokhina}\thanks{\href{mailto:anokhina@itep.ru}{anokhina@itep.ru}}}
\author[1,2,3,4]{{\bf E. Lanina}\thanks{\href{mailto:lanina.en@phystech.edu}{lanina.en@phystech.edu}}}
\author[1,2,3]{{\bf A. Morozov}\thanks{\href{mailto:morozov@itep.ru}{ morozov@itep.ru}}}

\vspace{5cm}

\affil[1]{Moscow Institute of Physics and Technology, 141700, Dolgoprudny, Russia}
\affil[2]{Institute for Information Transmission Problems, 127051, Moscow, Russia}
\affil[3]{NRC "Kurchatov Institute", 123182, Moscow, Russia}
\affil[4]{Saint Petersburg University, 199034, St. Petersburg, Russia}
\affil[5]{Institute for Theoretical and Experimental Physics, 117218, Moscow, Russia}
\renewcommand\Affilfont{\itshape\small}

\maketitle

\vspace{-7.5cm}

\begin{center}
	\hfill MIPT/TH-03/25 \\
	\hfill ITEP/TH-03/25 \\
	\hfill IITP/TH-03/25
\end{center}

\vspace{4.5cm}

\begin{abstract}
{
Bipartite calculus is a direct generalization of Kauffman planar expansion from $N=2$ to arbitrary $N$,
applicable to the restricted class of knots which are entirely made of antiparallel lock tangles.
Whenever applicable, it allows a straightforward generalization of the Khovanov calculus
without a need of the technically complicated matrix factorization
used for arbitrary $N$ in the Khovanov--Rozansky (KR) approach. The main object here is the $3^{n}$-dimensional hypercube with $n$ being the number of bipartite vertices. 
Maps, differentials, complex and Poincaré polynomials are straightforward and indeed reproduce
the Khovanov--Rozansky polynomials in the known cases.
This provides a great {\bf simplification of the Khovanov--Rozansky calculus} on the bipartite locus, what can make it an accessible tool for the study of superpolynomials.
}
\end{abstract}

\tableofcontents


\section{Introduction}

Knot calculus is the method to derive non-perturvative observables (averages of Wilson lines)
in the $3d$ topological Chern--Simons theory~\cite{Chern-Simons,Schwarz,Witten,MoSmi} -- the simplest non-trivial in the Yang-Mills class.
An amusing feature of this theory is that the answers are polynomials in the peculiar      
non-perturbative variables 
$q = \exp\left(\frac{2\pi}{g+N}\right)$ and $A=q^N$ (with $g$ being the coupling constant of the Chern--Simons theory),
known as (colored by a representation of the corresponding Lie algebra) HOMFLY-PT~\cite{freyd1985new,przytycki1987kobe}, Kauffman~\cite{kauffman1990invariant} and {\it universal} Vogel polynomials~\cite{vogel1999universal,vogel2011algebraic,mironov2016universal}, depending on the choice
of the gauge group and its representation.
An even more amusing fact is the existence of further deformation by the additional variable $T$,
leading to {\it superpolynomials}~\cite{gukov2005khovanov,dunfield2006superpolynomial,dunin2013superpolynomials}, which do not yet have a clear interpretation in terms of
quantum field theory. This makes the study of these quantities especially important and interesting. One of the suggestions is that superpolynomials, or {\it Khovanov--Rozansky polynomials}~\cite{khovanov2000categorification,bar2002khovanov,khovanov2004sl,khovanov2007virtual,khovanov2010categorifications,dolotin2014introduction}, are in turn observables in the refined Chern--Simons theory~\cite{hollowood2008matrix,iqbal2009refined,cherednik2013jones,aganagic2012refined}.

In the following years, the connection between the Chern–Simons theory and knot theory was actively developed. This led to great progress in both fields of research, and the identification of many surprising correspondences. For example, a connection was discovered between Chern–Simons theory and topological string theory. The hypothesis of topological string duality, proposed by Gopakumar--Vafa~\cite{gopakumar1999gauge,gopakumar1998m} and Ooguri--Vafa~\cite{ooguri2000knot}, connects the $U(N)$ Chern--Simons theory on a three-dimensional sphere with topological string theory on a resolved conifold. Moreover, it was shown~\cite{aganagic2015knot} that the partition function of the refined Chern—Simons theory on a 3d sphere equals to the partition function of the refined topological string.

This led to the reformulation of suitable combinations of Chern--Simons knot polynomials into invariants possessing integrality structures~\cite{labastida2001knots,labastida2001polynomial,labastida2002new,marino2001framed,mironov2017checks,mironov2017gaussian}, known as the Labastida, Marino, Ooguri, and Vafa condition. These integer invariants count the BPS states, or the spectra of M2-branes ending on M5-branes in M-theory compactified on a conifold. In particular, the Khovanov--Rozansky polynomials can be reformulated in terms of integers which capture the spectrum of BPS states in the string Hilbert space. They also determine the oriented topological amplitudes of strings. The task of finding these numbers requires an understanding of all the colored Khovanov--Rozansky polynomials for any knot or link. This connection with topological string theory has led to several interesting results. However, the knowledge of explicit formulas and their dependence on a representation for these polynomials is essential for further progress in this field.

One of the problems in calculation of the superpolynomials is that it is technically a very tedious work, with the single exception
of the fundamental representation of $\mathfrak{sl}_2$.
The corresponding superpolynomials were introduced by M. Khovanov with the help of
the peculiarly simple Kauffman approach, which substitutes the general Reshetikhin--Turaev calculus~\cite{Reshetikhin,guadagnini1990chern2,reshetikhin1990ribbon,turaev1990yang,reshetikhin1991invariants,mironov2012character}
for the HOMFLY and other ordinary knot polynomials.
Recently we explained~\cite{ALM, ALM2, ALM3} that the Kauffman calculus can be lifted from $N=2$ to an arbitrary $N$,
but for the restricted class of {\it bipartite links}~\cite{BipKnots,Lewark}, which have {\it link diagrams} made entirely of
antiparallel lock tangles, see Fig.\,\ref{fig:pladeco}.
In this case, Khovanov deformation is also straightforward, and this is what we are going to
explain in the present paper.
Surprisingly or not, it reproduces the answers, suggested by Khovanov and Rozansky~\cite{khovanov2007virtual}
with the help of a very difficult generalization to $N\neq 2$, based on matrix factorization.
Our calculus for bipartite knots does not need it and is as simple for arbitrary $N$,
as the Khovanov one for $N=2$.

In fact, the great simplification of the Khovanov--Rozansky calculus for bipartite links was first observed in~\cite{krasner2009computation}. In this work, Krasner used the planarization of a thick vertex in the matrix factorization approach and proposed the reduction of matrix factorizations of two antiparallel oriented vertices. Our algorithm is equivalent to the Krasner one, what will be shown in our forthcoming paper. However, the advantage of our method is that it deals with simple constructions -- $N$-dimensional spaces and differential operators, which are easily dealt by a computer. 

In this paper, in particular, we provide explicit view of the morphisms. They actually are direct generalizations of morphisms appearing in the case of $N=2$ in the reductions of $2^{2n}$ complexes for bipartite links to $3^n$ complexes with $n$ being the number of bipartite vertices. However, the story on simplifications of Khovanov complexes and the corresponding tangle calculus (for previous works, see~\cite{kaul1998chern,bar2005khovanov,mironov2015colored,morozov2018knot,mironov2018tangle,morozov2018knot,anokhina2021khovanov,anokhina2024towards}) deserves a separate paper.


\bigskip

\noindent This paper is organized as follows. We provide some basic notions in Section~\ref{sec:preliminaries}. In Section~\ref{sec:bip-HOMFLY-hypercube}, we rewrite the bipartite expansion of the HOMFLY polynomial in a way appropriate for the further $T$-deformation. In Section~\ref{sec:trunc-hyp}, we also provide examples of $N=2$ reduction to $3^n$ hypercube ($n$ is the number of bipartite vertices) which then generalizes to an arbitrary $N$. In Section~\ref{sec:method}, we give the algorithm for Khovanov--Rozansky cycle computation for an arbitrary $N$ for bipartite links. Then, in Section~\ref{sec:examples}, we provide examples of planar Khovanov--Rozansky calculus. In parallel with the duality between bipartite HOMFLY and precursor Jones polynomials~\cite{ALM} (see Section~\ref{sec:H-J-duality}), the Khovanov--Rozansky $3^n$-hypercube also gives rise to the Khovanov $2^n$ hypercube but for links obtained from the corresponding bipartite ones by shrinking each lock tangle to a single vertex (we call such links {\it precursors} or {\it precursor diagrams}). This subject is explained in Section~\ref{sec:precursors} and first illustrated by the examples in Section~\ref{sec:examples}. We formulate the results and open questions for future study in Section~\ref{sec:concl}.


\setcounter{equation}{0}
\section{Preliminaries}\label{sec:preliminaries}

In this section, we provide some preliminary information needed to explain further constructions. 

\subsection{Jones polynomial and hypercube of resolutions}\label{sec:Jones-2-hyp}

The Jones polynomial is the Wilson loop in the 3-dimensional Chern--Simons theory with $SU(2)$ gauge group and is the HOMFLY polynomial at the particular point $A = q^2$, i.e. $N=2$. It can be efficiently calculated with the use of the Kauffman bracket~\cite{Kauff}, see Fig.\,\ref{fig:Kauff}, which corresponds to a special choice of $U_q(\mathfrak{sl}_2)$ quantum $\cal R$-matrix.

\begin{figure}[h!]
\begin{picture}(100,200)(-200,-100)

\put(0,70){

\put(-60,-20){\line(1,1){40}}
\put(-20,-20){\line(-1,1){18}}
\put(-60,20){\line(1,-1){18}}

\put(10,-2){\mbox{$=$}}

\qbezier(30,20)(50,0)(70,20)
\qbezier(30,-20)(50,0)(70,-20)

\put(85,-2){\mbox{$- \ \ \ q $}}

\qbezier(125,20)(145,0)(125,-20)
\qbezier(150,20)(130,0)(150,-20)

\put(-68,22){\mbox{$i$}}  \put(-15,22){\mbox{$j$}}  \put(-68,-28){\mbox{$k$}}  \put(-15,-28){\mbox{$l$}}

\put(35,-45){\mbox{type $0$}}

\put(125,-45){\mbox{type $1$}}
}

\put(0,-20){

\put(-60,-20){\line(1,1){18}}
\put(-20,-20){\line(-1,1){40}}
\put(-37.5,2){\line(1,1){18}}

\put(-68,22){\mbox{$i$}}  \put(-15,22){\mbox{$j$}}  \put(-68,-28){\mbox{$k$}}  \put(-15,-28){\mbox{$l$}}

\put(10,-2){\mbox{$=$}}

\qbezier(30,20)(50,0)(70,20)
\qbezier(30,-20)(50,0)(70,-20)

\put(85,-2){\mbox{$- \ \ \ q^{-1} $}}

\qbezier(125,20)(145,0)(125,-20)
\qbezier(150,20)(130,0)(150,-20)

\put(35,-45){\mbox{type $1$}}

\put(125,-45){\mbox{type $0$}}
}

\put(60,-20){

\put(-100,-70){\circle{30}}
\put(-75,-72.5){\mbox{$=\ \ \ D_2 \ = \ q+q^{-1}$}}
}

\end{picture}
\caption{\footnotesize The  celebrated Kauffman bracket -- the planar decomposition
of the ${\cal R}$-matrix vertex for the fundamental representation of $U_q(\mathfrak{sl}_2)$.
In this case ($N=2$), the conjugate of the fundamental representation is isomorphic to it,
thus, tangles in the picture have no orientation. The resolutions are of two types which we denote 0 and 1. These numbers are values of $\alpha_i$, see the text below. 
}
\label{fig:Kauff}
\end{figure}

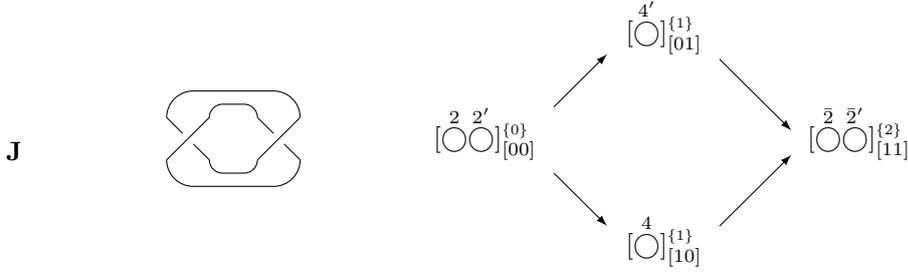
\begin{figure}[h!]
\centering
\begin{picture}(100,100)(-70,5)

\put(-240,45){\mbox{\bf J}}

\put(-80,20){

\put(-100,25){\line(1,1){16}}
\put(-100,41){\line(1,-1){6}}
\put(-90,31){\line(1,-1){6}}

\put(34,0){

\put(-100,25){\line(1,1){16}}
\put(-100,41){\line(1,-1){6}}
\put(-90,31){\line(1,-1){6}}
}


\put(-75,25){\oval(18,10)[b]}
\put(-75,41){\oval(18,10)[t]}

\put(-75,25){\oval(50,20)[b]}
\put(-75,41){\oval(50,20)[t]}
}

\put(-80,50){\mbox{$[\overset{2}{\bigcirc}\overset{2'}{\bigcirc}]_{[00]}^{\scaleto{\{0\}}{5.5pt}}$}}
\put(-35,65){\vector(1,1){20}}
\put(-35,40){\vector(1,-1){20}}

\put(-8,90){\mbox{$[\overset{4'}{\bigcirc}]_{[01]}^{\scaleto{\{1\}}{5.5pt}}$}}
\put(-8,10){\mbox{$[\overset{4}{\bigcirc}]_{[10]}^{\scaleto{\{1\}}{5.5pt}}$}}

\put(-5,0){

\put(65,50){\mbox{$[\overset{\bar 2}{\bigcirc}\overset{\bar 2'}{\bigcirc}]_{[11]}^{\scaleto{\{2\}}{5.5pt}}$}}
\put(32,83){\vector(1,-1){27}}
\put(32,20){\vector(1,1){27}}
}
    
\end{picture}
\caption{\footnotesize The Hopf link and its hypercube of resolutions.}
\label{fig:Hopf-hypercube}
\end{figure}

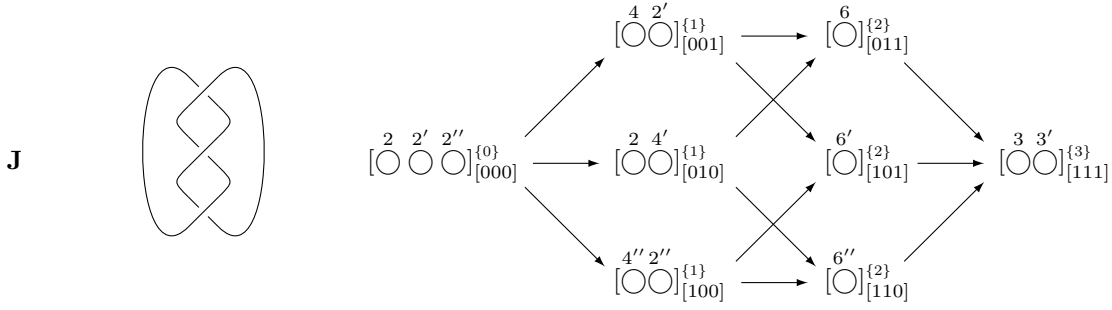
\begin{figure}[h!]
    \centering
\begin{picture}(300,120)(20,-10)

\put(-45,45){\mbox{\bf J}}

\put(90,45){\mbox{$[\overset{2}{\bigcirc}\overset{2'}{\bigcirc}\overset{2''}{\bigcirc}]_{[000]}^{\scaleto{\{0\}}{5.5pt}}$}}

\put(19,0){

\put(163,93){\mbox{$[\overset{4}{\bigcirc}\overset{2'}{\bigcirc}]_{[001]}^{\scaleto{\{1\}}{5.5pt}}$}}

\put(163,45){\mbox{$[\overset{2}{\bigcirc}\overset{4'}{\bigcirc}]_{[010]}^{\scaleto{\{1\}}{5.5pt}}$}}

\put(163,0){\mbox{$[\overset{4''}{\bigcirc}\overset{2''}{\bigcirc}]_{[100]}^{\scaleto{\{1\}}{5.5pt}}$}}

\put(130,57){\vector(1,1){30}}

\put(133,47){\vector(1,0){25}}

\put(130,38){\vector(1,-1){30}}

\put(17,0){

\put(225,93){\mbox{$[\overset{6}{\bigcirc}]_{[011]}^{\scaleto{\{2\}}{5.5pt}}$}}

\put(225,45){\mbox{$[\overset{6'}{\bigcirc}]_{[101]}^{\scaleto{\{2\}}{5.5pt}}$}}

\put(225,0){\mbox{$[\overset{6''}{\bigcirc}]_{[110]}^{\scaleto{\{2\}}{5.5pt}}$}}

\put(192,57){\vector(1,1){30}}

\put(192,10){\vector(1,1){30}}

\put(192,38){\vector(1,-1){30}}

\put(192,85){\vector(1,-1){30}}

\put(194,95){\vector(1,0){25}}

\put(194,2){\vector(1,0){25}}

\put(255,10){\vector(1,1){30}}

\put(255,85){\vector(1,-1){30}}

\put(290,45){\mbox{$[\overset{3}{\bigcirc}\overset{3'}{\bigcirc}]_{[111]}^{\scaleto{\{3\}}{5.5pt}}$}}

\put(260,47){\vector(1,0){25}}

}

}

\begin{tikzpicture}[scale=0.4]

\draw (0,0) -- (1,1); 
\draw (1,0) -- (0.65,0.35);
\draw (0,1) -- (0.35,0.65);

\draw (0,-2) -- (1,-1);
\draw (1,-2) -- (0.65,-1.65);
\draw (0,-1) -- (0.35,-1.35);

\draw (0,-4) -- (1,-3);
\draw (1,-4) -- (0.65,-3.65);
\draw (0,-3) -- (0.35,-3.35);

\draw (1,1) .. controls (3,3) and (3,-6) .. (1,-4);
\draw (1,0) .. controls (1.5,-0.5) .. (1,-1);
\draw (1,-2) .. controls (1.5,-2.5) .. (1,-3);

\draw (0,1) .. controls (-2,3) and (-2,-6) .. (0,-4);
\draw (0,0) .. controls (-0.5,-0.5) .. (0,-1);
\draw (0,-2) .. controls (-0.5,-2.5) .. (0,-3);

\end{tikzpicture}
\end{picture}
    \caption{\footnotesize The trefoil knot $3_1$ and its hypercube of resolutions.}
    \label{fig:trefoil-hypercube}
\end{figure}

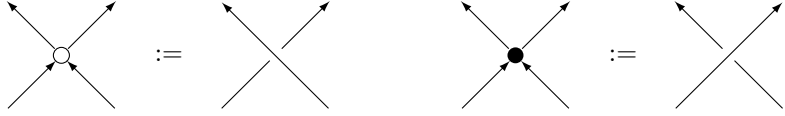
\begin{figure}[h!]
    \centering
\begin{picture}(300,65)(-135,-35)

\put(-80,0){

\put(-60,-20){\vector(1,1){18}}
\put(-20,-20){\vector(-1,1){18}}
\put(-42.5,2.5){\vector(-1,1){18}}
\put(-37.5,2.5){\vector(1,1){18}}

\put(-40,0){\circle{6}}

\put(-5,-2){\mbox{$:=$}}

}

\put(-60,-20){\line(1,1){18}}
\put(-20,-20){\vector(-1,1){40}}
\put(-37.5,2.5){\vector(1,1){18}}


\put(90,0){

\put(-60,-20){\vector(1,1){18}}
\put(-20,-20){\vector(-1,1){18}}
\put(-42.5,2.5){\vector(-1,1){18}}
\put(-37.5,2.5){\vector(1,1){18}}

\put(-40,0){\circle*{6}}

\put(-5,-2){\mbox{$:=$}}

}

\put(170,0){

\put(-60,-20){\vector(1,1){40}}
\put(-20,-20){\line(-1,1){18}}
\put(-42.5,2.5){\vector(-1,1){18}}

}

\end{picture}
    \caption{\footnotesize Denotations for a crossing and its mirror.}
    \label{fig:pos-neg-cr}
\end{figure}

The Kauffman bracket replaces each crossing by two resolutions of type 0 and type 1. Thus, a link with $n$ crossings has $2^n$ resolutions which can be organized in a hypercube. Different resolutions $r$ of a link stand in
vertices and are enumerated by words of zeros and unities $[\alpha_1 \ldots \alpha_n]$ with $\alpha_i = 0,\, 1$. 
Each resolution is associated with a set of $\nu(r)$ closed non-intersecting planar circles,
and these are the patterns which we associate with vertices. Actually, each circle consists of several segments between intersection points,
i.e. has a definite ``length''. The set of $\nu(r)$ lengths $l_a(r)$ can be also associated with the
vertex $r$ of the hypercube. Edges connect resolutions corresponding to elementary flips in which exactly one resolution of type $0$ is replaced by the resolution of type $1$. Resolutions of a link sharing the same vertical possess the same {\it height} $h := \sum_i \alpha_i$. Then, the Jones polynomial is
\begin{equation}\label{J-hypercube}
    J^{\cal L}(q) = (-)^{n_\circ} q^{n_\bullet - 2 n_\circ} \sum_{r}^{2^n} (-q)^{h(r)} D_2^{\nu(r)}
\end{equation}
where $\nu(r)$ is the number of cycles in the resolution $r$. Here the multiplier $(-)^{n_\circ} q^{n_\bullet - 2 n_\circ}$ is present to restore the topological invariance; $n_\circ$ and $n_\bullet$ are numbers of vertices depicted in Fig.\,\ref{fig:pos-neg-cr}. The examples of the hypercube of resolutions are in Figs.\,\ref{fig:Hopf-hypercube} and\,\ref{fig:trefoil-hypercube}. Here, in curly brackets we denote the heights and under cycles we write the corresponding lengths. 

We call such hypercubes 2-hypercubes because each edge of a cube consists of 2 vertices. In Section~\ref{sec:bip-HOMFLY-hypercube}, we will introduce the notion of 3-hypercube each edge of which consists of 3 vertices. We emphasise that these numbers 2 and 3 are not connected with the dimensions of the hypercubes. For example, the 2-hypercube for the Hopf link from Fig.\,\ref{fig:Hopf-hypercube} is 2-dimensional while the 2-hypercube from Fig.\,\ref{fig:trefoil-hypercube} is 3-dimensional. In general, a 2-hypercube for a link of $n$ crossings is $n$-dimensional.



\subsection{Cut-and-join operator and extended Jones polynomial}\label{sec:CAJ-ext-Jones}

Now, let us distinguish cycles in resolutions by weighting them by their own variables $p_a$ instead of the unique one $D_2 = q+q^{-1}$. It is convenient to label the variables by the corresponding cycle length. Then, for any resolution $r$, $\sum\limits_{a \in r} a = 2n$ with $n$ being the number of a link $\cal L$ crossings. We also replace $-q$ by $t$ and introduce the notion of the {\it extended Jones polynomial}\,:
\begin{equation}
    {\cal J}^{\cal L}\{t|p_a\} = \sum\limits_{r \in {\rm hypercube}} t^{h(r)} \prod\limits_{a \in r}^{\nu(r)} p_a\,.
\end{equation}
The topological invariant Jones polynomial arises after the restriction to the {\it topological locus}\,:
\begin{equation}
    J^{\cal L}(q) = (-)^{n_\circ} q^{n_\bullet - 2 n_\circ} {\cal J}^{\cal L}\left\{t=-q|p_a=D_2\right\}\,.
\end{equation}
Elementary flips $\horr \longleftrightarrow \; )\,(\;$ act on the set of cycles either gluing two into one, or splitting one into two, i.e. the corresponding operator is
\begin{equation}
    p_a \frac{\partial^2}{\partial p_b \partial p_c} + p_b p_c \frac{\partial}{\partial p_a}\,,\quad l_a = l_b + l_c
\end{equation}
where $l_a$, $l_b$, $l_c$ are the length of the corresponding cycles. Then, one can associate the {\it cut-and-join operator} (CAJ operator) to the whole hypercube:
\begin{equation}\label{CAJ}
    \hat{W} = \frac{1}{2} \sum\limits_{a,b,c} M_{bc}^a \left( p_a \frac{\partial^2}{\partial p_b \partial p_c} + p_b p_c \frac{\partial}{\partial p_a} \right).
\end{equation}
Note that it is independent of the starting point of a hypercube, i.e. on $\;\bullet \longleftrightarrow \circ\;$ crossing changes inside a link. In other words, the $\hat{W}$-operator is dependent only on an unoriented hypercube with all edges having no arrows. In~\eqref{CAJ}, $M^a_{bc}$ equals one, if a flip connecting the triple $(a|bc)$ exists, and zero otherwise. 

The cut-and-join operator acts on the extended Jones polynomial by changing one crossing to the mirror one, what is demonstrated by the example below. 

\paragraph{Hopf example.} As we see in Fig.\,\ref{fig:Hopf-hypercube}, we have the variables $p_2$ and $p'_2$ at $h=0$, $p_4$ and $p'_4$ at $h=1$, $\bar p_2$ and $\bar p'_2$ at $h=2$. The extended Jones polynomial for the Hopf link is
\begin{equation}
    {\cal J}^{\rm Hopf} = p_2 p'_2 + t(p_4 + p'_4) + t^2 \bar p_2 \bar p'_2\,.
\end{equation}
According to~\eqref{CAJ}, the $\hat{W}$-operator is
\begin{equation}
    \hat{W} = (p_4 + p'_4) \left( \frac{\partial^2}{\partial p_2 \partial p'_2} + \frac{\partial^2}{\partial \bar p_2 \partial \bar p'_2} \right) + (p_2 p'_2 + \bar p_2 \bar p'_2) \left( \frac{\partial}{\partial p_4} + \frac{\partial}{\partial p'_4} \right).
\end{equation}
It acts on the Jones polynomial for the Hopf link replacing it with the Jones polynomials for two unknots:
\begin{equation}
    \hat{W} {\cal J}^{\rm Hopf} = p_4 + p'_4 + 2 t(p_2 p'_2 + \bar p_2 \bar p'_2) + t^2 (p_4 + p'_4) = {\cal J}^{\bigcirc\bigcirc} + {\cal J}^{\overline{\bigcirc\bigcirc}}
\end{equation}
where
\begin{equation}
\begin{aligned}
    {\cal J}^{\bigcirc\bigcirc} &= p_4 + t(p_2 p'_2 + \bar p_2 \bar p'_2) + t^2 p'_4 \\
    {\cal J}^{\overline{\bigcirc\bigcirc}} &= p'_4 + t(p_2 p'_2 + \bar p_2 \bar p'_2) + t^2 p_4
\end{aligned}
\end{equation}
Now, $\hat{W}$-operator intertwines ${\cal J}^{\rm Hopf}$ and ${\cal J}^{\overline{\rm Hopf}}$\,:
\begin{equation}
    \hat{W} {\cal J}^{\bigcirc\bigcirc} = \hat{W} {\cal J}^{\overline{\bigcirc\bigcirc}} = {\cal J}^{\rm Hopf} + {\cal J}^{\overline{\rm Hopf}}
\end{equation}
with
\begin{equation}
    {\cal J}^{\overline{\rm Hopf}} = \bar p_2 \bar p'_2 + t(p_4 + p'_4) + t^2 p_2 p'_2\,.
\end{equation}


\subsection{Khovanov polynomial}\label{sec:Khovanov}

Note that the hypercube formula for the Jones polynomial~\eqref{J-hypercube} looks like the $q$-Euler characteristic of the corresponding graded complex. Such a construction allows for a {\it categorification} ({\it $T$-deformation}) which we describe in this subsection. The resulting polynomial is the celebrated {\it Khovanov polynomial}~\cite{khovanov2000categorification,bar2002khovanov} also known as the {\it Jones superpolynomial}. The method for its computation consists of the following steps.

\paragraph{Step 1: 2-hypercube.} Draw the $2$-hypercube consisting of $2^n$ vertices and $2^{n-1}n$ edges, as described in Section~\ref{sec:Jones-2-hyp}.

\paragraph{Step 2: spaces.} The formula for the Jones polynomial can be rewritten in the form 
\begin{equation}\label{J-2-hyp-qdim}
    J^{\cal L}(q) = (-)^{n_\circ} q^{n_\bullet - 2 n_\circ} \sum_{r}^{2^n} (-q)^{h(r)} \dim_q(V^{\smallotimes \nu(r)})\,, \quad \dim_q(V) = D_2 = q + q^{-1}\,.
\end{equation}
So that one associates $q$-graded 2-dimensional vector space $V = \langle \vth_1,\,\vth_2 \rangle$ to each closed cycle in a resolution. The basis vectors have the gradings ${\rm grad}(\vth_1)=q^{+1}$ and ${\rm grad}(\vth_2)=q^{-1}$. It is also convenient to label these spaces by the numbers of edges of the corresponding cycles.

\paragraph{Step 3: differentials.} In Section~\ref{sec:CAJ-ext-Jones}, to each arrow of the hypercube, we have associated the {\it cut operator}
\begin{equation}
    p_b p_c \frac{\partial}{\partial p_a}\,,\quad l_a = l_b + l_c\,,
\end{equation}
if a cycle splits into a pair of ones, and the {\it join operator}
\begin{equation}
    p_a \frac{\partial^2}{\partial p_b \partial p_c}\,,\quad l_a = l_b + l_c
\end{equation}
if a pair of cycles joins into one. Now, the morphisms are a kind of supersymmetrization of these operators:
\begin{equation}\label{Kh-morphisms}
\begin{aligned}
    \Delta &= \vth_2^b \vth_2^c \frac{\partial}{\partial \vth_2^a} + (\vth_2^b \vth_1^c + \vth_1^b \vth_2^c)\frac{\partial}{\partial \vth_1^a}\;:\quad V_a \; \mapsto \; V_b \otimes V_c \\
    m &= \vth_2^a \left( \frac{\partial^2}{\partial \vth_1^b \partial \vth_2^c} + \frac{\partial^2}{\partial \vth_2^b \partial \vth_1^c} \right) + \vth_1^a \frac{\partial^2}{\partial \vth_1^b \partial \vth_1^c}\;:\quad V_b \otimes V_c \; \mapsto \; V_a
\end{aligned}
\end{equation}
and are enumerated by $[\alpha_1\dots \alpha_{m-1}\,\star\, \alpha_{m+1}\dots \alpha_n]$ where the star stands at exactly the place where the mapped space label changes by unity. In order to provide the differentials property $\hat{d}_{i+1}\hat{d}_{i}=0$, with each edge, one also associates the sign factor
\begin{equation}\label{sign-morph}
    {\rm sign} = (-1)^{\alpha_1 + \dots + \alpha_{m-1}}\,.
\end{equation}
Differentials $\hat{d}_i$ are sums of operators~\eqref{Kh-morphisms} with appropriate signs~\eqref{sign-morph} acting between direct sums of spaces sharing the same fixed height $h$.

\paragraph{Step 4: kernels and images.} Now we calculate kernels, images, cohomologies\footnote{Here and in all cases below, we denote basis elements in triangular brackets.} of differentials and their quantum dimensions. For convenience, we present here kernels and images of the elementary maps~\eqref{Kh-morphisms}.

\medskip

\noindent $\bullet$ For the multiplication morphism $m\,$:
\begin{equation}
\begin{aligned}
    \Im(m) &= V \quad \Lra \quad \dim_q\Im(m) = D_2 \\
    \Ker(m) &= \langle \vth_2^b \vth_2^c,\, \vth_2^b \vth_1^c - \vth_1^b \vth_2^c  \rangle \quad \Lra \quad \dim_q\Ker(m) = q^{-2} + 1 = q^{-1} D_2
\end{aligned}
\end{equation}

\noindent $\bullet$ For the coproduct operator $\Delta\,$:
\ba 
\Im(\Delta) &= \langle \vth_2^b \vth_2^c,\, \vth_2^b \vth_1^c + \vth_1^b \vth_2^c  \rangle \quad \Lra \quad \dim_q \Im(\Delta) = q^{-1} D_2 \\
\Ker(\Delta) &= \varnothing \quad \Lra \quad \dim_q \Ker(\Delta) = 0
\ea 

\paragraph{Step 5: Khovanov polynomial.} The Khovanov polynomial is expressed through the quantum dimensions of cohomologies ${\cal H}_i = {\rm Ker}(\hat{d}_{i})\backslash {\rm Im}(\hat{d}_{i-1})$ as follows: 
\begin{equation}\label{Kh-def}
    {\rm Kh}^{\cal L}(q,T) = q^{n_\bullet}\cdot (T q^2)^{-n_\circ} \sum_{i=0}^n (qT)^i \dim_q \cH_i^{\cal L} = q^{n_\bullet}\cdot (T q^2)^{-n_\circ} \sum_{i=0}^n (qT)^i \left( \dim_q\Ker(\hat{d}_{i}^{\cal L}) - \dim_q\Im(\hat{d}_{i-1}^{\cal L}) \right)
\end{equation}
The Euler characteristic can be alternatively rewritten in terms of cohomologies, so that we return to the Jones polynomial at the particular point $T=-1\,$: $J^{\cal L}(q)={\rm Kh}^{\cal L}(q,T=-1)$.

\paragraph{Hopf example.} We consider the computation of the Khovanov polynomial for the Hopf link step by step as discussed above. The Jones 2-hypercube for the Hopf link is given in Fig.\,\ref{fig:Hopf-hypercube}. 

According to step 2, we attach a 2-dimensional vector space to each cycle. The labellings of these spaces are indicated above cycles. Recall that they also correspond to the number of edges of the cycles.

\begin{figure}[h!]
\centering
\begin{picture}(100,100)(-70,5)

\put(-240,45){\mbox{\bf Kh}}

\put(-80,20){

\put(-100,25){\line(1,1){16}}
\put(-100,41){\line(1,-1){6}}
\put(-90,31){\line(1,-1){6}}

\put(34,0){

\put(-100,25){\line(1,1){16}}
\put(-100,41){\line(1,-1){6}}
\put(-90,31){\line(1,-1){6}}
}


\put(-75,25){\oval(18,10)[b]}
\put(-75,41){\oval(18,10)[t]}

\put(-75,25){\oval(50,20)[b]}
\put(-75,41){\oval(50,20)[t]}
}

\put(-105,50){\mbox{$\varnothing\;\overset{0}{\longrightarrow} \; [\overset{2}{\bigcirc}\overset{2'}{\bigcirc}]_{[00]}^{\scaleto{\{0\}}{5.5pt}}$}}
\put(-35,65){\vector(1,1){20}}
\put(-35,40){\vector(1,-1){20}}
\put(-40,75){\mbox{$m$}}
\put(-40,25){\mbox{$m$}}

\put(-8,90){\mbox{$[\overset{4'}{\bigcirc}]_{[01]}^{\scaleto{\{1\}}{5.5pt}}$}}
\put(-8,10){\mbox{$[\overset{4}{\bigcirc}]_{[10]}^{\scaleto{\{1\}}{5.5pt}}$}}
\put(-3,50){\mbox{$\bigoplus$}}

\put(-5,0){

\put(65,50){\mbox{$[\overset{\bar 2}{\bigcirc}\overset{\bar 2'}{\bigcirc}]_{[11]}^{\scaleto{\{2\}}{5.5pt}}\; \overset{0}{\longrightarrow}\; \varnothing$}}
\put(32,83){\vector(1,-1){27}}
\put(32,20){\vector(1,1){27}}
\put(30,60){\mbox{$\Delta$}}
\put(25,35){\mbox{$-\Delta$}}
}
    
\end{picture}
\caption{\footnotesize The Hopf link and its Khovanov complex.}
\label{fig:Kh-Hopf-complex}
\end{figure}
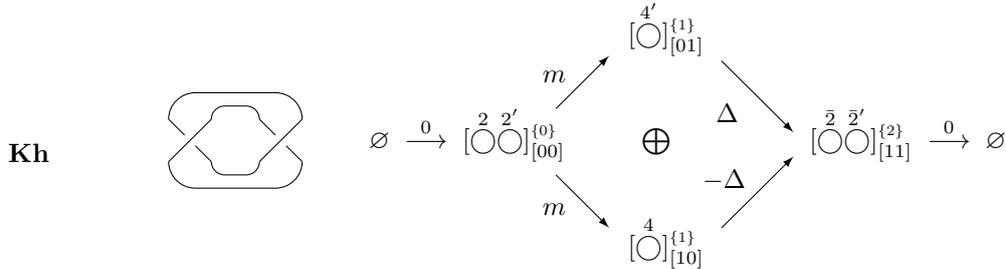

At the third step, we associate morphisms~\eqref{Kh-morphisms} to arrows. The operator $m$ joins two cycles into one, and the map $\Delta$ splits one cycle into a pair of cycles. As a result, we get the Khovanov hypercube in Fig.\,\ref{fig:Kh-Hopf-complex}. The differentials are
\begin{equation}
\begin{aligned}
    \hat{d}_0^{\,\rm Hopf} &= m + m = \left( \vth_2^{(4)} + \vth_2^{\prime(4)} \right) \left( \frac{\partial^2}{\partial \vth_1^{(2)} \partial \vth_2^{\prime (2)}} + \frac{\partial^2}{\partial \vth_2^{(2)} \partial \vth_1^{\prime (2)}} \right) + \left( \vth_1^{(4)} + \vth_1^{\prime(4)} \right) \frac{\partial^2}{\partial \vth_1^{(2)} \partial \vth_1^{\prime (2)}} \\
    \hat{d}_1^{\,\rm Hopf} &= \Delta + (-\Delta) = \bar\vth_2^{(2)} \bar\vth_2^{\prime (2)} \left(\frac{\partial}{\partial \vth_2^{\prime (4)}} - \frac{\partial}{\partial \vth_2^{(4)}} \right) + (\bar\vth_2^{(2)} \bar\vth_1^{\prime (2)} + \bar\vth_1^{(2)} \bar\vth_2^{\prime (2)})\left(\frac{\partial}{\partial \vth_1^{\prime (4)}} - \frac{\partial}{\partial \vth_1^{(4)}} \right)
\end{aligned}
\end{equation}

\noindent Now, we calculate kernels and images of the differentials. The zeroth operator acts on $V_2 \otimes V'_2$ as follows:
\begin{equation}
   \hat{d}_0^{\,\rm Hopf}\,: \qquad \begin{array}{ccc}
\left\langle \begin{array}{c}
        \vth_1^{\prime (2)} \vth_1^{(2)} \\
         \frac{1}{2}(\vth_1^{\prime (2)} \vth_2^{(2)} + \vth_2^{\prime (2)} \vth_1^{(2)})
    \end{array}\right\rangle & \begin{array}{c}
         \mapsto  \\
         \mapsto
    \end{array} & \left\langle \begin{array}{c}
         \vth_1^{(4)} + \vth_1^{\prime(4)} \\
         \vth_2^{(4)} + \vth_2^{\prime(4)}
    \end{array}\right\rangle \\ 
    \oplus & & \oplus \\
     \left\langle \begin{array}{c}
        \frac{1}{2}(\vth_1^{\prime (2)} \vth_2^{(2)} - \vth_2^{\prime (2)} \vth_1^{(2)}) \\
         \vth_2^{\prime (2)} \vth_2^{(2)} 
    \end{array}\right\rangle & \mapsto & 0
   \end{array} 
\end{equation}
so that
\ba 
\Im(\hat{d}_0^{\,\rm Hopf}) &= V = \langle \vth_1^{(4)} + \vth_1^{\prime(4)},\, \vth_2^{(4)} + \vth_2^{\prime(4)} \rangle \\
\Ker(\hat{d}_0^{\,\rm Hopf}) &= \Ker(m) = \langle \vth_2^{(2)} \vth_2^{\prime (2)},\, \vth_2^{(2)} \vth_1^{\prime (2)} - \vth_1^{(2)} \vth_2^{\prime (2)}  \rangle 
\ea 
The first differential acts on $V_4 \oplus V'_4$ as follows:
\begin{equation}
   \hat{d}_1^{\,\rm Hopf}\,: \qquad \begin{array}{ccc}
\left\langle \begin{array}{c}
        \frac{1}{2}(\vth_1^{(4)} - \vth_1^{\prime(4)}) \\
         \frac{1}{2}(\vth_2^{(4)} - \vth_2^{\prime(4)})
    \end{array}\right\rangle & \begin{array}{c}
         \mapsto  \\
         \mapsto
    \end{array} & \left\langle \begin{array}{c}
         \bar\vth_2^{(2)} \bar\vth_1^{\prime (2)} + \bar\vth_1^{(2)} \bar\vth_2^{\prime (2)} \\
         \bar\vth_2^{(2)} \bar\vth_2^{\prime (2)}
    \end{array}\right\rangle \\ 
    \oplus & & \oplus \\
     \left\langle \begin{array}{c}
        \frac{1}{2}(\vth_1^{(4)} + \vth_1^{\prime(4)}) \\
         \frac{1}{2}(\vth_2^{(4)} + \vth_2^{\prime(4)}) 
    \end{array}\right\rangle & \mapsto & 0
   \end{array} 
\end{equation}
so that
\begin{equation}
\begin{aligned}
    \Im(\hat{d}_1^{\,\rm Hopf}) &= \Im(\Delta) = \langle \bar\vth_2^{(2)} \bar\vth_2^{\prime (2)},\, \bar\vth_2^{(2)} \bar\vth_1^{\prime (2)} + \bar\vth_1^{(2)} \bar\vth_2^{\prime (2)}  \rangle \\
    \Ker(\hat{d}_1^{\,\rm Hopf}) &= V = \langle \vth_1^{(4)} + \vth_1^{\prime(4)},\, \vth_2^{(4)} + \vth_2^{\prime(4)} \rangle
\end{aligned}
\end{equation}
The cohomologies and their quantum dimensions are
\begin{equation}
\begin{aligned}
    \cH_0^{\,\rm Hopf} &= \Ker(\hat{d}_0^{\,\rm Hopf}) \quad \Lra \quad \dim_q \cH_0^{\,\rm Hopf} = q^{-1}D_2 \\
    \cH_1^{\,\rm Hopf} &= \Ker(\hat{d}_1^{\,\rm Hopf})\backslash \Im(\hat{d}_0^{\,\rm Hopf}) = \varnothing \quad \Lra \quad \dim_q \cH_1^{\,\rm Hopf} = 0 \\
    \cH_2^{\,\rm Hopf} &= \Ker(0)\backslash \Im(\hat{d}_1^{\,\rm Hopf}) = \langle \bar\vth_1^{(2)} \bar\vth_1^{\prime (2)},\, \bar\vth_2^{(2)} \bar\vth_1^{\prime (2)} - \bar\vth_1^{(2)} \bar\vth_2^{\prime (2)} \rangle \quad \Lra \quad \dim_q \cH_2^{\,\rm Hopf} = q^2 + 1 = q D_2
\end{aligned}
\end{equation}
According to equation~\eqref{Kh-def}, the Khovanov polynomial is
\begin{equation}
    \Kh^{\,\Hopf}(q,T) = q^2\left(\dim_q \cH_0^{\,\rm Hopf} + (qT)^2 \dim_q \cH_2^{\,\rm Hopf}\right) = q D_2 (1 + q^4 T^2)
\end{equation}

\setcounter{equation}{0}
\section{Bipartite HOMFLY polynomial from hypercube}
\label{sec:bip-HOMFLY-hypercube}

In this section, we recall the HOMFLY bipartite expansion in $\phi$, $\bphi$, $D_N$ variables (see Section~\ref{sec:bip-exp}). Then, in Section~\ref{sec:H-hypercubes}, we note that planar decomposition for the HOMFLY polynomial allows for 2-hypercube arrangement, as it was in the Jones case, see Section~\ref{sec:Jones-2-hyp}. However, this 2-hypercube must be expanded to 3-hypercube in order to provide the HOMFLY categorification. In Section~\ref{sec:H-J-duality}, we show that the bipartite 2-hypercube turns into the Jones 2-hypercube for the so-called precursor diagrams obtained from the initial bipartite link by shrinking lock vertices into single vertices, see Fig.\,\ref{fig:precursor-bip}. 


\subsection{HOMFLY polynomial for bipartite links}\label{sec:bip-exp}

\begin{figure}[h!]
\begin{picture}(100,170)(-110,-135)

\put(20,0){
\put(-105,15){\vector(1,-1){12}} \put(-87,3){\vector(1,1){12}}
\put(-93,-3){\vector(-1,-1){12}} \put(-75,-15){\vector(-1,1){12}}
\put(-90,0){\circle*{6}}  \put(-90,-9){\line(0,1){18}}

\put(-60,-2){\mbox{$:=$}}
}

\put(0,0){\put(-17,20){\line(1,-1){17}}\put(-17,20){\vector(1,-1){14}}   \put(0,3){\vector(1,1){17}}
 \put(0,-3){\vector(-1,-1){17}}   \put(17,-20){\line(-1,1){17}} \put(17,-20){\vector(-1,1){14}}
 \put(0,-3){\line(0,1){6}}}

\put(10,0){

\put(20,-2){\mbox{$:=$}}

\qbezier(50,20)(55,9)(58,4) \qbezier(63,-4)(85,-40)(110,20)
\put(56,8){\vector(1,-2){2}} \put(90,-13){\vector(1,1){2}} \put(109,18){\vector(1,2){2}}
\qbezier(50,-20)(75,40)(97,4)  \qbezier(102,-4)(105,-9)(110,-20)
\put(104,-8){\vector(-1,2){2}} \put(70,13){\vector(-1,-1){2}} \put(51,-18){\vector(-1,-2){2}}

}

\put(20,0){
\put(120,-2){\mbox{$=$}}
\put(138,-2){\mbox{$q^{-2N}$}}

\put(5,0){
\qbezier(163,-20)(153,0)(163,20)
\put(38,0){\qbezier(290,-20)(300,0)(290,20)}

\put(100,65){
\put(70,-60){\vector(1,0){40}}
\put(110,-70){\vector(-1,0){40}}
\put(120,-67){\mbox{$+\ \ \ (q^{N+1}-q^{N-1})$}}
\put(210,-50){\vector(0,-1){30}}
\put(220,-80){\vector(0,1){30}}

\put(75,-95){\mbox{type 0}}
\put(200,-95){\mbox{type 1}}

}
}
}

\put(0,-65){

\put(20,0){
\put(-105,15){\vector(1,-1){12}} \put(-87,3){\vector(1,1){12}}
\put(-93,-3){\vector(-1,-1){12}} \put(-75,-15){\vector(-1,1){12}}
\put(-90,0){\circle{6}}  \put(-90,-9){\line(0,1){18}}

\put(-60,-2){\mbox{$:=$}}
}

\put(0,0){
\put(0,0){\put(-17,20){\line(1,-1){17}}\put(-17,20){\vector(1,-1){14}}   \put(0,3){\vector(1,1){17}}
 \put(0,-3){\vector(-1,-1){17}}   \put(17,-20){\line(-1,1){17}} \put(17,-20){\vector(-1,1){14}}
\put(-1,-3){\line(0,1){6}} \put(1,-3){\line(0,1){6}}
}

\put(30,-2){\mbox{$:=$}}

\put(10,0){
\qbezier(50,20)(75,-40)(97,-4)  \qbezier(102,4)(105,9)(110,20)
\qbezier(50,-20)(55,-9)(58,-4) \qbezier(63,4)(85,40)(110,-20)
\put(55,9){\vector(1,-2){2}} \put(90,-13){\vector(1,1){2}} \put(109,18){\vector(1,2){2}}
\put(105,-9){\vector(-1,2){2}} \put(70,13){\vector(-1,-1){2}} \put(51,-18){\vector(-1,-2){2}}
}

\put(140,-2){\mbox{$=$}}

\put(25,0){
\put(137,-2){\mbox{$q^{2N}$}}

\put(100,65){
\put(67,-60){\vector(1,0){40}}
\put(107,-70){\vector(-1,0){40}}
\put(115,-67){\mbox{$-\ \ (q^{1-N}-q^{-N-1})$}}
\put(-25,0){
\put(233,-50){\vector(0,-1){30}}
\put(243,-80){\vector(0,1){30}}
}

\put(75,-95){\mbox{type 1}}
\put(200,-95){\mbox{type 0}}

}}}
\put(25,0){\qbezier(163,-20)(153,0)(163,20)
\put(40,0){\qbezier(290,-20)(300,0)(290,20)}}

}

\put(30,-50){

\put(-100,-70){\circle{30}}
\put(-75,-72.5){\mbox{$=\ \ \ D_N \ = \ \cfrac{q^N - q^{-N}}{q-q^{-1}}$}}
}

\end{picture}
\caption{\footnotesize
The planar decomposition of the positive (in the first line) and negative (in the second line) lock vertices in the \textit{topological} framing.
In \cite{ALM}, we used the {\it vertical} framing, but the {\it topological} one is more convenient for our considerations,
thus, we use it throughout the present paper, like we did also in \cite{ALM3}.
}\label{fig:pladeco}
\end{figure}

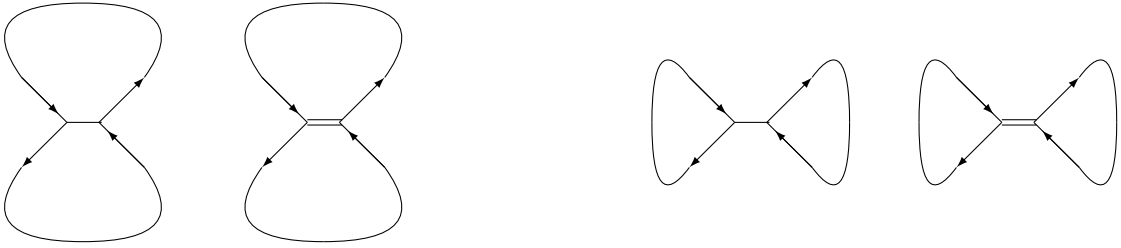
\begin{figure}[h!]
\begin{picture}(100,90)(-50,-45)

\put(100,0){

\put(-80,0){

\put(-6,0){\line(1,0){13}}
\put(-23,17){\line(1,-1){17}}\put(-23,17){\vector(1,-1){14}}   \put(6,0){\vector(1,1){17}}
\put(-6,0){\vector(-1,-1){17}}   \put(23,-17){\line(-1,1){17}} \put(23,-17){\vector(-1,1){14}}

\qbezier(-23,17)(-43,45)(0,45) \qbezier(23,17)(43,45)(0,45)
\qbezier(-23,-17)(-43,-45)(0,-45) \qbezier(23,-17)(43,-45)(0,-45)
}

\put(10,0){

\put(-6,-1){\line(1,0){13}}
\put(-6,1){\line(1,0){13}}
\put(-23,17){\line(1,-1){17}}\put(-23,17){\vector(1,-1){14}}   \put(6,0){\vector(1,1){17}}
\put(-6,0){\vector(-1,-1){17}}   \put(23,-17){\line(-1,1){17}} \put(23,-17){\vector(-1,1){14}}

\qbezier(-23,17)(-43,45)(0,45) \qbezier(23,17)(43,45)(0,45)
\qbezier(-23,-17)(-43,-45)(0,-45) \qbezier(23,-17)(43,-45)(0,-45)
}


\put(170,0){  

\put(-6,0){\line(1,0){13}}
\put(-23,17){\line(1,-1){17}}
\put(-23,17){\vector(1,-1){14}}   
\put(6,0){\vector(1,1){17}}
\put(-6,0){\vector(-1,-1){17}}   \put(23,-17){\line(-1,1){17}} 
\put(23,-17){\vector(-1,1){14}}

\qbezier(-23,17)(-37,36)(-37,0) \qbezier(-23,-17)(-37,-36)(-37,0)
\qbezier(23,17)(37,36)(37,0) \qbezier(23,-17)(37,-36)(37,0)

}

\put(270,0){  

\put(-6,1){\line(1,0){13}}
\put(-6,-1){\line(1,0){13}}
\put(-23,17){\line(1,-1){17}}
\put(-23,17){\vector(1,-1){14}}   
\put(6,0){\vector(1,1){17}}
\put(-6,0){\vector(-1,-1){17}}   \put(23,-17){\line(-1,1){17}} 
\put(23,-17){\vector(-1,1){14}}

\qbezier(-23,17)(-37,36)(-37,0) \qbezier(-23,-17)(-37,-36)(-37,0)
\qbezier(23,17)(37,36)(37,0) \qbezier(23,-17)(37,-36)(37,0)

}

}

\end{picture}
\caption{\footnotesize  Two closures of a single lock from Fig.\,\ref{fig:pladeco}: the unknot and its mirror on the left
and the Hopf link and its mirror on the right.
} \label{fig:one-bip-vert}
\end{figure}

\begin{figure}[h!]
\begin{picture}(100,100)(-30,-50)

\qbezier(17,10)(32,26)(40,19)

\qbezier(17,-10)(32,-26)(40,-19)

\qbezier(-13,10)(-21,20)(-13,30)

\qbezier(-13,-30)(-21,-20)(-13,-10)

\qbezier(60,19)(65,25)(60,30)

\qbezier(-13,30)(24,60)(60,30)

\qbezier(60,-30)(65,-24)(60,-19)

\qbezier(-13,-30)(24,-60)(60,-30)

\put(-13,10){\vector(1,-1){10}}

\put(7,0){\vector(1,1){10}}

\put(17,-10){\vector(-1,1){10}}

\put(-3,0){\vector(-1,-1){10}}

\put(-3,0){\line(1,0){10}}

\put(50,-9){\line(0,1){18}}

\put(60,-19){\vector(-1,1){10}}
\put(50,-9){\vector(-1,-1){10}}

\put(40,19){\vector(1,-1){10}}
\put(50,9){\vector(1,1){10}}


\put(100,0){

\qbezier(17,10)(32,26)(40,19)

\qbezier(17,-10)(32,-26)(40,-19)

\qbezier(-13,10)(-21,20)(-13,30)

\qbezier(-13,-30)(-21,-20)(-13,-10)

\qbezier(60,19)(65,25)(60,30)

\qbezier(-13,30)(24,60)(60,30)

\qbezier(60,-30)(65,-24)(60,-19)

\qbezier(-13,-30)(24,-60)(60,-30)

\put(-13,10){\vector(1,-1){10}}

\put(7,0){\vector(1,1){10}}

\put(17,-10){\vector(-1,1){10}}

\put(-3,0){\vector(-1,-1){10}}

\put(-3,1){\line(1,0){10}}
\put(-3,-1){\line(1,0){10}}

\put(49,-9){\line(0,1){18}}
\put(51,-9){\line(0,1){18}}

\put(60,-19){\vector(-1,1){10}}
\put(50,-9){\vector(-1,-1){10}}

\put(40,19){\vector(1,-1){10}}
\put(50,9){\vector(1,1){10}}

}


\put(200,0){

\qbezier(17,10)(32,26)(40,19)

\qbezier(17,-10)(32,-26)(40,-19)

\qbezier(-13,10)(-21,20)(-13,30)

\qbezier(-13,-30)(-21,-20)(-13,-10)

\qbezier(60,19)(65,25)(60,30)

\qbezier(-13,30)(24,60)(60,30)

\qbezier(60,-30)(65,-24)(60,-19)

\qbezier(-13,-30)(24,-60)(60,-30)

\put(-13,10){\vector(1,-1){10}}

\put(7,0){\vector(1,1){10}}

\put(17,-10){\vector(-1,1){10}}

\put(-3,0){\vector(-1,-1){10}}

\put(-3,1){\line(1,0){10}}
\put(-3,-1){\line(1,0){10}}

\put(50,-9){\line(0,1){18}}

\put(60,-19){\vector(-1,1){10}}
\put(50,-9){\vector(-1,-1){10}}

\put(40,19){\vector(1,-1){10}}
\put(50,9){\vector(1,1){10}}

}


\put(330,-90){

\put(0,100){
\qbezier(-3,0)(-13,-7)(-12,-10) \put(-12,-10){\vector(0,-1){2}}
\qbezier(7,0)(17,-7)(15,-10) \put(9,-2){\vector(-1,1){2}}
\put(-3,0){\line(1,0){10}}

\put(-13,10){\vector(1,-1){10}}
\put(7,0){\vector(1,1){10}}
}

\put(0,79){
\qbezier(-3,0)(-13,7)(-12,10) \put(-6,2){\vector(1,-1){2}}
\qbezier(7,0)(17,7)(15,10) \put(15,9){\vector(0,1){2}}
\put(-3,0){\line(1,0){10}}
}

\put(0,79){

\put(-3,0){\vector(-1,-1){10}}
\put(17,-10){\vector(-1,1){10}}
}

\qbezier(-13,110)(-30,120)(-30,89)
\qbezier(17,110)(34,120)(34,89)

\put(0,79){
\qbezier(-13,-10)(-30,-20)(-30,10)
\qbezier(17,-10)(34,-20)(34,10)
}
}

\put(410,-90){

\put(0,100){
\qbezier(-3,0)(-13,-7)(-12,-10) \put(-12,-10){\vector(0,-1){2}}
\qbezier(7,0)(17,-7)(15,-10) \put(9,-2){\vector(-1,1){2}}

\put(-3,1){\line(1,0){10}}
\put(-3,-1){\line(1,0){10}}

\put(-13,10){\vector(1,-1){10}}
\put(7,0){\vector(1,1){10}}
}

\put(0,79){
\qbezier(-3,0)(-13,7)(-12,10) \put(-6,2){\vector(1,-1){2}}
\qbezier(7,0)(17,7)(15,10) \put(15,9){\vector(0,1){2}}

\put(-3,1){\line(1,0){10}}
\put(-3,-1){\line(1,0){10}}
}

\put(0,79){

\put(-3,0){\vector(-1,-1){10}}
\put(17,-10){\vector(-1,1){10}}
}

\qbezier(-13,110)(-30,120)(-30,89)
\qbezier(17,110)(34,120)(34,89)

\put(0,79){
\qbezier(-13,-10)(-30,-20)(-30,10)
\qbezier(17,-10)(34,-20)(34,10)
}
}


\end{picture}
    \caption{\footnotesize From left to right: trefoil knot $3_1$, mirror trefoil knot $\bar 3_1$, figure-eight knot, $APT[2,4]$, $\overline{APT[2,4]}$.}
    \label{fig:two-bip-vert}
\end{figure}
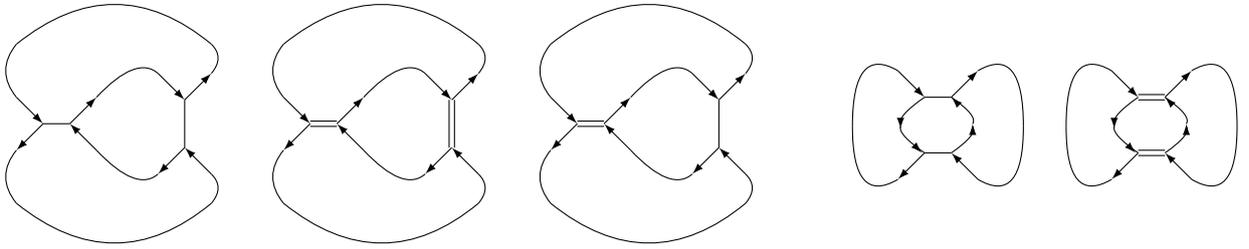

In~\cite{ALM}, we have shown that the Kauffman bracket for $N=2$ (Fig.\,\ref{fig:Kauff}) can be lifted to {\bf an arbitrary $N$} (i.e. to the HOMFLY polynomial in the fundamental representation) but for the {\bf special class of bipartite links} consisting only of antiparallel lock elements from Fig.\,\ref{fig:pladeco}. Examples of bipartite links are provided in Figs.\,\ref{fig:one-bip-vert} and \ref{fig:two-bip-vert}. Then, bipartite calculus of~\cite{ALM} represents the fundamental HOMFLY polynomial as the sum
\begin{equation}\label{bipexp}
\boxed{
H^{\cal L}(A=q^N,q) = {\rm Fr}\cdot \sum_{a,b,c}  D^a \phi^b \bar\phi^c = A^{-2(n_\bullet - n_\circ)} \sum_k \sum_{i=0}^{n_\bullet} \sum_{j=0}^{n_\circ} {\cal N}_{ijk} D_N^k \phi^i \bar \phi^j\,,
}
\end{equation}
where
\be
\phi = q^{N+1}-q^{N-1}, \ \ \ \ \bar\phi = -q^{1-N}+q^{-1-N} , \ \ \ \  D_N =\frac{ q^N-q^{-N} }{q-q^{-1}}\,.
\label{bippars}
\ee
The sum is over planar $2^{n_\bullet+ n_\circ}$ resolutions of $n_\bullet$ positive and $n_\circ$ negative antiparallel locks, see Fig.\,\ref{fig:pladeco}, and in the first equality all items come with unit coefficients.
However, some terms can coincide, and  if we  sum over different non-negative triples $i$, $j$, $k$,
then the coefficients are non-negative integers ${\cal N}_{ijk}$.

\subsection{2-hypercube vs 3-hypercube}\label{sec:H-hypercubes}

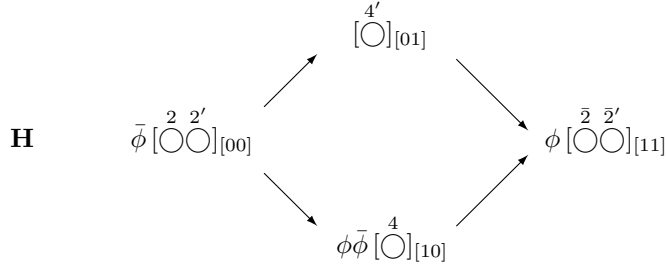
\begin{figure}[h!]
\centering
\begin{picture}(100,100)(-15,5)

\put(-130,50){\mbox{\bf H}}

\put(-85,50){\mbox{$\bphi\,[\overset{2}{\bigcirc}\overset{2'}{\bigcirc}]_{[00]}$}}
\put(-35,65){\vector(1,1){20}}
\put(-35,40){\vector(1,-1){20}}

\put(-2,90){\mbox{$[\overset{4'}{\bigcirc}]_{[01]}$}}
\put(-8,10){\mbox{$\phi\bphi\,[\overset{4}{\bigcirc}]_{[10]}$}}

\put(5,0){

\put(65,50){\mbox{$\phi\,[\overset{\bar 2}{\bigcirc}\overset{\bar 2'}{\bigcirc}]_{[11]}$}}
\put(32,83){\vector(1,-1){27}}
\put(32,20){\vector(1,1){27}}
}
    
\end{picture}
\caption{\footnotesize Bipartite HOMFLY hypercube for the figure-eight knot from Fig.\,\ref{fig:two-bip-vert}.}
\label{fig:H-tw-4-1-complex}
\end{figure}

Due to the reasoning of the previous subsection, for the HOMFLY polynomial, a construction of the 2-hypercube can be also introduced, see examples in Figs.\,\ref{fig:H-Hopf-complex}, \ref{fig:H-tw-trefoil-complex} and \ref{fig:H-tw-4-1-complex}. Here we again enumerate each resolution by a sequence $[\alpha_1 \ldots \alpha_n]$ with $\alpha_i=0,\,1$ according to Fig.\,\ref{fig:pladeco} and $n$ being the total number of lock vertices. To each cycle, we associate its number of edges written above. In the bipartite HOMFLY case, edges connect bipartite (double antiparallel) vertices, not single ones as it was in the Jones case. In these figures, we write $\phi$-multipliers appearing in the HOMFLY polynomial in front of the corresponding resolution.

\begin{figure}[h!]
    \begin{equation}\nn
    \textbf{H} \quad \qquad \quad [\overset{1}{\bigcirc} \overset{1'}{\bigcirc}]_{[0]} \quad \longrightarrow \quad \phi\,[\overset{2}{\bigcirc}]_{[1]}
\end{equation}
    \caption{\footnotesize Bipartite HOMFLY hypercube for the Hopf link from Fig.\,\ref{fig:one-bip-vert}.}
    \label{fig:H-Hopf-complex}
\end{figure}

\begin{figure}[h!]
\centering
\begin{picture}(100,100)(-15,5)

\put(-120,50){\mbox{\bf H}}

\put(-65,50){\mbox{$[\overset{4}{\bigcirc}]_{[00]}$}}
\put(-35,65){\vector(1,1){20}}
\put(-35,40){\vector(1,-1){20}}

\put(-5,90){\mbox{$\phi\,[\overset{2}{\bigcirc}\overset{2'}{\bigcirc}]_{[01]}$}}
\put(-5,10){\mbox{$\phi\,[\overset{\bar 2}{\bigcirc}\overset{\bar 2'}{\bigcirc}]_{[10]}$}}

\put(15,0){

\put(65,50){\mbox{$\phi^2\,[\overset{4'}{\bigcirc}]_{[11]}$}}
\put(32,83){\vector(1,-1){27}}
\put(32,20){\vector(1,1){27}}
}
    
\end{picture}
\caption{\footnotesize Bipartite HOMFLY hypercube for the trefoil knot from Fig.\,\ref{fig:two-bip-vert}.}
\label{fig:H-tw-trefoil-complex}
\end{figure}

\begin{figure}[h!]
\centering
\begin{picture}(100,180)(40,-35)

\put(-130,47){\mbox{\bf H}}

\put(15,0){

\put(-70,48){\mbox{$[\overset{2}{\bigcirc}\overset{2'}{\bigcirc}]_{[00]}^{\scaleto{\{-1-N\}}{5.5pt}}$}}
{\color{blue}\put(-35,65){\vector(1,1){20}}}
\put(-35,40){\vector(1,-1){20}}
}

\put(10,0){
\put(-10,90){\mbox{$[\overset{{\color{blue} 2}}{\bigcirc}\overset{{\color{blue} 2'}}{\bigcirc}]_{[01]}^{\scaleto{\{1-N\}}{5.5pt}}$}}
\put(-10,10){\mbox{$[\overset{4}{\bigcirc}]_{[10]}^{\scaleto{\{-2\}}{5.5pt}}$}}
}

\put(5,0){

\put(60,135){\mbox{$[\overset{4'}{\bigcirc}]_{[02]}^{\scaleto{\{0\}}{5.5pt}}$}}
\put(60,-35){\mbox{$[\overset{{\color{Green} 4}}{\bigcirc}]_{[20]}^{\scaleto{\{0\}}{5.5pt}}$}}
\put(25,105){\vector(1,1){30}}
{\color{Green} 
\put(25,0){\vector(1,-1){30}}}

\put(65,50){\mbox{$[\overset{{\color{Green} 4'}}{\bigcirc}]_{[11]}^{\scaleto{\{0\}}{5.5pt}}$}}
\put(32,83){\vector(1,-1){27}}
{\color{Green}
\put(32,20){\vector(1,1){27}}}

\put(5,0){

\put(91,130){\vector(1,-1){35}}
{\color{Green}
\put(91,-25){\vector(1,1){35}}}
\put(130,90){\mbox{$[\overset{\bar 2}{\bigcirc}\overset{\bar 2'}{\bigcirc}]_{[12]}^{\scaleto{\{N-1\}}{5.5pt}}$}}
\put(130,10){\mbox{$[\overset{{\color{Green} \bar 4}}{\bigcirc}]_{[21]}^{\scaleto{\{2\}}{5.5pt}}$}}
\put(100,57){\vector(1,1){27}}
{\color{Green} 
\put(100,50){\vector(1,-1){27}}}

\put(-5,0){

{\color{blue} \put(150,83){\vector(1,-1){27}}
}
\put(150,22){\vector(1,1){27}}
\put(182,50){\mbox{$[\overset{{\color{blue} \bar 2}}{\bigcirc}\overset{{\color{blue} \bar 2'}}{\bigcirc}]_{[22]}^{\scaleto{\{N+1\}}{5.5pt}}$}}

}
}
}
    
\end{picture}
\caption{\footnotesize The HOMFLY 3-hypercube for the figure-eight knot in the bipartite presentation.}
\label{fig:tw-8-H-complex}
\end{figure}
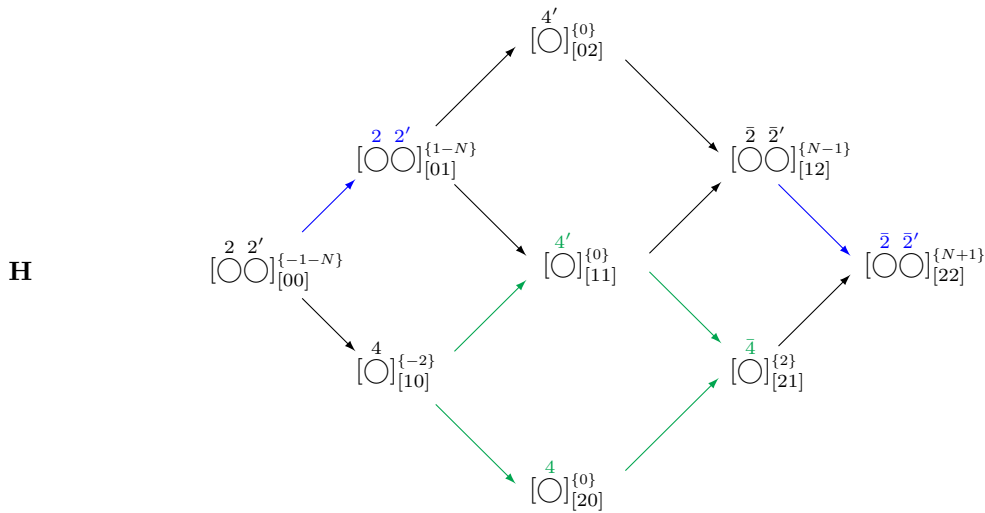

However, this HOMFLY 2-hypercube cannot be lifted to the Khovanov--Rozansky hypercube. Construction~\eqref{bipexp} is not the Euler characteristic of $q$-graded complex. First, the Khovanov--Rozansky hypercube calculus must represent the downgrade polynomial as an alternating sum, see the Jones hypercube calculations in Section~\ref{sec:Jones-2-hyp}. In other words, formula~\eqref{bipexp} lacks minus signs. Second, the $\phi$-multipliers are not monomials in $q$ as it must be to preserve differentials gradings. What we do now, we substitute explicit expressions for $\phi$ and $\bar\phi$ as in Fig.\,\ref{fig:pladeco}\,, i.e. split two summands $\;\horr + \phi \;)\,(\;$ into three ones $\;\horr - q^{N-1}\;)\,( \, +\, q^{N+1}\;)\,(\;$ (and analogously for the mirror lock), see Fig.\,\ref{fig:pladeco-3-hyp}:

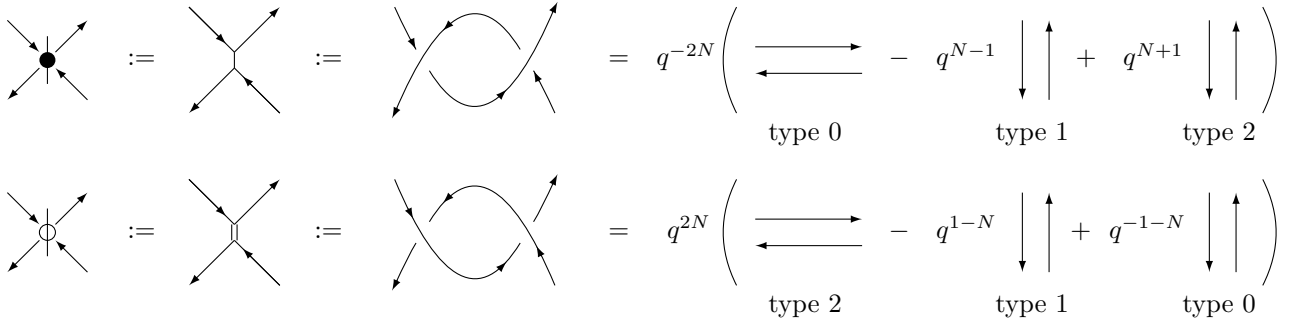
\begin{figure}[t]
\begin{picture}(100,120)(-90,-95)

\put(20,0){
\put(-105,15){\vector(1,-1){12}} \put(-87,3){\vector(1,1){12}}
\put(-93,-3){\vector(-1,-1){12}} \put(-75,-15){\vector(-1,1){12}}
\put(-90,0){\circle*{6}}  \put(-90,-9){\line(0,1){18}}

\put(-60,-2){\mbox{$:=$}}
}

\put(0,0){\put(-17,20){\line(1,-1){17}}\put(-17,20){\vector(1,-1){14}}   \put(0,3){\vector(1,1){17}}
 \put(0,-3){\vector(-1,-1){17}}   \put(17,-20){\line(-1,1){17}} \put(17,-20){\vector(-1,1){14}}
 \put(0,-3){\line(0,1){6}}}

\put(10,0){

\put(20,-2){\mbox{$:=$}}

\qbezier(50,20)(55,9)(58,4) \qbezier(63,-4)(85,-40)(110,20)
\put(56,8){\vector(1,-2){2}} \put(90,-13){\vector(1,1){2}} \put(109,18){\vector(1,2){2}}
\qbezier(50,-20)(75,40)(97,4)  \qbezier(102,-4)(105,-9)(110,-20)
\put(104,-8){\vector(-1,2){2}} \put(70,13){\vector(-1,-1){2}} \put(51,-18){\vector(-1,-2){2}}

}

\put(20,0){
\put(120,-2){\mbox{$=$}}
\put(138,-2){\mbox{$q^{-2N}$}}

\put(5,0){
\qbezier(163,-20)(153,0)(163,20)
\put(70,0){\qbezier(290,-20)(300,0)(290,20)}

\put(100,65){
\put(70,-60){\vector(1,0){40}}
\put(110,-70){\vector(-1,0){40}}
\put(120,-67){\mbox{$-\ \ \ q^{N-1}$}}

\put(-40,0){
\put(210,-50){\vector(0,-1){30}}
\put(220,-80){\vector(0,1){30}}
}

\put(30,0){
\put(210,-50){\vector(0,-1){30}}
\put(220,-80){\vector(0,1){30}}
}

\put(190,-67){\mbox{$+\ \ \ q^{N+1}$}}

\put(75,-95){\mbox{type 0}}
\put(160,-95){\mbox{type 1}}
\put(230,-95){\mbox{type 2}}

}
}
}

\put(0,-65){

\put(20,0){
\put(-105,15){\vector(1,-1){12}} \put(-87,3){\vector(1,1){12}}
\put(-93,-3){\vector(-1,-1){12}} \put(-75,-15){\vector(-1,1){12}}
\put(-90,0){\circle{6}}  \put(-90,-9){\line(0,1){18}}

\put(-60,-2){\mbox{$:=$}}
}

\put(0,0){
\put(0,0){\put(-17,20){\line(1,-1){17}}\put(-17,20){\vector(1,-1){14}}   \put(0,3){\vector(1,1){17}}
 \put(0,-3){\vector(-1,-1){17}}   \put(17,-20){\line(-1,1){17}} \put(17,-20){\vector(-1,1){14}}
\put(-1,-3){\line(0,1){6}} \put(1,-3){\line(0,1){6}}
}

\put(30,-2){\mbox{$:=$}}

\put(10,0){
\qbezier(50,20)(75,-40)(97,-4)  \qbezier(102,4)(105,9)(110,20)
\qbezier(50,-20)(55,-9)(58,-4) \qbezier(63,4)(85,40)(110,-20)
\put(55,9){\vector(1,-2){2}} \put(90,-13){\vector(1,1){2}} \put(109,18){\vector(1,2){2}}
\put(105,-9){\vector(-1,2){2}} \put(70,13){\vector(-1,-1){2}} \put(51,-18){\vector(-1,-2){2}}
}

\put(140,-2){\mbox{$=$}}

\put(25,0){
\put(137,-2){\mbox{$q^{2N}$}}

\put(100,65){
\put(70,-60){\vector(1,0){40}}
\put(110,-70){\vector(-1,0){40}}
\put(120,-67){\mbox{$-\ \ \ q^{1-N}$}}

\put(-40,0){
\put(210,-50){\vector(0,-1){30}}
\put(220,-80){\vector(0,1){30}}
}

\put(30,0){
\put(210,-50){\vector(0,-1){30}}
\put(220,-80){\vector(0,1){30}}
}

\put(188,-67){\mbox{$+ \ \ q^{-1-N}$}}

\put(75,-95){\mbox{type 2}}
\put(160,-95){\mbox{type 1}}
\put(230,-95){\mbox{type 0}}

}}}
\put(25,0){\qbezier(163,-20)(153,0)(163,20)
\put(70,0){\qbezier(290,-20)(300,0)(290,20)}}

}

\end{picture}
\caption{\footnotesize
The rearrangement of the planar decomposition of the positive (in the first line) and negative (in the second line) lock vertices in the \textit{topological} framing.
}\label{fig:pladeco-3-hyp}
\end{figure}

\begin{equation}\label{H-3-hyp}
    H^{\cal L}(A,q) = A^{-2(n_\bullet - n_\circ)} \sum_k \sum_{i=0}^{n_\bullet} \sum_{j=0}^{n_\circ} \sum_{n=0}^i \sum_{m=0}^j \begin{pmatrix}
        i \\ n
    \end{pmatrix}\begin{pmatrix}
        j \\ m
    \end{pmatrix}{\cal N}_{ijk} D_N^k \left(q^{N+1}\right)^{n-m}\left(-q^{N+1}\right)^{i-n-j+m}
\end{equation}
where $\begin{pmatrix}
        i \\ n
    \end{pmatrix} = \frac{i!}{n!(i-n)!}$ are binomial coefficients. Thus, we obtain the sum over vertices  of the peculiar 3-hypercube with $3^{n_\bullet + n_\circ}$ vertices. Examples of 3-hypercubes are shown in Figs.\,\ref{fig:Hopf-H-complex}, \ref{fig:tw-trefoil-H-complex} and \ref{fig:tw-8-H-complex}. Here we label each resolution by a sequence $[\alpha_1\ldots \alpha_n]$ with $\alpha_i = 0,\,1,\,2$ according to Fig.\,\ref{fig:pladeco-3-hyp} and $n$ being the number of bipartite vertices. In curly brackets we now denote the powers of $q$ appearing in the $q$-Euler characteristic. We again order vertices by the increasing height $h=\sum_i \alpha_i$ from left to right, and each resolution $r$ contribute to the HOMFLY polynomial with the $(-1)^{h(r)}$ sign, as it must be according to~\eqref{H-3-hyp}. 

    Note that black arrows can be again associated with the cut-and-join operators because they connect sets of cycles numbers of which differ by one. In other words, either one cycle splits to two ones, or a pair of cycles joins into one. We color in blue arrows which correspond to some new sewing operator which glues two vertices to form $\phi$ or $\bphi$ multipliers.

    It is instructive to demonstrate the calculation of the HOMFLY polynomial by hypercubes of resolution. For the Hopf link, we have\footnote{In what follows, we omit $N$ subscript of $D_N$.}:
    \begin{equation}\label{H-Hopf-deco}
        H^{\,\Hopf} \overset{{\rm Fig.}\,\ref{fig:H-Hopf-complex}}{=} A^{-2}(D^2 + \phi D) \overset{{\rm Fig.}\,\ref{fig:Hopf-H-complex}}{=} A^{-2}(D^2 - q^{N-1} D + q^{N+1} D)\,.
    \end{equation}
    For the trefoil knot, we get:
    \begin{equation}\label{H-trefoil-deco}
        H^{3_1} \overset{{\rm Fig.}\,\ref{fig:H-tw-trefoil-complex}}{=} A^{-4}(D+2\phi D^2 + \phi^2 D) \overset{{\rm Fig.}\,\ref{fig:tw-trefoil-H-complex}}{=} A^{-4}(D - 2 q^{N-1} D^2 + 2 q^{N-1} D^2 + q^{2(N-1)}D - 2 q^{2N} D + q^{2(N+1)}D)\,.
    \end{equation}
    For the figure-eight knot:
    \begin{equation}\label{H-8-deco}
        H^{3_1} \overset{{\rm Fig.}\,\ref{fig:H-tw-4-1-complex}}{=} \bphi D^2 + D + \phi \bphi D + \phi D^2 \overset{{\rm Fig.}\,\ref{fig:tw-8-H-complex}}{=} (q^{-1-N} - q^{1-N}) D^2 + D + (-q^{-2} - q^{2} + 2)D + (q^{N+1} - q^{N-1})D^2\,. 
    \end{equation}
    The two equalities in~\eqref{H-Hopf-deco},~\eqref{H-trefoil-deco} and~\eqref{H-8-deco} are obviously consistent because of $\phi = q^{N+1} - q^{N-1}$ and $\bphi = q^{-N-1} - q^{-N+1}$.



\begin{figure}[h!]
    \begin{equation}\nn
\begin{aligned}
    {\rm\bf H} \ \ \ \ \ \ \ \ \ \ \ \ \ \ \ [\overset{1}{\bigcirc} \overset{1'}{\bigcirc}]_{[0]}^{\scaleto{\{0\}}{5.5pt}} \quad \longrightarrow \quad \boxed{[\overset{2}{\bigcirc}]_{[1]}^{\scaleto{\{N-1\}}{5.5pt}} \quad {\color{Green} \longrightarrow} \quad [\overset{{\color{Green} 2}}{\bigcirc}]_{[2]}^{\scaleto{\{N+1\}}{5.5pt}}}_{\,\phi} 
\end{aligned}
\end{equation}
    \caption{\footnotesize The bipartite 3-hypercube for the Hopf link. We color in green the arrow for the sewing operator. The boxed elements connected by this blue arrow are shrinked when transferring to the 2-hypercube in Fig.\,\ref{fig:H-Hopf-complex}.} 
    \label{fig:Hopf-H-complex}
\end{figure}

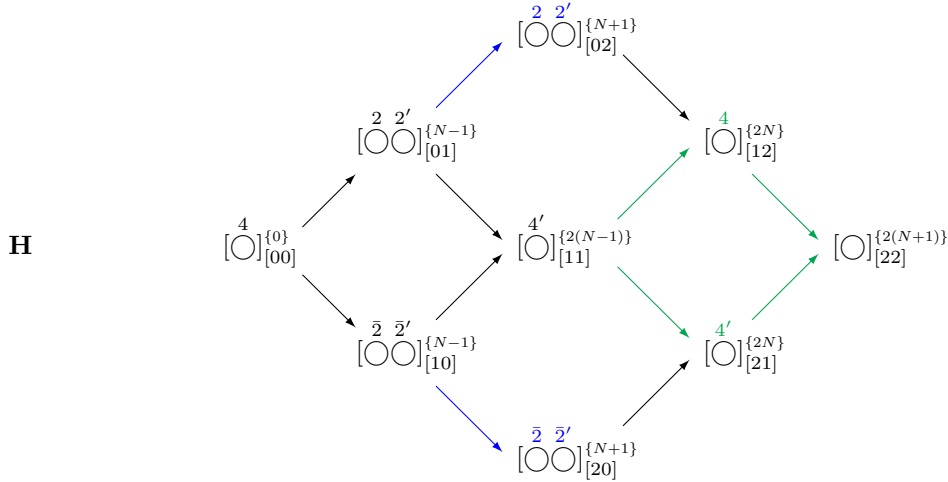
\begin{figure}[h!]
\centering
\begin{picture}(100,180)(40,-35)

\put(-140,50){\mbox{\bf H}}

\put(5,0){
\put(-65,50){\mbox{$[\overset{4}{\bigcirc}]_{[00]}^{\scaleto{\{0\}}{5.5pt}}$}}
\put(-35,60){\vector(1,1){20}}
\put(-35,42){\vector(1,-1){20}}
}

\put(-10,90){\mbox{$[\overset{2}{\bigcirc}\overset{2'}{\bigcirc}]_{[01]}^{\scaleto{\{N-1\}}{5.5pt}}$}}
\put(-10,10){\mbox{$[\overset{\bar 2}{\bigcirc}\overset{\bar 2'}{\bigcirc}]_{[10]}^{\scaleto{\{N-1\}}{5.5pt}}$}}

\put(10,0){

\put(40,130){\mbox{$[\overset{{\color{blue} 2}}{\bigcirc}\overset{{\color{blue} 2'}}{\bigcirc}]_{[02]}^{\scaleto{\{N+1\}}{5.5pt}}$}}
\put(40,-30){\mbox{$[\overset{{\color{blue} \bar 2}}{\bigcirc}\overset{{\color{blue} \bar 2'}}{\bigcirc}]_{[20]}^{\scaleto{\{N+1\}}{5.5pt}}$}}
{\color{blue} 
\put(10,105){\vector(1,1){25}}
\put(10,0){\vector(1,-1){25}}}

\put(40,50){\mbox{$[\overset{4'}{\bigcirc}]_{[11]}^{\scaleto{\{2(N-1)\}}{5.5pt}}$}}
\put(10,80){\vector(1,-1){25}}
\put(10,25){\vector(1,1){25}}

\put(10,0){

\put(70,125){\vector(1,-1){25}}
\put(70,-15){\vector(1,1){25}}
\put(100,90){\mbox{$[\overset{{\color{Green} 4}}{\bigcirc}]_{[12]}^{\scaleto{\{2N\}}{5.5pt}}$}}
\put(100,10){\mbox{$[\overset{{\color{Green} 4'}}{\bigcirc}]_{[21]}^{\scaleto{\{2N\}}{5.5pt}}$}}
{\color{Green} 
\put(68,63){\vector(1,1){27}}
\put(68,45){\vector(1,-1){27}}}

\put(-35,0){

{\color{Green} \put(150,80){\vector(1,-1){25}}
\put(150,25){\vector(1,1){25}}}
\put(180,50){\mbox{$[\bigcirc]_{[22]}^{\scaleto{\{2(N+1)\}}{5.5pt}}$}}

}
}
}
    
\end{picture}
\caption{\footnotesize The HOMFLY 3-hypercube for the trefoil knot in the bipartite presentation. We color in blue and green the arrows for the sewing operators and the corresponding labellings of spaces. Green arrows will be associated with the zero morphisms, see Section~\ref{sec:method}. Spaces are enumerated by $[\alpha_1\, \alpha_2]$ with $\alpha_1=0,\,1,\,2$ corresponding to smoothings shown in Fig.\,\ref{fig:pladeco-3-hyp}.}
\label{fig:tw-trefoil-H-complex}
\end{figure}

\subsection{Duality with precursor Jones polynomial}\label{sec:H-J-duality} 

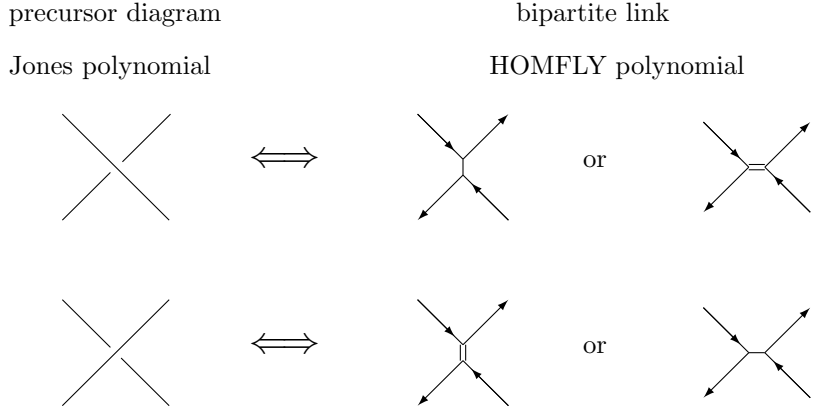
\begin{figure}[h!]
    \centering
\begin{picture}(300,170)(-60,-85)

\put(-80,75){\mbox{precursor diagram}}

\put(110,75){\mbox{bipartite link}}

\put(-80,55){\mbox{Jones polynomial}}

\put(100,55){\mbox{HOMFLY polynomial}}

\put(0,20){

\put(-60,-20){\line(1,1){18}}
\put(-20,-20){\line(-1,1){40}}
\put(-37.5,2){\line(1,1){18}}

}

\put(0,-50){

\put(-60,-20){\line(1,1){40}}
\put(-20,-20){\line(-1,1){18}}
\put(-60,20){\line(1,-1){18}}

}

\put(10,20){\mbox{\Large $\Longleftrightarrow$}}

\put(10,-50){\mbox{\Large $\Longleftrightarrow$}}

\put(90,20){

\put(-17,20){\line(1,-1){17}}\put(-17,20){\vector(1,-1){14}}   \put(0,3){\vector(1,1){17}}
\put(0,-3){\vector(-1,-1){17}}   \put(17,-20){\line(-1,1){17}} \put(17,-20){\vector(-1,1){14}}
\put(0,-3){\line(0,1){6}}

}

\put(90,-50){

\put(-17,20){\line(1,-1){17}}\put(-17,20){\vector(1,-1){14}}   \put(0,3){\vector(1,1){17}}
\put(0,-3){\vector(-1,-1){17}}   \put(17,-20){\line(-1,1){17}} \put(17,-20){\vector(-1,1){14}}
\put(-1,-3){\line(0,1){6}}
\put(1,-3){\line(0,1){6}}

}

\put(135,20){\mbox{or}}

\put(135,-50){\mbox{or}}

\put(200,20){

 \put(-20,17){\line(1,-1){17}}\put(-20,17){\vector(1,-1){14}}   \put(3,0){\vector(1,1){17}}
 \put(-3,0){\vector(-1,-1){17}}   \put(20,-17){\line(-1,1){17}} \put(20,-17){\vector(-1,1){14}}
 \put(-3,1){\line(1,0){6}}
  \put(-3,-1){\line(1,0){6}}
 
 }

 \put(200,-50){

 \put(-20,17){\line(1,-1){17}}\put(-20,17){\vector(1,-1){14}}   \put(3,0){\vector(1,1){17}}
 \put(-3,0){\vector(-1,-1){17}}   \put(20,-17){\line(-1,1){17}} \put(20,-17){\vector(-1,1){14}}
 \put(-3,0){\line(1,0){6}}
 
 }
    
\end{picture}
    \caption{\footnotesize If in the HOMFLY expansion in $\phi$, $\bphi$, $D_N$, one performs the change $\phi \rightarrow -q$, $\bphi \rightarrow -q^{-1}$, $D_N \rightarrow D_2$, the HOMFLY polynomial transforms to the Jones polynomial for the precursor diagram due to the equivalence of the planar expansion of bipartite vertices (see Fig.\,\ref{fig:pladeco}) and the Kauffmann bracket for single vertices (see Fig.\,\ref{fig:Kauff}).}
    \label{fig:precursor-bip}
\end{figure}

A precursor diagram is obtained from a bipartite link by shrinking lock vertices into single ones according to Fig.\,\ref{fig:precursor-bip}. Comparing Figs.\,\ref{fig:Kauff} and \ref{fig:pladeco}, one can then observe~\cite{ALM} that the HOMFLY polynomial for a bipartite link becomes the Jones polynomial for a corresponding precursor diagram under the substitution $\phi \rightarrow -q$, $\bphi \rightarrow -q^{-1}$, $D_N \rightarrow D_2$. Thus, a bipartite HOMFLY 2-hypercube transforms to the precursor Jones 2-hypercube under the same change. The same duality actually holds for bipartite Khovanov--Rozansky and precursor Khovanov polynomials, see Section~\ref{sec:precursors} and examples in Section~\ref{sec:examples}. Emphasize, that this downgrade holds for $2$-hypercubes, not for $3$-hypercubes, and for $N=2$, not for other values of $N$.

\newpage

\section{Truncation of hypercubes for $N=2$}\label{sec:trunc-hyp}

For the Khovanov polynomial $(N=2)$, one can explicitly observe a reduction from a $2^{2n_\circ + 2n_\bullet}$ hypercube to a $3^{n_\circ + n_\bullet}$ hypercube, what we will demonstrate in a separate text. Such a phenomenon can be shown already at the level of the Jones polynomial, see subsections below.   

\subsection{1-lock $\longrightarrow$ unknot}

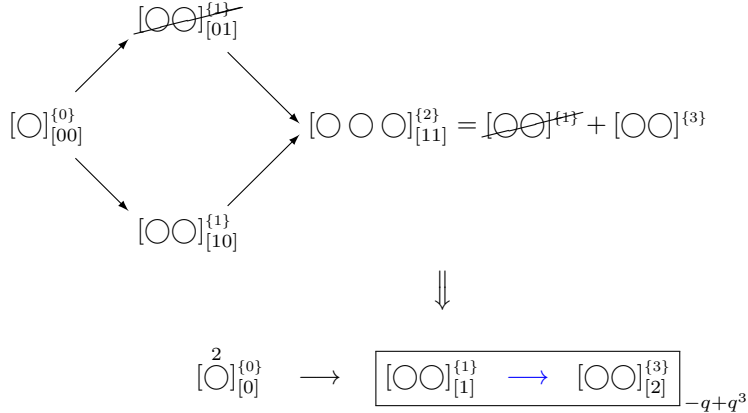
\begin{figure}[h!]
\centering
\begin{picture}(100,150)(50,-50)


\put(-60,50){\mbox{$[\bigcirc]_{[00]}^{\scaleto{\{0\}}{5.5pt}}$}}
\put(-35,65){\vector(1,1){20}}
\put(-35,40){\vector(1,-1){20}}

\put(-12,90){\mbox{$\cancel{[\bigcirc\bigcirc]_{[01]}^{\scaleto{\{1\}}{5.5pt}}}$}}
\put(-12,10){\mbox{$[\bigcirc\bigcirc]_{[10]}^{\scaleto{\{1\}}{5.5pt}}$}}

\put(-10,0){

\put(62,50){\mbox{$[\bigcirc\bigcirc\bigcirc]_{[11]}^{\scaleto{\{2\}}{5.5pt}}=\cancel{[\bigcirc\bigcirc]^{\scaleto{\{1\}}{5.5pt}}}+[\bigcirc\bigcirc]^{\scaleto{\{3\}}{5.5pt}}$}}
\put(32,80){\vector(1,-1){27}}
\put(32,22){\vector(1,1){27}}
}

\put(0,-45){\mbox{$\quad[\overset{2}{\bigcirc}]_{[0]}^{\scaleto{\{0\}}{5.5pt}} \quad \longrightarrow \quad \boxed{[\bigcirc\bigcirc]_{[1]}^{\scaleto{\{1\}}{5.5pt}} \quad {\color{blue} \longrightarrow} \quad  [\bigcirc\bigcirc]_{[2]}^{\scaleto{\{3\}}{5.5pt}}}_{\,-q+q^3}$}}

\put(100,0){\mbox{$\vertLra$}}
    
\end{picture}
\caption{\footnotesize Truncation of the Jones $2^2$-hypercube of resolutions to 3-hypercube for 2-unknot. One cycle $\bigcirc$ giving contribution $D_2$ can be expanded as $D_2 = q + q^{-1}$, so that $\bigcirc^{\scaleto{\{0\}}{5.5pt}} = \varnothing^{\scaleto{\{-1\}}{5.5pt}} + \varnothing^{\scaleto{\{+1\}}{5.5pt}}$ in our cycle notation. Here, this substitution is made in the right corner. After that, one of the middle pair of cycles added to the right one vanishes because middle cycles carry the minus sign.}
\label{fig:2-unknot-hypercube}
\end{figure}

For the 1-lock unknot (2-unknot), we have $2^2$ hypercube, see Fig.\,\ref{fig:2-unknot-hypercube}. {\bf Kauffman calculation for two vertices and $N=2$} gives:
\be
J^{\text{2-unknot}}=\frac{1}{q^4}\left(D_2 -2q D_2^2 + q^2 D_2^3\right) = \frac{D_2}{q^4}\Big(1-2q(q+q^{-1}) + q^2(q+q^{-1})^2\Big)=D_2
\ee
where $D_2 =D_{N=2} = q+q^{-1}$.
Of interest for us will be reordering of the r.h.s.:
\ba
J^{\text{2-unknot}}&=\frac{D_2}{q^4}\left(
\begin{array}{rcl}
1&- q(q+q^{-1}) &+q^3(q+q^{-1}) \\
\hline
& - q(q+q^{-1})& + q(q+q^{-1})
\end{array} \right)
= \\ 
&=\frac{D_2}{q^4}\Big(1- q(q+q^{-1})\underbrace{- q(q+q^{-1}) + q(q+q^{-1})}_0 +q^3(q+q^{-1})\Big)\,.
\label{reordunknot1}
\ea
It can be considered as the truncation of the $2^2$ hypercube to $3^1$ hypercube, see Fig.\ref{fig:2-unknot-hypercube},
and compared with the {\bf bipartite calculus}:

\be
H^{\text{2-unknot}} = \frac{D_N}{q^{2N}}\Big(1 + (-q^{N-1}+q^{N+1})D_N\Big) = D_N\,.
\ee
The first underlined row at the l.h.s. of (\ref{reordunknot1}) is exactly this expression at $N=2$,
while and in the second line we have a cancelation. 




\subsection{2-lock $\longrightarrow$ unknot}

\begin{figure}[h!]
\begin{picture}(300,480)(-20,-360)

\put(0,0){\mbox{$[\bigcirc]_{[0000]}^{\scaleto{\{0\}}{5.5pt}}$}}

\put(50,20){\mbox{$[\bigcirc\bigcirc]_{[0100]}^{\scaleto{\{1\}}{5.5pt}}$}}

\put(50,60){\mbox{$\cancel{[\bigcirc\bigcirc]_{[0001]}^{\scaleto{\{1\}}{5.5pt}}}$}}

\put(50,-20){\mbox{$[\bigcirc\bigcirc]_{[0010]}^{\scaleto{\{1\}}{5.5pt}}$}}

\put(50,-60){\mbox{$\cancel{[\bigcirc\bigcirc]_{[1000]}^{\scaleto{\{1\}}{5.5pt}}}$}}

\put(120,0){\mbox{$\cancel{[\bigcirc\bigcirc\bigcirc]_{[1001]}^{\scaleto{\{2\}}{5.5pt}}}$}}

\put(120,60){\mbox{$\equalto{[\bigcirc\bigcirc\bigcirc]_{[0101]}^{\scaleto{\{2\}}{5.5pt}}}{\cancel{[\bigcirc\bigcirc]^{\scaleto{\{1\}}{5pt}}}+[\bigcirc\bigcirc]^{\scaleto{\{3\}}{5pt}}}$}}

\put(120,100){\mbox{$\cancel{[\bigcirc\bigcirc\bigcirc]_{[0011]}^{\scaleto{\{2\}}{5.5pt}}}$}}

\put(120,-40){\mbox{$[\bigcirc\bigcirc\bigcirc]_{[0110]}^{\scaleto{\{2\}}{5.5pt}}$}}

\put(123,-80){\mbox{$\cancel{[\bigcirc\bigcirc\bigcirc]_{[1100]}^{\scaleto{\{2\}}{5.5pt}}}$}}

\put(120,-120){\mbox{$\equalto{[\bigcirc\bigcirc\bigcirc]_{[1010]}^{\scaleto{\{2\}}{5.5pt}}}{\cancel{[\bigcirc\bigcirc]^{\scaleto{\{1\}}{5pt}}}+[\bigcirc\bigcirc]^{\scaleto{\{3\}}{5pt}}}$}}

\put(200,35){\mbox{$\equalto{[\bigcirc\bigcirc\bigcirc\bigcirc]_{[1011]}^{\scaleto{\{3\}}{5.5pt}}}{\cancel{[\bigcirc\bigcirc\bigcirc]^{\scaleto{\{2\}}{5pt}}}+\cancel{[\bigcirc\bigcirc\bigcirc]^{\scaleto{\{4\}}{5pt}}}}$}}

\put(200,100){\mbox{$\equalto{[\bigcirc\bigcirc\bigcirc\bigcirc]_{[0111]}^{\scaleto{\{3\}}{5.5pt}}}{\cancel{[\bigcirc\bigcirc\bigcirc]^{\scaleto{\{2\}}{5pt}}}+[\bigcirc\bigcirc\bigcirc]^{\scaleto{\{4\}}{5pt}}}$}}

\put(200,-35){\mbox{$\equalto{[\bigcirc\bigcirc\bigcirc\bigcirc]_{[1101]}^{\scaleto{\{3\}}{5.5pt}}}{\cancel{[\bigcirc\bigcirc\bigcirc]^{\scaleto{\{2\}}{5pt}}}+\cancel{[\bigcirc\bigcirc\bigcirc]^{\scaleto{\{4\}}{5pt}}}}$}}

\put(200,-100){\mbox{$\equalto{[\bigcirc\bigcirc\bigcirc\bigcirc]_{[1110]}^{\scaleto{\{3\}}{5.5pt}}}{\cancel{[\bigcirc\bigcirc\bigcirc]^{\scaleto{\{2\}}{5pt}}}+[\bigcirc\bigcirc\bigcirc]^{\scaleto{\{4\}}{5pt}}}$}}

\put(320,0){\mbox{$\equalto{[\bigcirc\bigcirc\bigcirc\bigcirc\bigcirc]_{[1111]}^{\scaleto{\{4\}}{5.5pt}}}{\cancel{[\bigcirc\bigcirc\bigcirc]^{\scaleto{\{2\}}{5pt}}}+\cancel{2[\bigcirc\bigcirc\bigcirc]^{\scaleto{\{4\}}{5pt}}}+[\bigcirc\bigcirc\bigcirc]^{\scaleto{\{6\}}{5pt}}}$}}

\put(20,12){\vector(1,1){40}}

\put(20,-9){\vector(1,-1){40}}

\put(90,70){\vector(1,1){25}}

\put(100,60){\vector(1,0){20}}

\put(90,50){\vector(1,-1.5){27}}

\put(85,10){\vector(1,-2){40}}

\put(90,-10){\vector(1,2){50}}

\put(80,-70){\vector(1,-1){40}}

\put(100,-65){\vector(2,-1){20}}

\put(83,-47){\vector(1,1){40}}

\put(185,100){\vector(1,0){20}}

\put(175,90){\vector(1,-1){45}}

\put(180,50){\vector(1,-2){37}}

\put(170,12){\vector(1.5,1){33}}

\put(170,-10){\vector(1.5,-1){33}}

\put(185,-85){\vector(2,-1){20}}

\put(176,-70){\vector(1,1){30}}

\put(180,-105){\vector(1,3){44}}

\put(282,35){\vector(2,-1){60}}

\put(282,-33){\vector(2,1){60}}

\put(200,-150){\mbox{$\vertLra$}}

\put(130,-320){

\put(5,0){
\put(-65,50){\mbox{$[\bigcirc]_{[00]}^{\scaleto{\{0\}}{5.5pt}}$}}
\put(-35,60){\vector(1,1){20}}
\put(-35,42){\vector(1,-1){20}}
}

\put(-10,90){\mbox{$[\bigcirc\bigcirc]_{[01]}^{\scaleto{\{1\}}{5.5pt}}$}}
\put(-10,10){\mbox{$[\bigcirc\bigcirc]_{[10]}^{\scaleto{\{1\}}{5.5pt}}$}}

\put(10,0){

\put(43,130){\mbox{$[\bigcirc\bigcirc]_{[02]}^{\scaleto{\{3\}}{5.5pt}}$}}
\put(43,-30){\mbox{$[\bigcirc\bigcirc]_{[20]}^{\scaleto{\{3\}}{5.5pt}}$}}
{\color{blue} 
\put(10,105){\vector(1,1){25}}
\put(10,0){\vector(1,-1){25}}}

\put(40,50){\mbox{$[\bigcirc\bigcirc\bigcirc]_{[11]}^{\scaleto{\{2\}}{5.5pt}}$}}
\put(10,80){\vector(1,-1){25}}
\put(10,25){\vector(1,1){25}}

\put(20,0){

\put(70,125){\vector(1,-1){25}}
\put(70,-15){\vector(1,1){25}}
\put(100,90){\mbox{$[\bigcirc\bigcirc\bigcirc]_{[12]}^{\scaleto{\{4\}}{5.5pt}}$}}
\put(100,10){\mbox{$[\bigcirc\bigcirc\bigcirc]_{[21]}^{\scaleto{\{4\}}{5.5pt}}$}}
{\color{blue} 
\put(68,63){\vector(1,1){27}}
\put(69,40){\vector(1,-1){25}}}

\put(-30,0){

{\color{blue} \put(150,80){\vector(1,-1){25}}
\put(150,25){\vector(1,1){25}}}
\put(180,50){\mbox{$[\bigcirc\bigcirc\bigcirc]_{[22]}^{\scaleto{\{6\}}{5.5pt}}$}}

}
}
}

}

\thicklines

\put(25,10){\vector(2,1){20}}

\put(25,-8){\vector(2,-1){20}}

{\color{blue}
\put(90,30){\vector(1.5,1){30}}
\put(90,-30){\vector(1,-2){40}}
\put(180,-30){\vector(1,4){30}}
\put(180,-48){\vector(1,-1.5){27}}
\put(280,90){\vector(1,-1){80}}
\put(280,-93){\vector(1,1){60}}
}

\put(90,10){\vector(1,-1.5){27}}

\put(90,-28){\vector(3,-1){25}}

\put(180,70){\vector(1,1){25}}

\put(180,-110){\vector(2,1){23}}
    
\end{picture}
    \caption{\footnotesize Jones $2^4$-hypercube for the 4-unknot. It can be reduced to $3^2$-hypercube. In the above picture, the edges remaining after the reduction are thick. New enumerators $[\alpha_1^{(3)} \alpha_2^{(3)}]$ are expressed through old ones $[\alpha_1^{(2)}\alpha_2^{(2)}\alpha_3^{(2)}\alpha_4^{(2)}]$ in the following way: $\alpha_1^{(3)}=\alpha_1^{(2)}+\alpha_3^{(2)}$, $\alpha_2^{(3)}=\alpha_2^{(2)}+\alpha_4^{(2)}$. In this example, we again perform expansions $\bigcirc^{\scaleto{\{0\}}{5.5pt}} = \varnothing^{\scaleto{\{-1\}}{5.5pt}} + \varnothing^{\scaleto{\{+1\}}{5.5pt}}$ meaning $D_2 = q + q^{-1}$ substitution in terms of the Jones contribution. Then, some neighboring sets of cycles vanish and leave the $3$-hypercube as the result.}
    \label{fig:4-unknot-hypercube}
\end{figure}
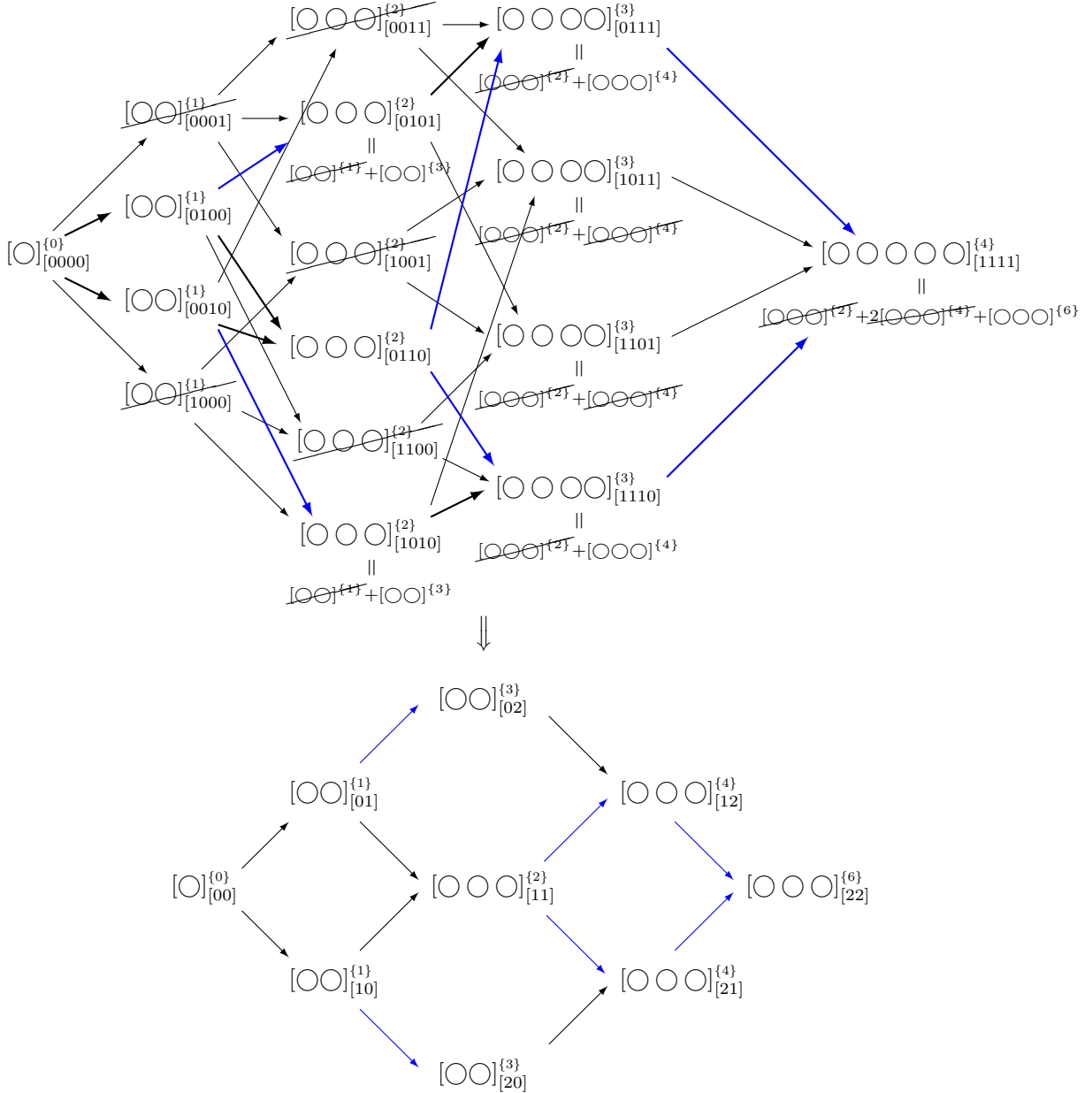


For the 2-lock unknot (4-unknot), we have $2^4$ hypercube, see Fig.\,\ref{fig:4-unknot-hypercube}. {\bf Kauffman calculation for four vertices and $N=2$} gives:
\ba
J^{\text{4-unknot}}&=\frac{D_2 -4qD_2^2 +6q^2D_2^3-4q^3D_2^4+ q^4D_2^5}{q^8}
= \\ 
&=\frac{D_2}{q^8}\Big(1-4q(q+q^{-1})+6q^2(q+q^{-1})^2-4q^3(q+q^{-1})^3 + q^4(q+q^{-1})^4\Big)=D_2\,.
\ea
It is a sum over $16$ vertices of a $2^4$ hypercube.
Of interest for is a reordering:
\be
J^{\text{4-unknot}}=\frac{D_2}{q^8}\Big(1-4q(q+q^{-1})+6q^2(q+q^{-1})^2-4q^3(q+q^{-1})^3 + q^4(q+q^{-1})^4\Big)=
\ \ \ \ \ \ \ \ \ \ \ \ \ \ \ \ \ \
\nn \\
=\frac{D_2}{q^8}\left(
\begin{array}{rcccc}
1&-2q(q+q^{-1})&+2q^3(q+q^{-1}) &-2q^4(q+q^{-1})^2 &+(q^2 + q^6)(q+q^{-1})^2 \\
\hline
 &-2q(q+q^{-1})& +2q(q+q^{-1})   &     &  \\
 && + 4q^2(q+q^{-1})^2 &-4q^2(q+q^{-1})^2& \\
 &&& -2q^4(q+q^{-1})^2  & + 2q^4(q+q^{-1})^2
\end{array}
\right)=D_2\,.
\ee
where in the last three lines we observe full cancellations.
This can be considered as truncation of the $2^4$ hypercube to $3^2$ hypercube, see Fig.\ref{fig:4-unknot-hypercube}.
The surviving first line now coincides with the {\bf bipartite calculus}:
\be
H^{\text{4-unknot}}=\frac{D_N}{q^{4N}}\Big(1 + 2(-q^{N-1}+q^{N+1})D_N+  (q^{2(N-1)}-2q^{2N}+q^{2(N+1)})D_N^2\Big)   = D_N\,.
\ee

\subsection{Another chiral 2-lock $\longrightarrow$ trefoil $3_1$}


For the 2-lock trefoil (see Fig.\,\ref{fig:two-bip-vert}), we have $2^4$ hypercube initially. {\bf Kauffman calculation for four vertices and $N=2$} gives:
\ba
J^{3_1} &= \frac{D_2 -4qD_2^2 +q^2(4D_2+2D_2^3)-4q^3D_2^2+ q^4D_2^3}{q^8}
= \\
&= \frac{D_2}{q^8}\Big(1-4q(q+q^{-1})+2q^2\Big(2+(q+q^{-1})^2\Big)-4q^3(q+q^{-1}) + q^4(q+q^{-1})^2\Big)
= q^6+q^2-1\,.
\ea
Let us {\bf reorder} this sum expanding some $D_2 = q + q^{-1}$:
\be
J^{3_1}=\frac{D_2}{q^8}\Big(1-4q(q+q^{-1})+2q^2\Big(2+(q+q^{-1})^2\Big)-4q^3(q+q^{-1}) + q^4(q+q^{-1})^2\Big)=
\nn \\
=\frac{D_2}{q^8}\left(
\begin{array}{rcccc}
1&-2q(q+q^{-1})&&+2q^3(q+q^{-1}) &+ (q^2 -2q^4+ q^6)  \\
\hline
 &-2q(q+q^{-1})&+2q(q+q^{-1})    \\
 && +2q^3(q+q^{-1})& -2 q^3(q+q^{-1})  \\
 &&+4q^2& -4 q^3(q+q^{-1})&+4q^4
\end{array}
\right).
 \ \
\ee
The last three lines vanish, and the resulting sum is a sum over 3-hypercube which is in correspondence with the {\bf bipartite calculus}, see Fig.\,\ref{fig:tw-trefoil-H-complex}\,:
\be
H^{3_1} = \frac{D_N}{q^{4N}}\Big(1 + 2(-q^{N-1}+q^{N+1})D_N+  (q^{2(N-1)}-2q^{2N}+q^{2(N+1)}) \Big)\,.
\ee

\subsection{Non-chiral 2-lock $\longrightarrow$ figure-eight knot $4_1$}


\noindent We again have the $2^4$ Jones hypercube for the $4_1$ knot, see Fig.\,\ref{fig:two-bip-vert}. {\bf Kauffman calculation for four vertices and $N=2$} gives:
\be
J^{4_1} = D_2\left( 5 - 4qD_2 - 4q^{-1}D_2 + q^2 D_2^2 + q^{-2}D_2^2 + D_2^2\right)\,.
\ee
\noindent One can {\bf reorder} the r.h.s. so that the remaining terms form the sum over 3-hypercube:
\be
J^{4_1} = D_2\left(
\begin{array}{rccccc}
	3 & -qD_2-q^2 & -q^{-1}D_2-q^{-2} & +q^3D_2 & +q^{-3}D_2 &   \\
	\hline
	2 & -1 & -1    \\
	& -2qD_2 & & +qD_2 & & +qD_2  \\
	&& -2q^{-1}D_2 & & +q^{-1}D_2 & +q^{-1}D_2
\end{array}
\right).
\ \
\ee
\noindent Here the last three lines vanish. This result is in correspondence with {\bf bipartite calculus} for the HOMFLY polynomial, see Fig.\,\ref{fig:tw-8-H-complex}\,:
\be
H^{4_1} = D_N\left(1+(-q^{N-1}+q^{N+1})(-q^{-N+1}+q^{-N-1})+(-q^{N-1}+q^{N+1}-q^{-N+1}+q^{-N-1})D_N\right)\,.
\ee

\setcounter{equation}{0}

\section{Khovanov--Rozansky polynomial for bipartite links}
\label{sec:method}



In this section, we categorify the 3-hypercube HOMFLY construction to the Khovanov--Rozansky polynomial. In Section~\eqref{sec:H-hypercubes}, we have already explained the introduction of the HOMFLY 3-hypercube. The formula~\eqref{H-3-hyp} (analogously to~\eqref{J-2-hyp-qdim}) can be rewritten in terms of quantum dimensions of $N$-dimensional spaces: 
\begin{equation}\label{H-3-hyp-qdim}
    H^{\cal L}(A,q) = A^{-2(n_\bullet - n_\circ)} \sum_k \sum_{i=0}^{n_\bullet} \sum_{j=0}^{n_\circ} \sum_{n=0}^i \sum_{m=0}^j \begin{pmatrix}
        i \\ n
    \end{pmatrix}\begin{pmatrix}
        j \\ m
    \end{pmatrix}{\cal N}_{ijk} \dim_q(V^{\smallotimes k}) \left(q^{N+1}\right)^{n-m}\left(-q^{N+1}\right)^{i-n-j+m}\,,
\end{equation}
and the HOMFLY polynomial is the $q$-Euler characteristic of a graded complex which is described below. After that, the $T$-deformation is similar to the construction in Section~\ref{sec:Khovanov}. The method for computation of the Khovanov--Rozansky polynomial consists of the following steps.

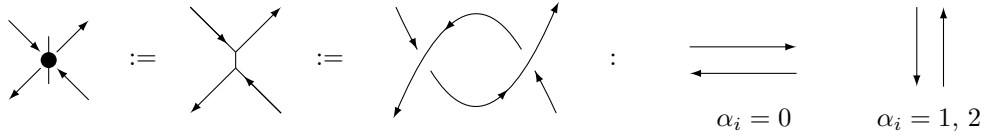
\begin{figure}[h!]
\begin{picture}(100,60)(-150,-30)

\put(20,0){
\put(-105,15){\vector(1,-1){12}} \put(-87,3){\vector(1,1){12}}
\put(-93,-3){\vector(-1,-1){12}} \put(-75,-15){\vector(-1,1){12}}
\put(-90,0){\circle*{6}}  \put(-90,-9){\line(0,1){18}}

\put(-60,-2){\mbox{$:=$}}
}

\put(0,0){\put(-17,20){\line(1,-1){17}}\put(-17,20){\vector(1,-1){14}}   \put(0,3){\vector(1,1){17}}
 \put(0,-3){\vector(-1,-1){17}}   \put(17,-20){\line(-1,1){17}} \put(17,-20){\vector(-1,1){14}}
 \put(0,-3){\line(0,1){6}}}

\put(10,0){

\put(20,-2){\mbox{$:=$}}

\qbezier(50,20)(55,9)(58,4) \qbezier(63,-4)(85,-40)(110,20)
\put(56,8){\vector(1,-2){2}} \put(90,-13){\vector(1,1){2}} \put(109,18){\vector(1,2){2}}
\qbezier(50,-20)(75,40)(97,4)  \qbezier(102,-4)(105,-9)(110,-20)
\put(104,-8){\vector(-1,2){2}} \put(70,13){\vector(-1,-1){2}} \put(51,-18){\vector(-1,-2){2}}

}

\put(0,0){
\put(140,-2){\mbox{$:$}}

\put(100,65){
\put(70,-60){\vector(1,0){40}}
\put(110,-70){\vector(-1,0){40}}
\put(80,-90){\mbox{$\alpha_i = 0$}}

\put(140,-90){\mbox{$\alpha_i = 1,\,2$}}
\put(155,-45){\vector(0,-1){30}}
\put(165,-75){\vector(0,1){30}}
}
}

\end{picture}
\caption{\footnotesize
The smoothings of the lock vertex.
}\label{fig:bip-vertex-smoothings}
\end{figure}

\begin{figure}[h!]
\begin{picture}(100,50)(-150,-20)

\put(20,0){
\put(-105,15){\vector(1,-1){12}} \put(-87,3){\vector(1,1){12}}
\put(-93,-3){\vector(-1,-1){12}} \put(-75,-15){\vector(-1,1){12}}
\put(-90,0){\circle{6}}  \put(-90,-9){\line(0,1){18}}

\put(-60,-2){\mbox{$:=$}}
}

\put(0,0){
\put(0,0){\put(-17,20){\line(1,-1){17}}\put(-17,20){\vector(1,-1){14}}   \put(0,3){\vector(1,1){17}}
 \put(0,-3){\vector(-1,-1){17}}   \put(17,-20){\line(-1,1){17}} \put(17,-20){\vector(-1,1){14}}
\put(-1,-3){\line(0,1){6}} \put(1,-3){\line(0,1){6}}
}

\put(30,-2){\mbox{$:=$}}

\put(10,0){

\qbezier(50,20)(75,-40)(97,-4)  \qbezier(102,4)(105,9)(110,20)
\qbezier(50,-20)(55,-9)(58,-4) \qbezier(63,4)(85,40)(110,-20)
\put(55,9){\vector(1,-2){2}} \put(90,-13){\vector(1,1){2}} \put(109,18){\vector(1,2){2}}
\put(105,-9){\vector(-1,2){2}} \put(70,13){\vector(-1,-1){2}} \put(51,-18){\vector(-1,-2){2}}
}

\put(140,-2){\mbox{$:$}}

\put(100,65){
\put(70,-60){\vector(1,0){40}}
\put(110,-70){\vector(-1,0){40}}
\put(80,-90){\mbox{$\alpha_i = 2$}}
\put(140,-90){\mbox{$\alpha_i = 0,\,1$}} \put(-25,0){
\put(180,-45){\vector(0,-1){30}}
\put(190,-75){\vector(0,1){30}}
}
}}

\end{picture}
\caption{\footnotesize The smothings of the "opposite" lock denoted by a double segment.
It is made from inverse vertices.
} \label{fig:oppvert}
\end{figure}
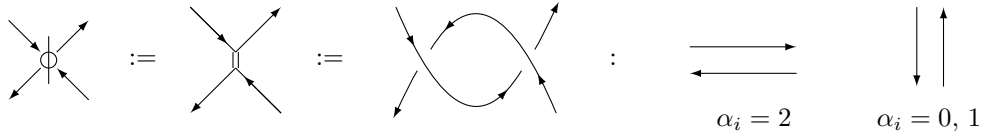

\paragraph{Step 1: 3-hypercube.} Draw the 3-hypercube consisting of $3^n$ vertices where $n$ is the number of AP locks in a link, as described in Section~\ref{sec:H-hypercubes}. Here we shortly repeat the construction. Each vertex of the hypercube is enumerated by a sequence of $n$ numbers $[\alpha_1,\,\alpha_2,\dots,\,\alpha_n]$ taking values in the following set: $\alpha_i\in\{0,\,1,\,2\}$. Vertices are ordered by increasing the height $h=|\alpha|=\sum_i \alpha_i$ from left to right. Each vertex of the hypercube is a complete smoothing of an initial link. The correspondence between $\alpha_i$ and the smoothings is shown in Figs.\,\ref{fig:bip-vertex-smoothings},\,\ref{fig:oppvert}.  

\paragraph{Step 2: spaces.} Each vertex of the hypercube is a disjoint sum of cycles. Each cycle is equipped with $N$-dimensional Grassmann graded vector space $V=\langle \vartheta_{i} \rangle$, $i=1,\dots,\,N$. A collection of $k$ circles corresponds to the tensor product of the vector spaces $V^{\smallotimes k}$. The basis vectors have the gradings ${\rm grad} (\vartheta_{i}) = q^{N-2i+1}$. We label each space with the number of edges of the corresponding cycle.

\paragraph{Step 3: differentials.} The maps act between spaces connected by one flip between the smoothings, i.e. the difference of labels is exactly between one $\alpha_i$ and equal to $1$. The morphisms are of four types -- the zero morphism and
{\small \begin{equation}\label{differentials}
\begin{aligned}
    \Delta &= \sum_{i=1}^N\left(\sum_{j=0}^{N-i}\vartheta^a_{N-j}\vartheta^b_{i+j}\right)\frac{\partial}{\partial \vartheta^c_i}\,:\quad V_c \; \mapsto \; V_a \otimes V_b \\
    m &= 
\sum_{\substack{1\le i,j\le N\\i+j \le N+1}}\vartheta_{i+j-1}^a\frac{\partial^2}{\partial\vartheta_i^b\partial\vartheta_j^c}\,:\quad V_b \otimes V_c \; \mapsto \; V_a \\
{\rm Sh} &= \sum_{i,j = 1}^{N-1}\left(\vartheta^{{\color{blue} a}}_{i+1}\vartheta^{{\color{blue} b}}_j-\vartheta^{{\color{blue} a}}_i\vartheta^{{\color{blue} b}}_{j+1}\right)\frac{\partial^2}{\partial\vartheta_i^a\partial\vartheta_j^b}+\sum_{i=1}^{N-1}\vartheta^{{\color{blue} a}}_{i+1}\vartheta^{{\color{blue} b}}_N\frac{\partial^2}{\partial\vartheta_i^a\partial\vartheta_N^b}-\sum_{j=1}^{N-1}\vartheta^{{\color{blue} a}}_N\vartheta^{{\color{blue} b}}_{j+1}\frac{\partial^2}{\partial\vartheta_N^a\partial\vartheta_j^b}: \; V_a \otimes V_b \; \mapsto \; V_{{\color{blue} a}} \otimes V_{{\color{blue} b}}
\end{aligned}
\end{equation}}

\noindent and are enumerated with $[\alpha_1\dots\alpha_{m-1}\star\alpha_{m+1}\dots \alpha_n]$ where the star stands at exactly the place where the mapped space label changes by one. 
One also supplies these morphisms with the following signs 
\begin{equation}
    {\rm sign} = (-1)^{\alpha_1+\dots+\alpha_{m-1}}
\end{equation}
There appear zero morphisms but they can be hidden in ${\rm Sh}$. First, always $V\; \overset{0}{\mapsto} \; V$ because ${\rm Sh}$ acts on $V$ as zero (as containing double theta-derivatives). Sh also acts as zero when a pair of cycles joins at the previous step, because in this case, $V_a = V_b$ and $V_{\color{blue} a} = V_{\color{blue} b}$. Note that the operator Sh can be written in an invariant form if taking ${\vartheta}_{N+1}^a\equiv 0\,$:
\begin{equation}
\begin{aligned}
    {\rm Sh} = \sum_{i,j=1}^{N}\left(\vartheta^{{\color{blue} a}}_{i+1}\vartheta^{{\color{blue} b}}_j-\vartheta^{{\color{blue} a}}_i\vartheta^{{\color{blue} b}}_{j+1}\right)\frac{\partial^2}{\partial\vartheta_i^a\partial\vartheta_j^b}
\end{aligned}
\end{equation}
Differentials are sums of operators~\eqref{differentials} and act between direct sums of spaces of the same fixed height $h$.  

\paragraph{Step 4: grading shifts.} Note that the morphisms $\Delta$ and $m$ shift the grading by $q^{-N+1}$, and ${\rm Sh}$ shifts the grading by $q^{-2}$: 
\begin{equation}
\begin{aligned}
    \Delta \vartheta_1^c = \vartheta_N^a \vartheta_1^b + \vartheta_{N-1}^a \vartheta_2^b + \dots + \vartheta_1^a \vartheta_N^b\,: \qquad q^{N-1} \quad &\longrightarrow \quad q^{-N+1} \cdot q^{N-1}\,, \\
    \Delta \vartheta_2^c = \vartheta_N^a \vartheta_2^b + \dots + \vartheta_2^a \vartheta_N^b\,: \qquad q^{N-3} \quad &\longrightarrow \quad q^{-N+1} \cdot q^{N-3}\,, \\
    &\ldots \\
    \Delta \vartheta_N^c = \vartheta_N^a \vartheta_N^b\,: \qquad q^{-N+1} \quad &\longrightarrow \quad q^{-N+1} \cdot q^{-N+1}\,,
\end{aligned}
\end{equation}
\begin{equation}
\begin{aligned}
    m \,\vartheta_1^b \vartheta_1^c = \vartheta_1^a\,: \qquad q^{N-1} \cdot q^{N-1} \quad &\longrightarrow \quad q^{N-1}\,, \\
    m \,\vartheta_1^b \vartheta_2^c = \vartheta_2^a\,: \qquad q^{N-1} \cdot q^{N-3} \quad &\longrightarrow \quad q^{N-3}\,, \\
    &\ldots \\ 
    m \,\vartheta_N^b \vartheta_N^c = 0\,: \qquad q^{-N+1} \cdot q^{-N+1} \quad &\longrightarrow \quad 0\,,
\end{aligned}
\end{equation}
\begin{equation}
\begin{aligned}
    {\rm Sh}\, \vartheta^a_1 \vartheta^b_1 = \vartheta^a_2 \vartheta^b_1 - \vartheta^a_1 \vartheta^b_2\,: \qquad q^{N-1} \cdot q^{N-1} \quad &\longrightarrow \quad q^{N-3} \cdot q^{N-1}\,, \\
    {\rm Sh}\, \vartheta^a_1 \vartheta^b_2 = \vartheta^a_2 \vartheta^b_2 - \vartheta^a_1 \vartheta^b_3\,: \qquad q^{N-1} \cdot q^{N-3} \quad &\longrightarrow \quad q^{N-3} \cdot q^{N-3}\,, \\
    &\ldots \\ 
    {\rm Sh}\, \vartheta^a_N \vartheta^b_N = 0\,: \qquad q^{-N+1} \cdot q^{-N+1} \quad &\longrightarrow \quad 0\,.
\end{aligned}
\end{equation}
Thus, to calculate the Poincaré polynomial, one should restore the gradings by the corresponding $\{\text{degree shifts} \}$:
\begin{equation}\label{gradings}
\begin{aligned}
    \Delta&:\quad V^{\scaleto{\{0\}}{5.5pt}}\; \longrightarrow \; (V\otimes V)^{\scaleto{\{N-1\}}{5.5pt}}\,, \\
    m&:\quad (V\otimes V)^{\scaleto{\{0\}}{5.5pt}}\; \longrightarrow \; V^{\scaleto{\{N-1\}}{5.5pt}}\,, \\
    {\rm Sh}&:\quad (V\otimes V)^{\scaleto{\{0\}}{5.5pt}} \; \longrightarrow \; (V\otimes V)^{\scaleto{\{2\}}{5.5pt}}
\end{aligned}
\end{equation}
where ${\rm dim}_q V^{\scaleto{\{l\}}{5.5pt}}=q^l {\rm dim}_q V$. Here we hide the $q$-multipliers into quantum dimensions of the vector spaces in order to write down the general formula~\eqref{Kh-R-poly}. Note that these degree shifts are not a characteristic of a complex. They are placed for convenience to restore the powers of $q$ in the calculation of the Khovanov--Rozansky polynomial.

\paragraph{Step 5: kernels and images.}\mbox{}\\

\noindent Then, we calculate kernels and images of the differentials. For convenience, we present here kernels, images and quantum dimensions of the basic morphisms~\eqref{differentials}.

\smallskip
\noindent $\bullet$ For the $m$ operator
{\small \begin{equation}
\begin{aligned}
    {\rm Im}(m) &= V 
    \; \Rightarrow \; \boxed{\dim_q {\rm Im}(m) = [N]\,,} \\
    {\rm Ker}(m) &= \langle {\vartheta}_i^b {\vartheta}_j^c \rangle \backslash \langle {\vartheta}_1^b {\vartheta}_1^c,\, {\vartheta}_1^b {\vartheta}_2^c + {\vartheta}_2^b {\vartheta}_1^c,\dots,\, {\vartheta}_1^b {\vartheta}_{N}^c + {\vartheta}_2^b {\vartheta}_{N-1}^c + \dots + {\vartheta}_N^b {\vartheta}_1^c \rangle \; \Rightarrow \; \boxed{\dim_q {\rm Ker}(m) = [N]^2 - q^{N-1}[N] = q^{-1}[N][N-1]\,.}
\end{aligned}
\end{equation}}

\noindent $\bullet$ For the $\Delta$ morphism
{\small \begin{equation}
\begin{aligned}
    {\rm Im}(\Delta) &= \langle \vartheta_1^a \vartheta_N^b + \vartheta_2^a \vartheta_{N-1}^b + \dots + \vartheta_N^a \vartheta_1^b,\, \vartheta_2^a \vartheta_N^b + \dots + \vartheta_N^a \vartheta_2^b,\dots,\, \vartheta_N^a \vartheta_{N-1}^b + \vartheta_{N-1}^a \vartheta_N^b,\, \vartheta_{N}^a \vartheta_N^b \rangle \; \Rightarrow \; \boxed{\dim_q {\rm Im}(\Delta) = q^{-N+1}[N]\,,} \\
    {\rm Ker}(\Delta) &= \varnothing \; \Rightarrow \; \boxed{\dim_q {\rm Ker}(\Delta) = 0\,.}
\end{aligned}
\end{equation}}

\noindent $\bullet$ For the Sh operator

{\small \begin{equation}
\begin{aligned}
    {\rm Im}({\rm Sh}) &= \langle \vartheta^a_{i+1}\vartheta^b_j-\vartheta^a_i\vartheta^b_{j+1} \rangle = {\rm Ker}(m),\, i,\,j = 1,\dots,\, N,\, \vartheta_{N+1}\equiv 0 \; \Rightarrow \; \boxed{\dim_q {\rm Im}({\rm Sh}) = q^{-1}[N][N-1]\,,} \\
    {\rm Ker}({\rm Sh}) &= {\rm Im}(\Delta) \; \Rightarrow \; \boxed{\dim_q {\rm Ker}({\rm Sh}) = q^{-N+1}[N]\,.} 
\end{aligned}
\end{equation}}

\paragraph{Properties.} These calculations show that the differential properties are satisfied:
\begin{equation}
    m\, {\rm Sh} = 0\,,\quad {\rm Sh}\, \Delta = 0\,. 
\end{equation}

\paragraph{Step 6: Khovanov--Rozansky polynomial.} The Khovanov--Rozansky polynomial is expressed through the quantum dimensions of cohomologies ${\cal H}_i = {\rm Ker}(\hat{d}_{i})\backslash {\rm Im}(\hat{d}_{i-1})$ as follows:
\begin{equation}\label{Kh-R-poly}
\begin{aligned}
    P^{\cal L}(T,q,A=q^N) &= q^{-2N(n_\bullet - n_\circ)}T^{-2n_\bullet}\cdot q^{-(N+1)n_\circ} \sum_{i=0}^{2n+1} T^i \dim_q {\cal H}_i^{\cal L} = \\
    &= q^{-2N(n_\bullet - n_\circ)}T^{-2n_\bullet}\cdot q^{-(N+1)n_\circ}\sum_{i=0}^{2n+1} T^i \left(\dim_q {\rm Ker}(\hat{d}^{\,\cal L}_{i}) - \dim_q{\rm Im}(\hat{d}^{\,\cal L}_{i-1})\right)
\end{aligned}
\end{equation}
where $n$ is the total number of lock vertices in a link ${\cal L}$, $n_\bullet$ is the number of positive lock tangles and $n_\circ$ is the number of negative ones, see Figs.\,\ref{fig:bip-vertex-smoothings} and \ref{fig:oppvert}. The framing factor $q^{-2N(n_\bullet - n_\circ)}T^{-2n_\bullet}\cdot q^{-(N+1)n_\circ}$ is present to restore the topological invariance.

\setcounter{equation}{0}

\section{Examples of bipartite calculus}
\label{sec:examples}

In this section, we provide examples of the calculation of the Khovanov--Rozansky polynomials following the algorithm in the previous section. We also provide computation of the $q$-Euler characteristic and the corresponding precursor Khovanov polynomials.

\subsection{One bipartite vertex}

Here we calculate Khovanov--Rozansky cohomologies of all the 1-lock links from Fig.\,\ref{fig:one-bip-vert}.












\subsubsection{Hopf link}\label{sec:Hopf-bip}

This is our first example of cycle calculus for the Khovanov--Rozansky polynomial. In the Hopf link example, there appear the new zero morphism absent in the Khovanov ($N=2$) planar techique for single vertices. We also demonstrate how to get the precursor Khovanov complex. The general output is given in Section~\ref{sec:precursors}.

We have already drawn the 3-hypercube for the Hopf link in Fig.\,\ref{fig:Hopf-H-complex}. Now, according to steps 2--4, we associate a $N$-dimensional graded vector space to each cycle, morphisms to arrows and write the grading shifts in the curly brackets. So that the resulting complex for the Hopf link looks like
\begin{figure}[h!]
	\begin{equation}\nn
		\begin{aligned}
			{\rm\bf KhR} \ \ \ \ \ \ \ \ \ \ \ \ \ \ \ \varnothing \quad \overset{0}{\longrightarrow} \quad [\overset{1}{\bigcirc} \overset{1'}{\bigcirc}]_{[0]}^{\scaleto{\{0\}}{5.5pt}} \quad \overset{m}{\longrightarrow} \quad \boxed{[\overset{2}{\bigcirc}]_{[1]}^{\scaleto{\{N-1\}}{5.5pt}} \quad \overset{0}{{\color{Green} \longrightarrow}} \quad [\overset{{\color{Green} 2}}{\bigcirc}]_{[2]}^{\scaleto{\{N+1\}}{5.5pt}}}_{\,\phi} \quad \overset{0}{\longrightarrow} \quad \varnothing
		\end{aligned}
	\end{equation}
	\caption{\footnotesize The complex for the Hopf link. We color in green the arrow for the differential Sh which is zero in this case because Sh acts on {\it pairs} of spaces (contains two $\theta$-derivatives) and vanishes when acting on a single one. The second reason of this morphism to be zero is that it is coming after the $m$-morphism, and this will be the case when there will be more than one cycles in resolutions, see Fig.\,\ref{fig:Tw-4-complex}. The boxed elements connected by this green arrow are shrinked when transferring to the precursor complex.} 
	\label{fig:Hopf-complex}
\end{figure}

\noindent Here the subscripts of smoothings are their enumerators $[\alpha=0,\,1,\,2]$. The numbers under the circles are numbers of edges and label the spaces. The non-zero differential is
\begin{equation}
	\hat{d}_0^{\,\rm Hopf} = m = \sum_{\substack{1\le i,j\le N\\i+j \le N+1}}\vartheta_{i+j-1}^{(2)}\frac{\partial^2}{\partial\vartheta_i^{(1)}\partial\vartheta_j^{\prime (1)}}\,:\quad V^{(1)} \otimes V^{\prime (1)} \; \mapsto \; V^{(2)}\,. 
\end{equation}
Its explicit action on a special basis of $V^{(1)} \otimes V^{\prime (1)}$ is
\begin{equation}
	\hat{d}_0^{\,\rm Hopf}\,: \qquad \begin{array}{ccc}
		\vth_1^{\prime (1)} \vth_1^{(1)} & \mapsto & \vth_1^{(2)} \\
		\frac{1}{2}(\vth_1^{\prime (1)} \vth_2^{(1)} + \vth_2^{\prime (1)} \vth_1^{(1)}) & \mapsto & \vth_2^{(2)} \\
		\frac{1}{3}(\vth_1^{\prime (1)} \vth_3^{(1)} + \vth_2^{\prime (1)} \vth_2^{(1)} + \vth_3^{\prime (1)} \vth_1^{(1)}) & \mapsto & \vth_3^{(2)} \\
		\dots \\
		\frac{1}{N}(\vth_1^{\prime (1)} \vth_3^{(1)} + \vth_2^{\prime (1)} \vth_2^{(1)} + \vth_3^{\prime (1)} \vth_1^{(1)}) & \mapsto & \vth_N^{(2)} \\
		\vth_n^{\prime (1)} \vth_m^{(1)}\,, \; m + n > N+1 & \mapsto & 0 \\
		\vth_{i+1}^{\prime (1)} \vth_j^{(1)} - \vth_i^{\prime (1)} \vth_{j+1}^{(1)}\,, \; i + j \leq N+1 & \mapsto & 0
	\end{array} 
\end{equation}
Thus, at step 5, we get the following kernel and image of the differential\footnote{Recall that we denote grading shifts in curly brackets, see step 4 in Section~\ref{sec:method}.}:
\ba \nn 
\Ker(\hat{d}_0^{\,\rm Hopf}) &= {\rm Ker}(m)^{\scaleto{\{0\}}{5.5pt}} = \langle \vartheta_n^{(1)}\vartheta_m^{\prime (1)},\, \vartheta_{i+1}^{(1)}\vartheta_j^{\prime (1)} - \vartheta_i^{(1)}\vartheta_{j+1}^{\prime (1)} \rangle^{\scaleto{\{0\}}{5.5pt}},\, m+n > N+1,\, i+j \leq N+1\,, \\
\Im(\hat{d}_0^{\,\rm Hopf}) &= V^{\scaleto{\{N-1\}}{5.5pt}}\,.
\ea 
According to step 6, we calculate the quantum dimensions of the cohomologies:
\begin{equation} \nn
		\begin{aligned}
			{\cal H}_0^{\rm Hopf} &= \Ker(\hat{d}_0^{\,\rm Hopf})^{\scaleto{\{0\}}{5.5pt}} \; \Longrightarrow \; \boxed{{\rm dim}_q {\cal H}_0^{\rm Hopf} = q^{-1}[N][N-1]\,,} \\
			{\cal H}_1^{\rm Hopf} &= ({\rm Ker}(0) \backslash {\rm Im}(\hat{d}_0^{\,\rm Hopf}))^{\scaleto{\{N-1\}}{5.5pt}} \overset{{\rm Ker}(0) = {\rm Im}(\hat{d}_0^{\,\rm Hopf})}{=} \varnothing \quad \Longrightarrow \quad \boxed{{\rm dim}_q {\cal H}_1^{\rm Hopf} = 0\,,} \\
			{\cal H}_2^{\rm Hopf} &= {\rm Ker}(0)^{\scaleto{\{N+1\}}{5.5pt}} = V^{\scaleto{\{N+1\}}{5.5pt}} \quad \Longrightarrow \quad \boxed{{\rm dim}_q {\cal H}_2^{\rm Hopf} = q^{N+1}[N]\,.}
		\end{aligned}
\end{equation}

\noindent The Khovanov--Rozansky polynomial is
\begin{equation}
	\boxed{\boxed{P^{\rm Hopf}(A,q,T) = (q^N T)^{-2}\left({\rm dim}_q {\cal H}_0^{\rm Hopf} + T^2 {\rm dim}_q {\cal H}_2^{\rm Hopf}\right) = q^{-2N-1}[N] \left( T^{-2}[N-1] + q^{N+2}\right)}} 
\end{equation}
what reproduces the known answer in \cite{carqueville2014computing,anokhina2014towards-R}.

The unnormalized HOMFLY polynomial is the graded Euler characteristic of a complex. Thus,
\begin{equation}\label{H-Hopf}
	H^{\rm Hopf}(A,q) \overset{{\rm Fig.\,}\ref{fig:Hopf-complex}}{=} A^{-2}(D^2 - q^{N-1} D + q^{N+1} D) \overset{\eqref{2-hyp-Hopf}}{=} A^{-2} \cdot D (D+\phi) = D^2\left(1-A^{-1}\cdot\frac{\{Aq\}\{A/q\}}{\{A\}}\right)\,.
\end{equation}
In the second equality we have just substituted $\phi = q^{N+1} - q^{N-1}$. It also can be obtained from the HOMFLY 2-hypercube in~\eqref{2-hyp-Hopf}. 


\paragraph{Precursor.}

The precursor diagram for the Hopf link is the unknot. To make the construction of the precursor complex clearer, we do it in 2 steps. First, we formally shrink the boxed items in Fig.\,\ref{fig:Hopf-complex}:
\begin{equation}\label{2-hyp-Hopf}
	\begin{aligned}
		\varnothing \quad \overset{0}{\longrightarrow} \quad [\overset{1}{\bigcirc} \overset{1'}{\bigcirc}]_{[0]} \quad \overset{m}{\longrightarrow} \quad \phi\,[\overset{2}{\bigcirc}]_{[1]} \quad \overset{0}{\longrightarrow} \quad \varnothing
	\end{aligned}
\end{equation}
Here $\phi$ in parallel with the grading shift multiplies the quantum dimension by $\phi$, so that the weighted sum~\eqref{H-Hopf} remains the same. Note that at this stage we have got rid of $[2]$-element of the initial complex and of the green morphism. Second, in order to pass to the precursor diagram and its Khovanov polynomial, we put $N=2$ and change $\phi$ to $-q$, what we rewrite as the grading $\{1\}$:
\begin{figure}[h!]
	\begin{picture}(300,30)(-200,15)
		
		\put(30,-5){
			
			\put(-100,25){\line(1,1){16}}
			\put(-100,41){\line(1,-1){6}}
			\put(-90,31){\line(1,-1){6}}
			
			\put(-92,25){\oval(16,16)[b]}
			\put(-92,41){\oval(16,16)[t]}
		}
		
		\put(-130,25){\mbox{${\bf Kh}$}}
		
		\put(-10,25){\mbox{$\varnothing \quad \overset{0}{\longrightarrow} \quad [\overset{1}{\bigcirc} \overset{1'}{\bigcirc}]_{[0]}^{\scaleto{\{0\}}{5.5pt}} \quad \overset{m}{\longrightarrow} \quad [\overset{2}{\bigcirc}]_{[1]}^{\scaleto{\{1\}}{5.5pt}} \quad \overset{0}{\longrightarrow} \quad \varnothing$}}
		
	\end{picture}
	\caption{\footnotesize The unknot being the precursor diagram of the Hopf link and its Khovanov complex.}
	\label{fig:enter-label}
\end{figure}

\noindent Let us calculate the cohomologies:
\begin{equation}
	\begin{aligned}
		{\cal H}_0^{\text{1-unknot}} &= {\rm Ker}(m) = \langle \vartheta_2^{(1)} \vartheta_2^{\prime (1)},\, \vartheta_1^{(1)} \vartheta_2^{\prime (1)} - \vartheta_2^{(1)} \vartheta_1^{\prime (1)} \rangle \quad \Longrightarrow \quad \dim_q{\cal H}_0^{\text{1-unknot}} = q^{-2} + 1\,, \\
		{\cal H}_1^{\text{1-unknot}} &= {\rm Ker}(0) \backslash {\rm Im}(m) = \varnothing \quad \Longrightarrow \quad \dim_q{\cal H}_1^{\text{1-unknot}} = 0\,.
	\end{aligned}
\end{equation}
Thus, the Khovanov polynomial is
\begin{equation}
	{\rm Kh}^{\text{1-unknot}}(q,T) = q \cdot \dim_q{\cal H}_0^{\text{1-unknot}} = D_2\,.
\end{equation}

\paragraph{Summary.} By this example, we have demonstrated the appearance of the zero morphism in the Khovanov--Rozansky complex. We have also shown the transition to precursor complex and its Khovanov polynomial.

\subsubsection{Mirror Hopf link}\label{sec:mirr-Hopf-bip}

The complex for the mirror Hopf link looks like shown in Fig.\,\ref{fig:mirr-Hopf-3-complex}.
\begin{figure}[h!]
\begin{equation}
    {\bf KhR} \ \ \ \ \ \ \ \ \ \varnothing \quad \overset{0}{\longrightarrow} \quad \boxed{[\overset{2}{\bigcirc}]_{[0]}^{\scaleto{\{0\}}{5.5pt}} \quad \overset{0}{{\color{Green} \longrightarrow}} \quad [\overset{{\color{Green} 2}}{\bigcirc}]_{[1]}^{\scaleto{\{2\}}{5.5pt}}}_{\,q^{N+1}\bphi} \quad \overset{\Delta}{\longrightarrow} \quad [\overset{1}{\bigcirc}\overset{1'}{\bigcirc}]_{[2]}^{\scaleto{\{N+1\}}{5.5pt}} \quad \overset{0}{\longrightarrow} \quad \varnothing
\end{equation}
    \caption{\footnotesize The complex for the mirror Hopf link. We color in green the arrow for the zero differential Sh. The boxed elements connected by this green arrow are shrinked when transferring to the precursor complex.}
    \label{fig:mirr-Hopf-3-complex}
\end{figure}

\noindent The non-zero differential is
\begin{equation}
     \hat{d}_0^{\overline{\,\rm Hopf}} = \Delta = \sum_{i=1}^N\left(\sum_{j=0}^{N-i}\vartheta^{(1)}_{N-j}\vartheta^{\prime (1)}_{i+j}\right)\frac{\partial}{\partial \vartheta^{{\color{Green} (2)}}_i}\,:\quad V^{{\color{Green} (2)}} \; \mapsto \; V^{(1)} \otimes V^{\prime (1)}\,. 
\end{equation}
Calculate the quantum dimensions of the cohomologies:
\ba \nn
{\cal H}_0^{\overline{\rm Hopf}} &= \Ker(0)^{\scaleto{\{0\}}{5.5pt}} = V^{\scaleto{\{0\}}{5.5pt}} \quad \Lra \quad \boxed{\dim_q {\cal H}_0 = [N]\,,} \\ 
{\cal H}_1^{\overline{\rm Hopf}} &= (\Ker(\Delta)\backslash \Im(0))^{\scaleto{\{2\}}{5.5pt}} = \varnothing \quad \Lra \quad \boxed{\dim_q {\cal H}_1 = 0\,,} \\
{\cal H}_2^{\overline{\rm Hopf}} &= (\Ker(0)\backslash \Im(\Delta))^{\scaleto{\{N+1\}}{5.5pt}} \quad \Lra \quad \boxed{\dim_q {\cal H}_2 = q^{N+1}([N]^2 - q^{-N+1}[N]) = q^{N+2}[N] [N-1]\,.} 
\ea
The Khovanov--Rozansky polynomial is
\begin{equation}
    \boxed{\boxed{P^{\,\overline{\rm Hopf}}(A,q,T) = q^{N-1} \left({\rm dim}_q {\cal H}_0^{\rm Hopf} + T^2 {\rm dim}_q {\cal H}_2^{\rm Hopf}\right) = q^{N}[N] \left( q^{N+1} T^{2}[N-1] + q^{-1}\right)}} 
\end{equation}
what reproduces the known answer in \cite{carqueville2014computing,anokhina2014towards-R}.

Note that $P^{\,\overline{\rm Hopf}}(A,q,T) = P^{\,\rm Hopf}(A^{-1},q^{-1},T^{-1})$. This actually holds for an arbitrary link $\cal L$ and its mirror $\overline{\cal L}$: $P^{\overline{\cal L}}(A,q,T) = P^{\cal L}(A^{-1},q^{-1},T^{-1})$. 

The unnormalized HOMFLY polynomial is the graded Euler characteristic of a complex. Thus,
\begin{equation}
    H^{\overline{\rm Hopf}}(A,q) \overset{{\rm Fig.\,}\ref{fig:mirr-Hopf-3-complex}}{=} A q^{-1} (D - q^{2} D + q^{N+1} D^2) \overset{\eqref{2-hyp-mirr-Hopf}}{=} A^2 \cdot D (D+\bar\phi) = D^2\left(1-A^{-1}\cdot\frac{\{Aq\}\{A/q\}}{\{A\}}\right)\,.
\end{equation}
The second equality is obtained by gathering the monomials into $\bphi = q^{-N-1} - q^{-N+1}$. It can also be evaluated from the HOMFLY 2-hypercube~\eqref{2-hyp-mirr-Hopf}.

\paragraph{Precursor.}

The precursor diagram for the mirror Hopf link is the unknot. To make the construction of the precursor complex clearer, we do it in 2 steps. First, we formally shrink the boxed items in Fig.\,\ref{fig:Hopf-complex}:
\begin{equation}\label{2-hyp-mirr-Hopf}
    \varnothing \quad \overset{0}{\longrightarrow} \quad \bphi \, [\overset{2}{\bigcirc}]_{[0]}^{\scaleto{\{N+1\}}{5.5pt}} \quad \overset{\Delta}{\longrightarrow} \quad [\overset{1}{\bigcirc}\overset{1'}{\bigcirc}]_{[2]}^{\scaleto{\{N+1\}}{5.5pt}} \quad \overset{0}{\longrightarrow} \quad \varnothing
\end{equation}
Here $\bphi$ in parallel with the grading shift additionally multiplies the quantum dimension by $\bphi$, so that the weighted sum remains the same (in comparison with the $q$-Euler characteristic). Note that at this stage we have got rid of $[1]$-element of the initial complex and of the blue morphism. Second, in order to pass to the precursor diagram and its Khovanov polynomial, we put $N=2$ and change $\bphi$ to $-q^{-1}$ and the label $[2]$ to $[1]$:
\begin{figure}[h!]
\begin{picture}(300,30)(-200,15)

\put(30,-5){

\put(-84,25){\line(-1,1){16}}
\put(-84,41){\line(-1,-1){6}}
\put(-94,31){\line(-1,-1){6}}

\put(-92,25){\oval(16,16)[b]}
\put(-92,41){\oval(16,16)[t]}
}

\put(-130,25){\mbox{${\bf Kh}$}}

\put(-10,25){\mbox{$\varnothing \quad \overset{0}{\longrightarrow} \quad [\overset{2}{\bigcirc}]_{[0]}^{\scaleto{\{-1\}}{5.5pt}} \quad \overset{\Delta}{\longrightarrow} \quad [\overset{1}{\bigcirc}\overset{1'}{\bigcirc}]_{[1]}^{\scaleto{\{0\}}{5.5pt}} \quad \overset{0}{\longrightarrow} \quad \varnothing$}}
    
\end{picture}
    \caption{\footnotesize The unknot being the precursor diagram of the mirror Hopf link and its Khovanov complex.}
    \label{fig:enter-label}
\end{figure}

\noindent Here we have also got rid of the same grading shifts $\{N+1\}$, as they factorize in the unique multiplier. Let us calculate the cohomologies:
\begin{equation}
\begin{aligned}
    {\cal H}_0^{\overline{\text{1-unknot}}} &= {\rm Ker}(\Delta) = \varnothing \quad \Longrightarrow \quad \dim_q{\cal H}_0^{\overline{\text{1-unknot}}} = 0\,, \\
    {\cal H}_1^{\overline{\text{1-unknot}}} &= {\rm Ker}(0) \backslash {\rm Im}(\Delta) = \langle \vartheta_1^{(1)} \vartheta_1^{\prime (1)},\, \vartheta_1^{(1)} \vartheta_2^{\prime (1)} - \vartheta_2^{(1)} \vartheta_1^{\prime (1)} \rangle \quad \Longrightarrow \quad \dim_q{\cal H}_1^{\overline{\text{1-unknot}}} = 1 + q^2\,. 
\end{aligned}
\end{equation}
Thus, the Khovanov polynomial is
\begin{equation}
    {\rm Kh}^{\overline{\text{1-unknot}}}(q,T) = (qT)^{-1} \cdot T\dim_q{\cal H}_1^{\overline{\text{1-unknot}}} = D_2\,.
\end{equation}

\paragraph{Summary.} This example shows the equality
\begin{equation}
 P^{\overline{\cal L}}(A,q,T) = P^{\cal L}(A^{-1},q^{-1},T^{-1})
\end{equation}
for an arbitrary link $\cal L$ and its mirror.

\subsubsection{2-unknot}

In this example we show the appearance of the Sh morphism coming after the $\Delta$ operator. We also get the precursor Khovanov complex.

The complex for the 2-unknot is shown in Fig.\,\ref{fig:2-unknot-complex}.
\begin{figure}[h!]
    \begin{equation}\nn
    {\bf KhR} \ \ \ \ \ \ \ \ \ \ \ \ \ \varnothing \quad \overset{0}{\longrightarrow} \quad [\overset{2}{\bigcirc}]_{[0]}^{\scaleto{\{0\}}{5.5pt}} \quad \overset{\Delta}{\longrightarrow} \quad \boxed{[\overset{1}{\bigcirc}\overset{ 1'}{\bigcirc}]_{[1]}^{\scaleto{\{N-1\}}{5.5pt}} \quad \overset{\rm Sh}{{\color{blue} \longrightarrow}} \quad  [\overset{{\color{blue} 1}}{\bigcirc}\overset{{\color{blue} 1'}}{\bigcirc}]_{[2]}^{\scaleto{\{N+1\}}{5.5pt}}}_{\,\phi} \quad \overset{0}{\longrightarrow} \quad \varnothing
\end{equation}
    \caption{\footnotesize The complex for the 2-unknot. We color in blue the arrow for the differential Sh. It is not zero in this case because it acts after the $\Delta$ morphism. The boxed elements connected by this blue arrow are shrinked when transferring to the precursor complex.}
    \label{fig:2-unknot-complex}
\end{figure}

\noindent Here the subscripts of smoothings are their enumerators $[\alpha=0,\,1,\,2]$, and curly brackets denote grading shifts. The numbers under the circles are numbers of edges and label the spaces. The differentials are
\begin{equation}
\begin{aligned}
    \hat{d}_0^{\,\text{2-unknot}} &= \Delta = \sum_{i=1}^N\left(\sum_{j=0}^{N-i}\vartheta^{(1)}_{N-j}\vartheta^{\prime (1)}_{i+j}\right)\frac{\partial}{\partial \vartheta^{(2)}_i}\,:\quad V^{(2)} \; \mapsto \; V^{(1)} \otimes V^{\prime (1)}\,, \\
    \hat{d}_1^{\,\text{2-unknot}} &= {\rm Sh} = \sum_{i,j=1}^{N}\left(\vartheta^{{\color{blue} (1)}}_{i+1}\vartheta^{{\color{blue} \prime (1)}}_j-\vartheta^{{\color{blue} (1)}}_i\vartheta^{{\color{blue} \prime (1)}}_{j+1}\right)\frac{\partial^2}{\partial\vartheta_i^{( 1)}\partial\vartheta_j^{\prime ( 1)}},\quad \vartheta_{N+1} \equiv 0 \,:\quad V^{(1)} \otimes V^{\prime (1)} \; \mapsto \; V^{{\color{blue} (1)}} \otimes V^{{\color{blue} \prime (1)}}\,.
\end{aligned}
\end{equation}
Note that here black and blue indices are different. The cohomologies and their quantum dimensions are
{\small \begin{equation} \nn
\begin{aligned}
    {\cal H}_0^{\text{2-unknot}} &= {\rm Ker}(\Delta) = \varnothing \quad \Longrightarrow \quad \boxed{\dim_q {\cal H}_0^{\text{2-unknot}} = 0\,,} \\
    {\cal H}_1^{\text{2-unknot}} &= ({\rm Ker}({\rm Sh}) \backslash {\rm Im}(\Delta))^{\scaleto{\{N-1\}}{5.5pt}} \overset{{\rm Ker}({\rm Sh}) = {\rm Im}(\Delta)}{=} \varnothing \quad \Longrightarrow \quad \boxed{\dim_q {\cal H}_1^{\text{2-unknot}} = 0\,,} \\
    {\cal H}_2^{\text{2-unknot}} &= ({\rm Ker}(0) \backslash {\rm Im}({\rm Sh}))^{\scaleto{\{N+1\}}{5.5pt}} \\ &\Longrightarrow \quad \boxed{\dim_q {\cal H}_2^{\text{2-unknot}} = q^{N+1}\left( \dim_q {\rm Ker}(0) - \dim_q {\rm Im}({\rm Sh}) \right) = q^{N+1}\left([N]^2 - q^{-1}[N][N-1] \right) = q^{2N}[N]\,.}
\end{aligned}
\end{equation}}
The Khovanov--Rozansky polynomial is
\begin{equation}
    \boxed{\boxed{P^{\,\text{2-unknot}}(A,q,T) = (q^N T)^{-2} \cdot T^2 {\rm dim}_q {\cal H}_2^{\text{2-unknot}} = (q^N T)^{-2} \cdot q^{2N}T^2[N] = [N]\,.}} 
\end{equation}
The unreduced HOMFLY polynomial is
\begin{equation}
    H^{\text{2-unknot}}(A,q) \overset{{\rm Fig.\,}\ref{fig:2-unknot-complex}}{=} A^{-2}(D - q^{N-1} D^2 + q^{N+1} D^2) \overset{\eqref{2-hyp-2-unknot}}{=} A^{-2} \cdot D (1+\phi D) = D\,.
\end{equation}

\paragraph{Precursor.}

Again, we firstly shrink the boxed cycles in Fig.\,\ref{fig:2-unknot-complex}\,:
\begin{equation}\label{2-hyp-2-unknot}
    \varnothing \quad \overset{0}{\longrightarrow} \quad [\overset{2}{\bigcirc}]_{[0]} \quad \overset{\Delta}{\longrightarrow} \quad \phi \, [\overset{1}{\bigcirc}\overset{ 1'}{\bigcirc}]_{[1]} \quad \overset{0}{\longrightarrow} \quad \varnothing
\end{equation}
In the above complex we do not have the enumerator $\alpha=2$ and the Sh morphism. Thus, it is a good candidate to become the complex of the precursor diagram. As in Section~\ref{sec:Hopf-bip}, we transfer to $2$-dimensional spaces and substitute $\phi \rightarrow -q$, and get the following complex for the unknot:
\begin{figure}[h!]
\begin{picture}(300,30)(-200,15)

\put(30,-5){

\put(-100,25){\line(1,1){16}}
\put(-100,41){\line(1,-1){6}}
\put(-90,31){\line(1,-1){6}}

\put(-84,33){\oval(16,16)[r]}
\put(-100,33){\oval(16,16)[l]}
}

\put(-140,25){\mbox{${\bf Kh}$}}

\put(-10,25){\mbox{$\varnothing \quad \overset{0}{\longrightarrow} \quad [\overset{2}{\bigcirc}]_{[0]}^{\scaleto{\{0\}}{5.5pt}} \quad \overset{\Delta}{\longrightarrow} \quad [\overset{1}{\bigcirc}\overset{ 1'}{\bigcirc}]_{[1]}^{\scaleto{\{1\}}{5.5pt}} \quad \overset{0}{\longrightarrow} \quad \varnothing$}}
    
\end{picture}
    \caption{\footnotesize The unknot being the precursor diagram of the 2-unknot and its Khovanov complex.}
    \label{fig:enter-label}
\end{figure}

\noindent Calculate the cohomologies:
\begin{equation}
\begin{aligned}
    {\cal H}_0^{\overline{\text{1-unknot}}} &= {\rm Ker}(\Delta) = \varnothing \quad \Longrightarrow \quad \dim_q {\cal H}_0^{\overline{\text{1-unknot}}} = 0\,, \\
    {\cal H}_1^{\overline{\text{1-unknot}}} &= {\rm Ker}(0) \backslash \Im(\Delta) = \langle \vth^{(1)}_1\vth^{\prime (1)}_1,\, \vth^{(1)}_1\vth^{\prime (1)}_2 - \vth^{(1)}_2\vth^{\prime (1)}_1 \rangle \quad \Longrightarrow \quad \dim_q {\cal H}_1^{\overline{\text{1-unknot}}} = q^2 + 1\,.
\end{aligned}
\end{equation}
The Khovanov polynomial is
\begin{equation}
    {\rm Kh}^{\overline{\text{1-unknot}}}(q,T) = q^{-2}T^{-1}\cdot qT \dim_q{\cal H}_1^{\overline{\text{1-unknot}}} = D_2\,.
\end{equation}

\paragraph{Summary.} In this subsection, we have shown the appearance of the Sh morphism. Emphasize that it is present only after the action of the coproduct $\Delta$. If at the step before, the multiplication $m$ acts, then the zero morphism appears.

\subsubsection{Mirror 2-unknot}
The complex for the mirror 2-unknot is
\begin{figure}[h!]
    \begin{equation}\nn
    {\bf KhR} \ \ \ \ \ \ \ \ \ \ \ \ \ \ \ \ \varnothing \quad \overset{0}{\longrightarrow} \quad \boxed{[\overset{1}{\bigcirc}\overset{ 1'}{\bigcirc}]_{[0]}^{\scaleto{\{0\}}{5.5pt}} \quad \overset{\rm Sh}{{\color{blue} \longrightarrow}} \quad  [\overset{{\color{blue} 1}}{\bigcirc}\overset{{\color{blue} 1'}}{\bigcirc}]_{[1]}^{\scaleto{\{2\}}{5.5pt}}}_{\,q^{N+1}\bphi}\quad \overset{m}{\longrightarrow} \quad [\overset{2}{\bigcirc}]_{[2]}^{\scaleto{\{N+1\}}{5.5pt}} \quad \overset{0}{\longrightarrow} \quad \varnothing
\end{equation}
    \caption{\footnotesize The complex for the 2-unknot. We color in blue the arrow for the differential Sh. The boxed elements connected by this blue arrow are shrinked when transferring to the precursor complex.}
    \label{fig:2-unknot-complex-mirr}
\end{figure}

\noindent Here the subscripts of smoothings are their enumerators $[\alpha=0,\,1,\,2]$, and curly brackets denote grading shifts. The numbers under the circles are numbers of edges and label the spaces. The differentials are
\begin{equation}
\begin{aligned}
    \hat{d}_0^{\,\overline{\text{2-unknot}}} &= m = \sum_{\substack{1\le i,j\le N\\i+j \le N+1}}\vartheta_{i+j-1}^{(2)}\frac{\partial^2}{\partial\vartheta_i^{{\color{blue}(1)}}\partial\vartheta_j^{{\color{blue}\prime (1)}}}\,:\quad V^{{\color{blue}(1)}} \otimes V^{{\color{blue}\prime (1)}} \; \mapsto \; V^{(2)}\,,  \\
    \hat{d}_1^{\,\overline{\text{2-unknot}}} &= {\rm Sh} = \sum_{i,j=1}^{N}\left(\vartheta^{{\color{blue} (1)}}_{i+1}\vartheta^{{\color{blue} \prime (1)}}_j-\vartheta^{{\color{blue} (1)}}_i\vartheta^{{\color{blue} \prime (1)}}_{j+1}\right)\frac{\partial^2}{\partial\vartheta_i^{( 1)}\partial\vartheta_j^{\prime ( 1)}},\quad \vartheta_{N+1} \equiv 0 \,:\quad V^{(1)} \otimes V^{\prime (1)} \; \mapsto \; V^{{\color{blue} (1)}} \otimes V^{{\color{blue} \prime (1)}}\,.
\end{aligned}
\end{equation}
Note that here black and blue indices are different. The cohomologies and their quantum dimensions are
\begin{equation} \nn
\begin{aligned}
    {\cal H}_0^{\overline{\text{2-unknot}}} &= {\rm Ker}({\rm Sh}) \quad \Longrightarrow \quad \boxed{\dim_q {\cal H}_0^{\overline{\text{2-unknot}}} = q^{-N+1}[N]\,, } \\
    {\cal H}_1^{\overline{\text{2-unknot}}} &= ({\rm Ker}(m) \backslash {\rm Im}({\rm Sh}))^{\scaleto{\{2\}}{5.5pt}} \overset{{\rm Im}({\rm Sh}) = {\rm Ker}(m)}{=} \varnothing \quad \Longrightarrow \quad \boxed{\dim_q {\cal H}_1^{\overline{\text{2-unknot}}} = 0\,,} \\
    {\cal H}_2^{\overline{\text{2-unknot}}} &= ({\rm Ker}(0) \backslash {\rm Im}(m))^{\scaleto{\{N+1\}}{5.5pt}} =\varnothing \quad \Longrightarrow \quad \boxed{\dim_q {\cal H}_2^{\overline{\text{2-unknot}}} = 0\,.}
\end{aligned}
\end{equation}
The Khovanov--Rozansky polynomial is
\begin{equation}
    \boxed{\boxed{P^{\,\overline{\text{2-unknot}}}(A,q,T) = q^{N-1} \cdot {\rm dim}_q {\cal H}_0^{\overline{\text{2-unknot}}} = [N]\,.}} 
\end{equation}
The unreduced HOMFLY polynomial is
\begin{equation}
    H^{\overline{\text{2-unknot}}}(A,q) \overset{{\rm Fig.\,}\ref{fig:2-unknot-complex-mirr}}{=} A q^{-1}(D^2 - q^{2} D^2 + q^{N+1} D) \overset{\eqref{2-hyp-2-unknot-mirr}}{=} A^{2} \cdot D (1+\bphi D) = D=P^{\,\overline{\text{2-unknot}}}(A,q,T=-1)\,.
\end{equation}

\paragraph{Precursor.}

Again, we firstly shrink the boxed cycles in Fig.\,\ref{fig:2-unknot-complex-mirr}\,:
\begin{equation}\label{2-hyp-2-unknot-mirr}
    \varnothing \quad \overset{0}{\longrightarrow} \quad \bphi\,[\overset{1}{\bigcirc}\overset{ 1'}{\bigcirc}]_{[0]}^{\scaleto{\{N+1\}}{5.5pt}}\quad \overset{m}{\longrightarrow} \quad [\overset{2}{\bigcirc}]_{[2]}^{\scaleto{\{N+1\}}{5.5pt}} \quad \overset{0}{\longrightarrow} \quad \varnothing
\end{equation}
In the above complex, we do not have the Sh morphism and have 2 spaces instead of 3. Thus, it is a good candidate to become the complex of the precursor diagram. As in Section~\ref{sec:mirr-Hopf-bip}, we transfer to $2$-dimensional spaces and substitute $\bphi \rightarrow -q^{-1}$, and get the complex for the unknot in Fig.\,\ref{fig:unk-prec-unk}. 
\begin{figure}[h!]
\begin{picture}(300,30)(-200,15)

\put(30,-5){

\put(-84,25){\line(-1,1){16}}
\put(-84,41){\line(-1,-1){6}}
\put(-94,31){\line(-1,-1){6}}

\put(-84,33){\oval(16,16)[r]}
\put(-100,33){\oval(16,16)[l]}
}

\put(-140,25){\mbox{${\bf Kh}$}}

\put(-10,25){\mbox{$\varnothing \quad \overset{0}{\longrightarrow} \quad [\overset{1}{\bigcirc}\overset{ 1'}{\bigcirc}]_{[0]}^{\scaleto{\{-1\}}{5.5pt}}\quad \overset{m}{\longrightarrow} \quad [\overset{2}{\bigcirc}]_{[2]}^{\scaleto{\{0\}}{5.5pt}} \quad \overset{0}{\longrightarrow} \quad \varnothing$}}
    
\end{picture}
    \caption{\footnotesize The unknot being the precursor diagram of the 2-unknot and its Khovanov complex.}
    \label{fig:unk-prec-unk}
\end{figure}
Here we again get rid of the grading shifts $\{N+1\}$. Calculate the cohomologies:
\begin{equation} \nn
\begin{aligned}
    {\cal H}_0^{\text{1-unknot}} &= {\rm Ker}(m) = \langle \vartheta_2^{(1)} \vartheta_2^{\prime (1)},\, \vartheta_1^{(1)} \vartheta_2^{\prime (1)} - \vartheta_2^{(1)} \vartheta_1^{\prime (1)} \rangle \quad \Longrightarrow \quad \dim_q{\cal H}_0^{\text{1-unknot}} = q^{-2} + 1\,, \\
    {\cal H}_1^{\text{1-unknot}} &= {\rm Ker}(0) \backslash {\rm Im}(m) = \varnothing \quad \Longrightarrow \quad \dim_q{\cal H}_1^{\text{1-unknot}} = 0\,.
\end{aligned}
\end{equation}
The Khovanov polynomial is
\begin{equation}
    {\rm Kh}^{\text{1-unknot}}(q,T) = q^{2} \cdot q^{-1} \dim_q{\cal H}_0^{\text{1-unknot}} = D_2\,.
\end{equation}


\subsection{Two bipartite vertices}

In this subsection, we calculate Khovanov--Rozansky cohomologies of the 2-lock links from Fig.\,\ref{fig:two-bip-vert}. In the figure below, we present examples of 3-hypercubes of 2-locks bipartite diagrams.

\begin{figure}[h!]
    \centering
    \includegraphics[width=0.55\linewidth]{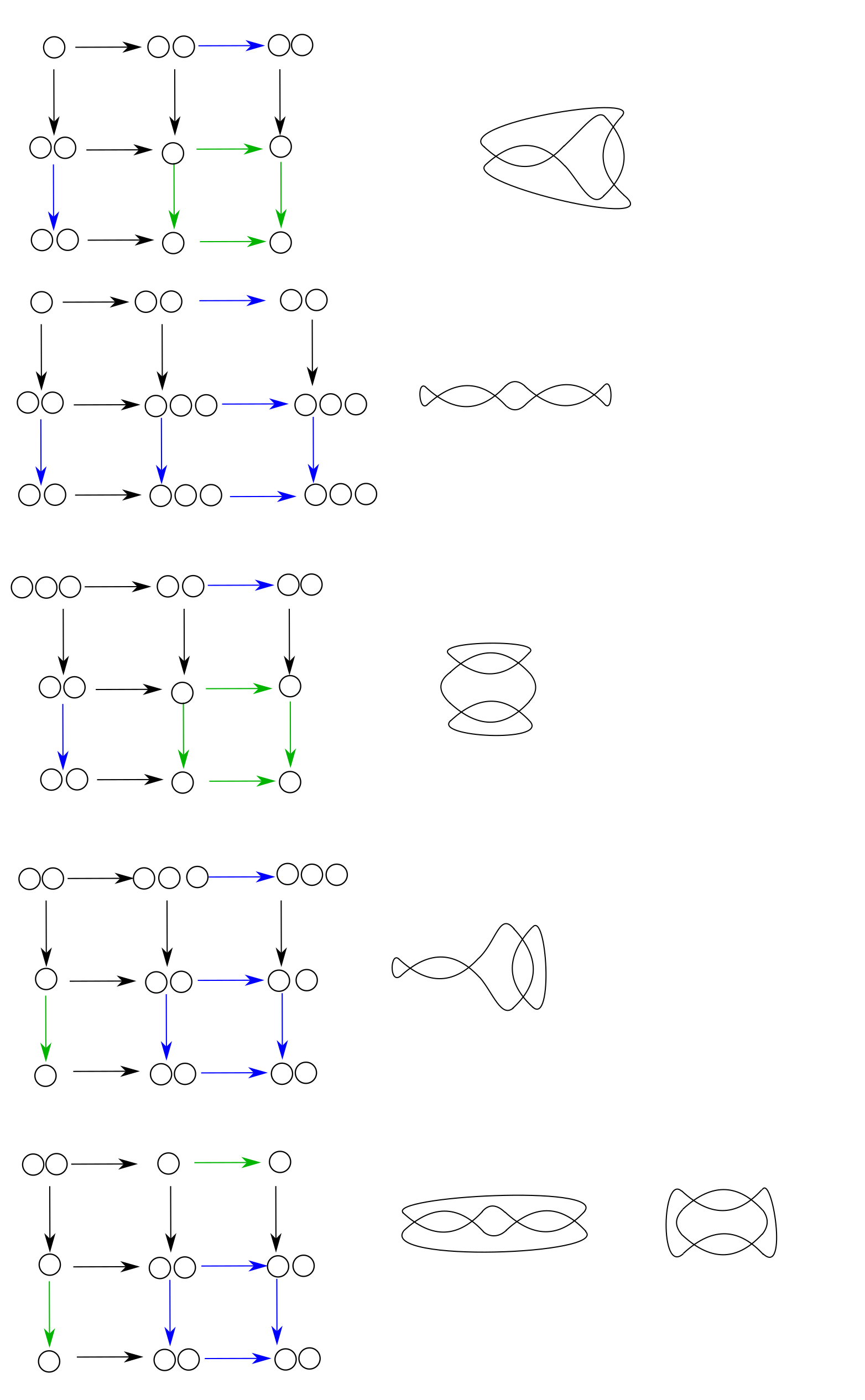}
    \caption{\footnotesize 3-hypercubes of 2-locks bipartite diagrams.}
    \label{fig:2-locks-BD}
\end{figure}

\subsubsection{Twist trefoil knot: Tw$_2$}\label{sec:tw-tref}

This is our first two-lock example. In such cases, the both morphisms zero and Sh can appear.

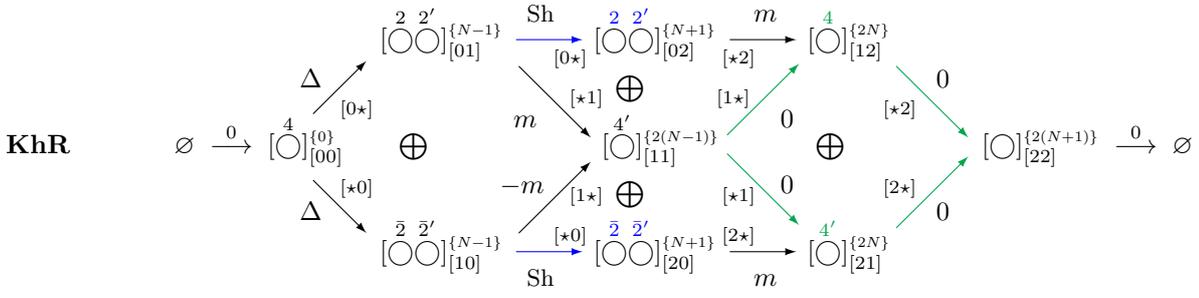
\begin{figure}[h!]
\centering
\begin{picture}(100,110)(40,-5)

\put(-150,50){\mbox{\bf KhR}}

\put(-87,50){\mbox{$\varnothing\;\overset{0}{\longrightarrow} \; [\overset{4}{\bigcirc}]_{[00]}^{\scaleto{\{0\}}{5.5pt}}$}}
\put(-35,65){\vector(1,1){20}}
\put(-35,40){\vector(1,-1){20}}
\put(-40,75){\mbox{$\Delta$}}
\put(-25,65){\mbox{\scriptsize $[0\star]$}}
\put(-40,25){\mbox{$\Delta$}}
\put(-25,35){\mbox{\scriptsize $[\star 0]$}}

\put(-10,90){\mbox{$[\overset{2}{\bigcirc}\overset{2'}{\bigcirc}]_{[01]}^{\scaleto{\{N-1\}}{5.5pt}}$}}
\put(-10,10){\mbox{$[\overset{\bar 2}{\bigcirc}\overset{\bar 2'}{\bigcirc}]_{[10]}^{\scaleto{\{N-1\}}{5.5pt}}$}}
\put(-3,50){\mbox{$\bigoplus$}}

\put(10,0){

\put(60,90){\mbox{$[\overset{{\color{blue} 2}}{\bigcirc}\overset{{\color{blue} 2'}}{\bigcirc}]_{[02]}^{\scaleto{\{N+1\}}{5.5pt}}$}}
\put(60,10){\mbox{$[\overset{{\color{blue} \bar 2}}{\bigcirc}\overset{{\color{blue} \bar 2'}}{\bigcirc}]_{[20]}^{\scaleto{\{N+1\}}{5.5pt}}$}}
{\color{blue} \put(31,93){\vector(1,0){25}}
\put(31,13){\vector(1,0){25}}}
\put(35,100){\mbox{\small Sh}}
\put(45,17){\mbox{\scriptsize $[\star 0]$}}
\put(45,84){\mbox{\scriptsize $[0\star]$}}
\put(35,0){\mbox{\small Sh}}

\put(68,32){\mbox{$\bigoplus$}}
\put(68,72){\mbox{$\bigoplus$}}
\put(63,50){\mbox{$[\overset{4'}{\bigcirc}]_{[11]}^{\scaleto{\{2(N-1)\}}{5.5pt}}$}}
\put(32,83){\vector(1,-1){27}}
\put(30,60){\mbox{$m$}}
\put(51,69){\mbox{\scriptsize $[\star 1]$}}
\put(32,20){\vector(1,1){27}}
\put(25,35){\mbox{$-m$}}
\put(51,32){\mbox{\scriptsize $[1\star]$}}

\put(10,0){

\put(110,0){\mbox{$m$}}
\put(110,100){\mbox{$m$}}
\put(101,93){\vector(1,0){25}}
\put(101,13){\vector(1,0){25}}
\put(130,90){\mbox{$[\overset{{\color{Green} 4}}{\bigcirc}]_{[12]}^{\scaleto{\{2N\}}{5.5pt}}$}}
\put(130,10){\mbox{$[\overset{{\color{Green} 4'}}{\bigcirc}]_{[21]}^{\scaleto{\{2N\}}{5.5pt}}$}}
\put(133,50){\mbox{$\bigoplus$}}
{\color{Green} 
\put(100,57){\vector(1,1){27}}
\put(100,50){\vector(1,-1){27}}}
\put(98,17){\mbox{\scriptsize $[2\star]$}}
\put(98,84){\mbox{\scriptsize $[\star 2]$}}
\put(120,35){\mbox{$0$}}
\put(120,60){\mbox{$0$}}
\put(98,32){\mbox{\scriptsize $[\star 1]$}}
\put(96,69){\mbox{\scriptsize $[1 \star]$}}

\put(10,0){

{\color{Green} 
\put(150,83){\vector(1,-1){27}}
\put(150,22){\vector(1,1){27}}}
\put(182,50){\mbox{$[\bigcirc]_{[22]}^{\scaleto{\{2(N+1)\}}{5.5pt}}\;\overset{0}{\longrightarrow} \; \varnothing$}}

\put(165,25){\mbox{$0$}}
\put(165,75){\mbox{$0$}}
\put(145,65){\mbox{\scriptsize $[\star 2]$}}
\put(145,35){\mbox{\scriptsize $[2 \star]$}}
}
}
}
    
\end{picture}
\caption{\footnotesize The complex for the trefoil knot in the bipartite presentation. We color in blue the arrows for the morphisms Sh and the corresponding labellings of spaces. The Sh morphisms go after the $\Delta$ morphisms, while the zero morphisms go after the $m$ morphisms. The arrows and labels in green correspond to zero morphisms. Spaces are enumerated by $[\alpha_1\, \alpha_2]$ with $\alpha_1=0,\,1,\,2$ corresponding to smoothings shown in Fig.\,\ref{fig:bip-vertex-smoothings}. Arrows are labelled by $[\alpha_1 \star]$ or $[\star \alpha_2]$ depending on the change of the enumerators of spaces.}
\label{fig:tw-trefoil-complex}
\end{figure}

The complex is shown in Fig.\,\ref{fig:tw-trefoil-complex}.
\noindent The subscripts of smoothings are their enumerators $[\alpha_1 \; \alpha_2]$, and curly brackets denote gradings. Note that the 3-hypercube structure is better noticeable in the first line of Fig.\,\ref{fig:2-locks-BD}. Here we squeeze the hypercube in Fig.\,\ref{fig:tw-trefoil-H-complex} in vertical direction to save space. The numbers under the circles are numbers of edges and label the spaces. The full differentials act between direct sums of spaces sharing the same vertical.

\medskip

\noindent $\bullet$ The $0$-th differential is
\begin{equation}
    \hat{d}_0 = \Delta + \Delta = \sum_{i=1}^N\left(\sum_{j=0}^{N-i}\left(\vartheta^{(2)}_{N-j}\vartheta^{\prime (2)}_{i+j} + \bar\vartheta^{(2)}_{N-j}\bar \vartheta^{\prime (2)}_{i+j}\right)\right)\frac{\partial}{\partial \vartheta^{(4)}_i} = \Delta\,:\quad V^{(4)} \; \mapsto \; V^{(2) + (\bar 2)} \otimes V^{\prime (2) + (\bar 2)}
\end{equation}
where the basis in $V^{(2) + (\bar 2)} \otimes V^{\prime (2) + (\bar 2)}$ is formed by vectors
\begin{equation} \nn
    \vartheta_i^{\scaleto{(2)+(\bar 2)}{5.5pt}} \vartheta_j^{\prime \scaleto{(2)+(\bar 2)}{5.5pt}} = \vartheta_i^{(2)} \vartheta_j^{\prime (2)} + \bar\vartheta_i^{(2)} \bar\vartheta_j^{\prime (2)}\,,\quad 1\leq i,\,j \leq N\,.
\end{equation}
This differential kernel and image are
{\footnotesize \begin{equation} \nn
\begin{aligned}
    {\rm Im}(\hat{d}_0) &= \langle \vartheta_1^{\scaleto{(2)+(\bar 2)}{5.5pt}} \vartheta_N^{\prime \scaleto{(2)+(\bar 2)}{5.5pt}} + \vartheta_2^{\scaleto{(2)+(\bar 2)}{5.5pt}} \vartheta_{N-1}^{\prime \scaleto{(2)+(\bar 2)}{5.5pt}} + \dots + \vartheta_N^{\scaleto{(2)+(\bar 2)}{5.5pt}} \vartheta_1^{\prime \scaleto{(2)+(\bar 2)}{5.5pt}},\, \vartheta_2^{\scaleto{(2)+(\bar 2)}{5.5pt}} \vartheta_N^{\prime \scaleto{(2)+(\bar 2)}{5.5pt}} + \dots + \vartheta_N^{\scaleto{(2)+(\bar 2)}{5.5pt}} \vartheta_2^{\prime \scaleto{(2)+(\bar 2)}{5.5pt}},\dots,\, \vartheta_{N}^{\scaleto{(2)+(\bar 2)}{5.5pt}} \vartheta_N^{\prime \scaleto{(2)+(\bar 2)}{5.5pt}} \rangle^{\scaleto{\{N-1\}}{5.5pt}}\,, \\
    {\rm Ker}(\hat{d}_0) &= \varnothing\,. 
\end{aligned}
\end{equation}}

\noindent $\bullet$ The $1$-st differential is
\begin{equation}
\begin{aligned}
    \hat{d}_1 &= {\rm Sh} + m + (-m) + {\rm Sh} =  \left(\sum_{i,j=1}^{N}\left(\vartheta^{{\color{blue} (2)}}_{i+1}\vartheta^{{\color{blue} \prime (2)}}_j-\vartheta^{{\color{blue} (2)}}_i\vartheta^{{\color{blue} \prime (2)}}_{j+1}\right)+\sum_{\substack{1\le i,j\le N\\i+j \le N+1}}\vartheta_{i+j-1}^{\prime (4)}\right)\frac{\partial^2}{\partial\vartheta_i^{(2)}\partial\vartheta_j^{\prime (2)}}+ \\
    &+\left(\sum_{i,j=1}^{N}\left(\bar\vartheta^{{\color{blue} (2)}}_{i+1}\bar\vartheta^{{\color{blue} \prime (2)}}_j- \bar\vartheta^{{\color{blue} (2)}}_i\bar\vartheta^{{\color{blue} \prime (2)}}_{j+1}\right)-\sum_{\substack{1\le i,j\le N\\i+j \le N+1}}\vartheta_{i+j-1}^{\prime (4)}\right)\frac{\partial^2}{\partial\bar\vartheta_i^{(2)}\partial\bar \vartheta_j^{\prime (2)}}\,,\quad \vartheta_{N+1} \equiv 0\,.
\end{aligned} 
\end{equation}
Its kernel and image are 
{\footnotesize \begin{equation} \nn
\begin{aligned}
    {\rm Im}(\hat{d}_1) &= \langle  \vartheta_{i+1}^{{\color{blue} (2)}}\vartheta_j^{{\color{blue}\prime (2)}} - \vartheta_i^{{\color{blue} (2)}}\vartheta_{j+1}^{{\color{blue} \prime (2)}}+\vartheta_{i+j-1}^{\prime (4)}\rangle^{\scaleto{\{N+1\}}{5.5pt}} \oplus \langle \vartheta^{\prime (4)}_N \rangle^{\scaleto{\{2(N-1)\}}{5.5pt}} \oplus \langle \bar\vartheta_{i+1}^{{\color{blue} (2)}}\bar\vartheta_j^{{\color{blue}\prime (2)}} - \bar\vartheta_i^{{\color{blue} (2)}}\bar\vartheta_{j+1}^{{\color{blue} \prime (2)}}-\vartheta_{i+j-1}^{\prime (4)}\rangle^{\scaleto{\{N+1\}}{5.5pt}},\, \vartheta_{N+1} \equiv 0\,, \\
    {\rm Ker}(\hat{d}_1) &= \langle \vartheta_1^{\scaleto{(2)+(\bar 2)}{5.5pt}} \vartheta_N^{\prime \scaleto{(2)+(\bar 2)}{5.5pt}} + \vartheta_2^{\scaleto{(2)+(\bar 2)}{5.5pt}} \vartheta_{N-1}^{\prime \scaleto{(2)+(\bar 2)}{5.5pt}} + \dots + \vartheta_N^{\scaleto{(2)+(\bar 2)}{5.5pt}} \vartheta_1^{\prime \scaleto{(2)+(\bar 2)}{5.5pt}},\, \vartheta_2^{\scaleto{(2)+(\bar 2)}{5.5pt}} \vartheta_N^{\prime \scaleto{(2)+(\bar 2)}{5.5pt}} + \dots + \vartheta_N^{\scaleto{(2)+(\bar 2)}{5.5pt}} \vartheta_2^{\prime \scaleto{(2)+(\bar 2)}{5.5pt}},\dots,\, \vartheta_{N}^{\scaleto{(2)+(\bar 2)}{5.5pt}} \vartheta_N^{\prime \scaleto{(2)+(\bar 2)}{5.5pt}} \rangle^{\scaleto{\{N-1\}}{5.5pt}} \oplus \\
    &\oplus \underbrace{\langle \vartheta_2^{(2)} \vartheta_N^{\prime (2)} + \dots + \vartheta_N^{(2)} \vartheta_2^{\prime (2)},\dots,\, \vartheta_N^{(2)} \vartheta_{N-1}^{\prime (2)} + \vartheta_{N-1}^{(2)} \vartheta_N^{\prime (2)},\, \vartheta_{N}^{(2)} \vartheta_N^{\prime (2)} \rangle^{\scaleto{\{N-1\}}{5.5pt}}}_{{\rm Ker}({\rm Sh}) \cap {\rm Ker}(m)}\,. \\
\end{aligned}
\end{equation}}

\noindent $\bullet$ The $2$-nd differential is
\begin{equation}
    \hat{d}_2 = m + 0 + 0 + m = \sum_{\substack{1\le i,j\le N\\i+j \le N+1}}\vartheta_{i+j-1}^{{\color{Green} (4)}}\frac{\partial^2}{\partial\vartheta_i^{{\color{blue}(2)}}\partial\vartheta_j^{{\color{blue} \prime (2)}}} + \sum_{\substack{1\le i,j\le N\\i+j \le N+1}}\vartheta_{i+j-1}^{{\color{Green} \prime (4)}}\frac{\partial^2}{\partial\bar\vartheta_i^{{\color{blue}(2)}}\partial\bar \vartheta_j^{{\color{blue}\prime (2)}}}\,.
\end{equation}
Its kernel and image are
\begin{equation} \nn
\begin{aligned}
    {\rm Im}(\hat{d}_2) &= V^{\scaleto{\{2N\}}{5.5pt}} \oplus V^{\scaleto{\{2N\}}{5.5pt}} \\
    {\rm Ker}(\hat{d}_2) &= \langle  \vartheta_{i+1}^{{\color{blue} (2)}}\vartheta_j^{{\color{blue}\prime (2)}} - \vartheta_i^{{\color{blue} (2)}}\vartheta_{j+1}^{{\color{blue} \prime (2)}}\rangle^{\scaleto{\{N+1\}}{5.5pt}} \oplus \langle \bar\vartheta_{i+1}^{{\color{blue} (2)}}\bar\vartheta_j^{{\color{blue}\prime (2)}} - \bar\vartheta_i^{{\color{blue} (2)}}\bar\vartheta_{j+1}^{{\color{blue} \prime (2)}}\rangle^{\scaleto{\{N+1\}}{5.5pt}}\oplus  \langle \vartheta^{\prime (4)}_i \rangle^{\scaleto{\{2(N-1)\}}{5.5pt}},\, \vartheta_{N+1} \equiv 0
\end{aligned}
\end{equation}

\noindent $\bullet$ The $3$-rd differential is
\begin{equation}
\begin{aligned}
    \hat{d}_3 &= 0 + 0\,.
\end{aligned}
\end{equation}
Its kernel and image are
\begin{equation} \nn
\begin{aligned}
    {\rm Im}(\hat{d}_3) &= \varnothing\,, \\
    {\rm Ker}(\hat{d}_3) &= V^{\scaleto{\{2N\}}{5.5pt}} \oplus V^{\scaleto{\{2N\}}{5.5pt}}\,.
\end{aligned}
\end{equation}
The cohomologies and their quantum dimensions are
\begin{equation} \nn
\begin{aligned}
    {\cal H}_0^{3_1} &= {\rm Ker}(\hat{d}_0) = \varnothing \quad \Longrightarrow \quad \boxed{\dim_q {\cal H}_0^{3_1} = 0\,,} \\
    {\cal H}_1^{3_1} &= {\rm Ker}(\hat{d}_1)\backslash {\rm Im}(\hat{d}_0) = ({\rm Ker}({\rm Sh}) \cap {\rm Ker}(m))^{\scaleto{\{N-1\}}{5.5pt}} \cong \langle \sum_{i=k}^N \vartheta_i^{(2)} \vartheta_{N+k-i}^{\prime (2)} \rangle^{\scaleto{\{N-1\}}{5.5pt}} ,\, k > 1 \\
    &\Longrightarrow \quad \boxed{\dim_q {\cal H}_1^{3_1} = [N]-q^{N-1}=q^{-1}[N-1]\,,} \\
    {\cal H}_2^{3_1} &= {\rm Ker}(\hat{d}_2)\backslash {\rm Im}(\hat{d}_1) = \langle \vartheta^{\prime (4)}_{i<N} \rangle^{\scaleto{\{2(N-1)\}}{5.5pt}} \quad \Longrightarrow \quad \boxed{\dim_q {\cal H}_2^{3_1} = q^{2(N-1)}([N]-q^{-N+1})=q^{2N-1}[N-1]\,,} \\
    {\cal H}_3^{3_1} &= {\rm Ker}(\hat{d}_3)\backslash {\rm Im}(\hat{d}_2) \overset{{\rm Ker}(\hat{d}_3)={\rm Im}(\hat{d}_2)}{=} \varnothing \quad \Longrightarrow \quad \boxed{\dim_q {\cal H}_3^{3_1} = 0} \\
    {\cal H}_4^{3_1} &= {\rm Ker}(0) \backslash {\rm Im}(\hat{d}_3) = V^{\scaleto{\{2(N+1)\}}{5.5pt}} \quad \Longrightarrow \quad \boxed{\dim_q {\cal H}_4^{3_1} = q^{2N+2}[N]\,.}
\end{aligned}
\end{equation}
The Khovanov--Rozansky polynomial is
{\small \begin{equation} \nn
    \boxed{\boxed{P^{3_1}(A,q,T) = (q^{N}T)^{-4} \left(T {\rm dim}_q {\cal H}_1^{3_1} + T^2 {\rm dim}_q {\cal H}_2^{3_1} + T^4 {\rm dim}_q {\cal H}_4^{3_1}\right) = q^{-2N-1}T^{-2}[N-1](1+q^{-2N}T^{-1})  + q^{-2N+2} [N]}}
\end{equation}}
what reproduces the known answer in \cite{carqueville2014computing,anokhina2014towards-R}.

The graded Euler characteristic is
\begin{equation}
\begin{aligned}
    H^{3_1}(A,q) & \overset{{\rm Fig.\,}\ref{fig:tw-trefoil-complex}}{=} A^{-4}\left(D - 2q^{N-1} D^2 + 2 q^{N+1} D^2 + q^{2(N-1)} D - 2 q^{2 N} D + q^{2(N+1)} D\right) = \\
    & \overset{{\rm Fig.\,}\ref{fig:tw-trefoil-2-complex}}{=} A^{-4} \cdot D (1 + 2\phi D + \phi^2) = D\left(1-\left(1-A^{-4}\right)\frac{\{Aq\}\{A/q\}}{\{A\}^2}\right) = P^{3_1}(A,q,T=-1).
\end{aligned}
\end{equation}

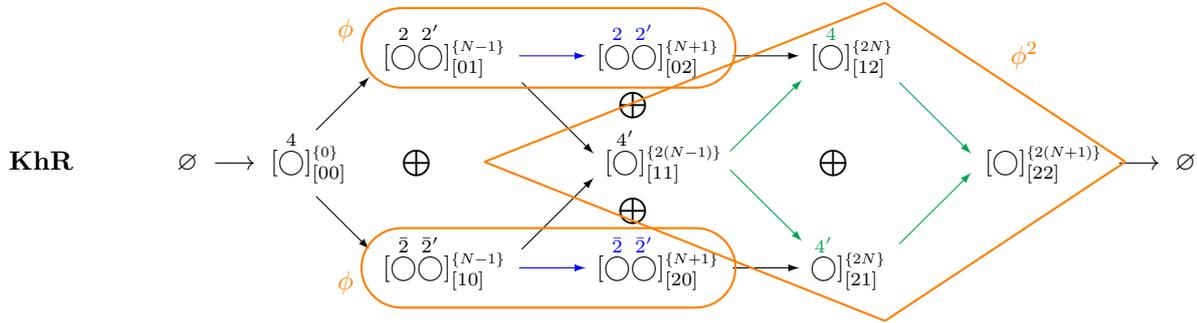
\begin{figure}[h!]
\centering
\begin{picture}(100,120)(40,-5)

\put(-150,50){\mbox{\bf KhR}}

\put(-87,50){\mbox{$\varnothing\;\longrightarrow \; [\overset{4}{\bigcirc}]_{[00]}^{\scaleto{\{0\}}{5.5pt}}$}}
\put(-35,65){\vector(1,1){20}}
\put(-35,40){\vector(1,-1){20}}

\put(-10,90){\mbox{$[\overset{2}{\bigcirc} \overset{2'}{\bigcirc}]_{[01]}^{\scaleto{\{N-1\}}{5.5pt}}$}}
\put(-10,10){\mbox{$[\overset{\bar 2}{\bigcirc} \overset{\bar 2'}{\bigcirc}]_{[10]}^{\scaleto{\{N-1\}}{5.5pt}}$}}
\put(-3,50){\mbox{$\bigoplus$}}

\put(10,0){

\put(60,90){\mbox{$[\overset{{\color{blue} 2}}{\bigcirc} \overset{{\color{blue} 2'}}{\bigcirc}]_{[02]}^{\scaleto{\{N+1\}}{5.5pt}}$}}
\put(60,10){\mbox{$[\overset{{\color{blue} \bar 2}}{\bigcirc} \overset{{\color{blue} \bar 2'}}{\bigcirc}]_{[20]}^{\scaleto{\{N+1\}}{5.5pt}}$}}
{\color{blue} \put(31,93){\vector(1,0){25}}
\put(31,13){\vector(1,0){25}}}

\put(68,32){\mbox{$\bigoplus$}}
\put(68,72){\mbox{$\bigoplus$}}
\put(63,50){\mbox{$[\overset{4'}{\bigcirc}]_{[11]}^{\scaleto{\{2(N-1)\}}{5.5pt}}$}}
\put(32,83){\vector(1,-1){27}}
\put(32,20){\vector(1,1){27}}

\put(10,0){

\put(101,93){\vector(1,0){25}}
\put(101,13){\vector(1,0){25}}
\put(130,90){\mbox{$[\overset{{\color{Green} 4}}{\bigcirc}]_{[12]}^{\scaleto{\{2N\}}{5.5pt}}$}}
\put(130,10){\mbox{$\overset{{\color{Green} 4'}}{\bigcirc}]_{[21]}^{\scaleto{\{2N\}}{5.5pt}}$}}
\put(133,50){\mbox{$\bigoplus$}}
{\color{Green} 
\put(100,57){\vector(1,1){27}}
\put(100,50){\vector(1,-1){27}}}

\put(10,0){

{\color{Green} \put(150,83){\vector(1,-1){27}}
\put(150,22){\vector(1,1){27}}}
\put(182,50){\mbox{$[\bigcirc]_{[22]}^{\scaleto{\{2(N+1)\}}{5.5pt}}\;\longrightarrow \; \varnothing$}}

}
}
}

\thicklines

{\color{orange}

\put(52,96){\oval(140,30)}

\put(52,13){\oval(140,30)}

\put(-50,0){

\put(78,53){\line(2.5,1){150}}
\put(78,53){\line(2.5,-1){150}}
\put(228,113){\line(1.5,-1){90}}
\put(228,-7){\line(1.5,1){90}}

\put(275,90){\mbox{$\phi^2$}}
}

\put(-27,100){\mbox{$\phi$}}
\put(-27,5){\mbox{$\phi$}}

}
    
\end{picture}
\caption{\footnotesize The complex for the trefoil knot in the bipartite presentation. We color in blue the arrows for the morphisms Sh and in green the arrows for the zero morphisms. The elements in rhomb and ovals connected by these colored arrows are shrunk when transferring to the precursor complex.}
\label{fig:tw-trefoil-complex-min}
\end{figure}

\paragraph{Precursor.} Let us rewrite the complex in Fig.\,\ref{fig:tw-trefoil-complex} in the minimalistic form, see Fig.\,\ref{fig:tw-trefoil-complex-min}. 

\noindent The circled spaces can be shrunk to single ones without changing the HOMFLY polynomial. Instead of ordinary gradings, these spaces have powers of $\phi$ as multipliers:

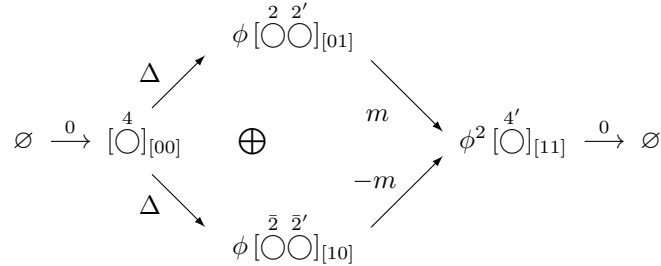
\begin{figure}[h!]
\centering
\begin{picture}(100,100)(-15,5)

\put(-87,50){\mbox{$\varnothing\;\overset{0}{\longrightarrow} \; [\overset{4}{\bigcirc}]_{[00]}$}}
\put(-35,65){\vector(1,1){20}}
\put(-35,40){\vector(1,-1){20}}
\put(-40,75){\mbox{$\Delta$}}
\put(-40,25){\mbox{$\Delta$}}

\put(-5,90){\mbox{$\phi\,[\overset{2}{\bigcirc}\overset{2'}{\bigcirc}]_{[01]}$}}
\put(-5,10){\mbox{$\phi\,[\overset{\bar 2}{\bigcirc}\overset{\bar 2'}{\bigcirc}]_{[10]}$}}
\put(-3,50){\mbox{$\bigoplus$}}

\put(15,0){

\put(65,50){\mbox{$\phi^2\,[\overset{4'}{\bigcirc}]_{[11]}\; \overset{0}{\longrightarrow}\; \varnothing$}}
\put(32,83){\vector(1,-1){27}}
\put(32,20){\vector(1,1){27}}
\put(30,60){\mbox{$m$}}
\put(25,35){\mbox{$-m$}}
}
    
\end{picture}
\caption{\footnotesize The 2-hypercube for the trefoil knot in the bipartite presentation with shrunk colored edges. This complex cannot give the Khovanov--Rozansky polynomial but its weighted (with $\phi$ multipliers) sum is the HOMFLY polynomial for the trefoil knot.}
\label{fig:tw-trefoil-2-complex}
\end{figure}

\noindent If we put $\phi \rightarrow -q$ and consider $N=2$, we get the precursor complex:
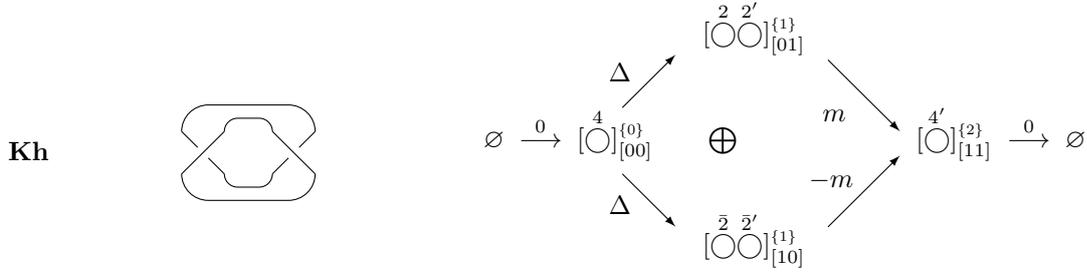
\begin{figure}[h!]
    \centering
\begin{picture}(300,100)(-50,0)

\put(-125,45){\mbox{\bf Kh}}

\put(40,15){

\put(-100,25){\line(1,1){16}}
\put(-100,41){\line(1,-1){6}}
\put(-90,31){\line(1,-1){6}}

\put(-50,25){\line(-1,1){16}}
\put(-50,41){\line(-1,-1){6}}
\put(-60,31){\line(-1,-1){6}}

\put(-75,25){\oval(18,10)[b]}
\put(-75,41){\oval(18,10)[t]}

\put(-75,25){\oval(50,20)[b]}
\put(-75,41){\oval(50,20)[t]}
}

\put(140,0){

\put(-87,50){\mbox{$\varnothing\;\overset{0}{\longrightarrow} \; [\overset{4}{\bigcirc}]_{[00]}^{\scaleto{\{0\}}{5.5pt}}$}}
\put(-35,65){\vector(1,1){20}}
\put(-35,40){\vector(1,-1){20}}
\put(-40,75){\mbox{$\Delta$}}
\put(-40,25){\mbox{$\Delta$}}

\put(-5,90){\mbox{$[\overset{2}{\bigcirc}\overset{2'}{\bigcirc}]_{[01]}^{\scaleto{\{1\}}{5.5pt}}$}}
\put(-5,10){\mbox{$[\overset{\bar 2}{\bigcirc}\overset{\bar 2'}{\bigcirc}]_{[10]}^{\scaleto{\{1\}}{5.5pt}}$}}
\put(-3,50){\mbox{$\bigoplus$}}

\put(10,0){

\put(65,50){\mbox{$[\overset{4'}{\bigcirc}]_{[11]}^{\scaleto{\{2\}}{5.5pt}}\; \overset{0}{\longrightarrow}\; \varnothing$}}
\put(32,83){\vector(1,-1){27}}
\put(32,20){\vector(1,1){27}}
\put(30,60){\mbox{$m$}}
\put(25,35){\mbox{$-m$}}
}
}

\end{picture}
    \caption{\footnotesize The two unknots being the precursor diagram of the bipartite trefoil knot and its Khovanov complex.}
    \label{fig:enter-label}
\end{figure}

\medskip

\noindent $\bullet$ The zeroth differential is
\begin{equation}
    \hat{d}_0^{\,\bigcirc \bigcirc} = \underbrace{\left( \vartheta_2^{(2)} \vartheta_2^{\prime (2)} + \bar\vartheta_2^{(2)} \bar\vartheta_2^{\prime (2)}\right)}_{\vartheta_2^{\scaleto{(2)+(\bar 2)}{5.5pt}} \vartheta_2^{\prime \scaleto{(2)+(\bar 2)}{5.5pt}}}\frac{\partial}{\partial \vartheta_2^{(4)}} + \Big( \underbrace{\vartheta_1^{(2)} \vth_2^{\prime (2)} + \bar\vartheta_1^{(2)} \bar\vth_2^{\prime (2)}}_{\vartheta_1^{\scaleto{(2)+(\bar 2)}{5.5pt}} \vth_2^{\prime \scaleto{(2)+(\bar 2)}{5.5pt}}} + \underbrace{\vth_2^{(2)} \vartheta_1^{\prime (2)} + \bar\vth_2^{(2)} \bar\vartheta_1^{\prime (2)}}_{\vth_2^{\scaleto{(2)+(\bar 2)}{5.5pt}} \vartheta_1^{\prime \scaleto{(2)+(\bar 2)}{5.5pt}}} \Big) \frac{\partial}{\partial \vth_1^{(4)}}\,.
\end{equation}
Its image and kernel are
\begin{equation} \nn
\begin{aligned}
    \Im(\hat{d}_0^{\,\bigcirc \bigcirc}) &= \langle \vartheta_2^{\scaleto{(2)+(\bar 2)}{5.5pt}} \vartheta_2^{\prime \scaleto{(2)+(\bar 2)}{5.5pt}},\,  \vartheta_1^{\scaleto{(2)+(\bar 2)}{5.5pt}} \vth_2^{\prime \scaleto{(2)+(\bar 2)}{5.5pt}} + \vth_2^{\scaleto{(2)+(\bar 2)}{5.5pt}} \vartheta_1^{\prime \scaleto{(2)+(\bar 2)}{5.5pt}} \rangle\,, \\
    \Ker(\hat{d}_0^{\,\bigcirc \bigcirc}) &= \varnothing\,. 
\end{aligned}
\end{equation}
$\bullet$ The first differential is
\begin{equation}
    \hat{d}_1^{\,\bigcirc \bigcirc} = \vartheta_2^{\prime (4)} \Bigg( \underbrace{\frac{\partial}{\partial\vth_1^{(2)}} \frac{\partial}{\partial\vartheta_2^{\prime (2)}} - \frac{\partial}{\partial\bar \vth_1^{(2)}} \frac{\partial}{\partial\bar \vartheta_2^{\prime (2)}}}_{\frac{\partial}{\partial\vth_1^{\scaleto{(2)-(\bar 2)}{5pt}}} \frac{\partial}{\partial\vth_2^{\prime \scaleto{(2)-(\bar 2)}{5pt}}}} + \underbrace{\frac{\partial}{\partial\vartheta_2^{(2)}} \frac{\partial}{\partial\vth_1^{\prime (2)}} - \frac{\partial}{\partial\bar \vartheta_2^{(2)}} \frac{\partial}{\partial\bar \vth_1^{\prime (2)}}}_{\frac{\partial}{\partial\vth_2^{\scaleto{(2)-(\bar 2)}{5pt}}} \frac{\partial}{\partial\vth_1^{\prime \scaleto{(2)-(\bar 2)}{5pt}}}} \Bigg) + \vth_1^{\prime (4)} \underbrace{\left(\frac{\partial}{\partial\vth_1^{(2)}} \frac{\partial}{\partial\vth_1^{\prime (2)}} - \frac{\partial}{\partial\bar\vth_1^{(2)}} \frac{\partial}{\partial\bar\vth_1^{\prime (2)}} \right)}_{\frac{\partial}{\partial\vth_1^{\scaleto{(2)-(\bar 2)}{5pt}}} \frac{\partial}{\partial\vth_1^{\prime \scaleto{(2)-(\bar 2)}{5pt}}}}.
\end{equation}
Its image and kernel are
{\small \ba \nn
\Im(\hat{d}_1^{\,\bigcirc \bigcirc})&=V\,, \\
\Ker(\hat{d}_1^{\,\bigcirc \bigcirc})&= \langle \vartheta_2^{\scaleto{(2)+(\bar 2)}{5.5pt}} \vartheta_2^{\prime \scaleto{(2)+(\bar 2)}{5.5pt}},\, \vartheta_1^{\scaleto{(2)+(\bar 2)}{5.5pt}} \vartheta_1^{\prime \scaleto{(2)+(\bar 2)}{5.5pt}},\, \vartheta_1^{\scaleto{(2)+(\bar 2)}{5.5pt}} \vartheta_2^{\prime \scaleto{(2)+(\bar 2)}{5.5pt}},\, \vartheta_2^{\scaleto{(2)+(\bar 2)}{5.5pt}} \vartheta_1^{\prime \scaleto{(2)+(\bar 2)}{5.5pt}} \rangle \oplus \langle \vartheta_2^{\scaleto{(2)-(\bar 2)}{5.5pt}} \vartheta_2^{\prime \scaleto{(2)-(\bar 2)}{5.5pt}},\, \vartheta_1^{\scaleto{(2)-(\bar 2)}{5.5pt}} \vartheta_2^{\prime \scaleto{(2)-(\bar 2)}{5.5pt}} - \vartheta_2^{\scaleto{(2)-(\bar 2)}{5.5pt}} \vartheta_1^{\prime \scaleto{(2)-(\bar 2)}{5.5pt}} \rangle\,.
\ea}

\noindent Now, let us compute the cohomologies and their quantum dimensions:
{\small \ba \nn
{\cal H}_0 &= \Ker(\hat{d}_0^{\,\bigcirc \bigcirc}) = \varnothing \quad \Longrightarrow \quad \boxed{\dim_q {\cal H}_0 = 0\,,} \\
{\cal H}_1 &= \Ker(\hat{d}_1^{\,\bigcirc \bigcirc}) \backslash \Im(\hat{d}_0^{\,\bigcirc \bigcirc}) = \langle \vartheta_1^{\scaleto{(2)+(\bar 2)}{5.5pt}}\vartheta_1^{\prime \scaleto{(2)+(\bar 2)}{5.5pt}},\,  \vartheta_1^{\scaleto{(2)+(\bar 2)}{5.5pt}} \vth_2^{\prime \scaleto{(2)+(\bar 2)}{5.5pt}} - \vth_2^{\scaleto{(2)+(\bar 2)}{5.5pt}} \vartheta_1^{\prime \scaleto{(2)+(\bar 2)}{5.5pt}} \rangle \oplus \langle \vartheta_2^{\scaleto{(2)-(\bar 2)}{5.5pt}} \vartheta_2^{\prime \scaleto{(2)-(\bar 2)}{5.5pt}},\, \vartheta_1^{\scaleto{(2)-(\bar 2)}{5.5pt}} \vartheta_2^{\prime \scaleto{(2)-(\bar 2)}{5.5pt}} - \vartheta_2^{\scaleto{(2)-(\bar 2)}{5.5pt}} \vartheta_1^{\prime \scaleto{(2)-(\bar 2)}{5.5pt}} \rangle \quad \Longrightarrow \quad \dim_q {\cal H}_0 = 0 \\
&\Lra \quad \boxed{\dim_q {\cal H}_1 = q^2 + q^{-2} + 2 = (q+q^{-1})^2\,.}
\ea} 

\noindent The Khovanov polynomial is
\begin{equation}
    {\rm Kh}^{\bigcirc\bigcirc} = (qT)^{-1} \cdot qT \dim_q {\cal H}_1 = D_2^2\,.
\end{equation}

\paragraph{Summary.} This example is our first 2-lock example. It shows that there can appear both the Sh and zero morphisms in a complex. It is also the first example when four spaces are shrunk in one with $\phi^2$ multiplier to transfer to the precursor complex.


\subsubsection{Mirror trefoil knot: $\overline{\text{Tw}}_2$}

\begin{figure}[h!]
\centering
\begin{picture}(100,110)(40,-5)

\put(-150,50){\mbox{\bf KhR}}

\put(-87,50){\mbox{$\varnothing\;\overset{0}{\longrightarrow} \; [\overset{4}{\bigcirc}]_{[00]}^{\scaleto{\{0\}}{5.5pt}}$}}
{\color{Green}\put(-35,65){\vector(1,1){20}}
\put(-35,40){\vector(1,-1){20}}}
\put(-40,75){\mbox{$0$}}
\put(-25,65){\mbox{\scriptsize $[0\star]$}}
\put(-40,25){\mbox{$0$}}
\put(-25,35){\mbox{\scriptsize $[\star 0]$}}

\put(-10,90){\mbox{$[\overset{\bar 4}{\bigcirc}]_{[01]}^{\scaleto{\{2\}}{5.5pt}}$}}
\put(-10,10){\mbox{$[\overset{\bar 4'}{\bigcirc}]_{[10]}^{\scaleto{\{2\}}{5.5pt}}$}}
\put(-3,50){\mbox{$\bigoplus$}}

\put(-10,0){

\put(60,90){\mbox{$[\overset{2}{\bigcirc}\overset{2'}{\bigcirc}]_{[02]}^{\scaleto{\{N+1\}}{5.5pt}}$}}
\put(60,10){\mbox{$[\overset{\bar 2}{\bigcirc}\overset{\bar 2'}{\bigcirc}]_{[20]}^{\scaleto{\{N+1\}}{5.5pt}}$}}
\put(31,93){\vector(1,0){25}}
\put(31,13){\vector(1,0){25}}
\put(35,100){\mbox{$\Delta$}}
\put(45,17){\mbox{\scriptsize $[\star 0]$}}
\put(45,84){\mbox{\scriptsize $[0\star]$}}
\put(35,0){\mbox{$\Delta$}}

\put(70,32){\mbox{$\bigoplus$}}
\put(70,72){\mbox{$\bigoplus$}}
\put(70,50){\mbox{$[\overset{{\color{Green} 4}}{\bigcirc}]_{[11]}^{\scaleto{\{4\}}{5.5pt}}$}}
{\color{Green}
\put(32,83){\vector(1,-1){27}}
\put(32,20){\vector(1,1){27}}}
\put(35,60){\mbox{$0$}}
\put(51,69){\mbox{\scriptsize $[\star 1]$}}
\put(35,35){\mbox{$0$}}
\put(51,32){\mbox{\scriptsize $[1\star]$}}

\put(10,0){

\put(110,0){\mbox{\small \Sh}}
\put(110,100){\mbox{\small \Sh}}
{\color{blue}\put(101,93){\vector(1,0){25}}
\put(101,13){\vector(1,0){25}}}
\put(130,90){\mbox{$[\overset{{\color{blue} 2}}{\bigcirc}\overset{{\color{blue} 2'}}{\bigcirc}]_{[12]}^{\scaleto{\{N+3\}}{5.5pt}}$}}
\put(130,10){\mbox{$[\overset{{\color{blue} \bar 2}}{\bigcirc}\overset{{\color{blue} \bar 2'}}{\bigcirc}]_{[21]}^{\scaleto{\{N+3\}}{5.5pt}}$}}
\put(133,50){\mbox{$\bigoplus$}}
\put(100,57){\vector(1,1){27}}
\put(100,50){\vector(1,-1){27}}
\put(98,17){\mbox{\scriptsize $[2\star]$}}
\put(98,84){\mbox{\scriptsize $[\star 2]$}}
\put(120,35){\mbox{$\Delta$}}
\put(115,60){\mbox{$-\Delta$}}
\put(98,32){\mbox{\scriptsize $[\star 1]$}}
\put(96,69){\mbox{\scriptsize $[1 \star]$}}

\put(10,0){

\put(150,83){\vector(1,-1){27}}
\put(150,22){\vector(1,1){27}}
\put(182,50){\mbox{$[\overset{4'}{\bigcirc}]_{[22]}^{\scaleto{\{2(N+1)\}}{5.5pt}}\;\overset{0}{\longrightarrow} \; \varnothing$}}

\put(165,25){\mbox{$m$}}
\put(165,75){\mbox{$m$}}
\put(145,65){\mbox{\scriptsize $[\star 2]$}}
\put(145,35){\mbox{\scriptsize $[2 \star]$}}
}
}
}
    
\end{picture}
\caption{\footnotesize The complex for the mirror trefoil knot in the bipartite presentation. We color in blue and green the arrows for the the morphisms Sh and zero, respectively, and the corresponding labellings of spaces. Spaces are enumerated by $[\alpha_1\, \alpha_2]$ with $\alpha_1=0,\,1,\,2$ corresponding to smoothings shown in Fig.\,\ref{fig:bip-vertex-smoothings}. Arrows are labelled by $[\alpha_1 \star]$ or $[\star \alpha_2]$ depending on the change of the enumerators of spaces.}
\label{fig:mirr-tw-trefoil-complex}
\end{figure}
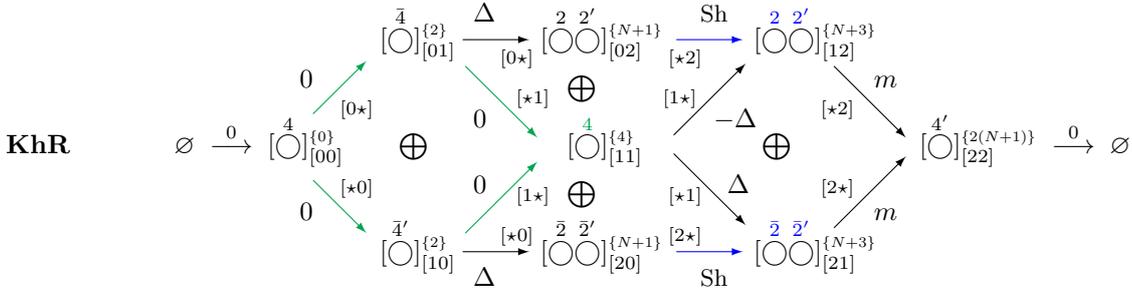

The complex is shown in Fig.\,\ref{fig:mirr-tw-trefoil-complex}.
\noindent Here the subscripts of smoothings are their enumerators $[\alpha_1 \; \alpha_2]$, and curly brackets denote grading shifts. The numbers under the circles are numbers of edges and label the spaces. The full differentials act between direct sums of spaces sharing the same vertical.

\medskip

\noindent $\bullet$ The $0$-th differential is
\begin{equation}
    \hd_0 = 0 + 0\,.
\end{equation}
Its kernel and image are
\begin{equation} \nn
\begin{aligned}
    \Ker(\hd_0) &= V^{\scaleto{\{0\}}{5.5pt}}\,, \\
    \Im(\hd_0) &= \varnothing\,.
\end{aligned}
\end{equation}

\noindent $\bullet$ The $1$-st differential is
\begin{equation}
    \hd_1 = \Delta + 0 + 0 + \Delta = \sum_{i=1}^N\sum_{j=0}^{N-i} \vartheta^{(2)}_{N-j} \vartheta^{\prime (2)}_{i+j}\frac{\partial}{\partial \bar \vartheta^{(4)}_i} + \sum_{i=1}^N\sum_{j=0}^{N-i} \bar \vartheta^{(2)}_{N-j} \bar \vartheta^{\prime (2)}_{i+j}\frac{\partial}{\partial \bar \vartheta^{\prime (4)}_i}\,. 
\end{equation}
Its kernel and image are
\ba \nn
\Ker(\hd_1) &= \varnothing\,, \\
\Im(\hd_1) &= \Im(\Delta)^{\scaleto{\{2\}}{5.5pt}} \oplus \Im(\Delta)^{\scaleto{\{2\}}{5.5pt}}\,.
\ea 

\noindent $\bullet$ The $2$-nd differential is
\begin{equation}
\begin{aligned}
    \hd_2 &= \Sh + (-\Delta) + \Delta + \Sh = \sum_{i,j=1}^{N}\left(\vartheta^{{\color{blue} (2)}}_{i+1}\vartheta^{{\color{blue}\prime (2)}}_j-\vartheta^{{\color{blue} (2)}}_i\vartheta^{{\color{blue}\prime (2)}}_{j+1}\right)\frac{\partial^2}{\partial\vartheta_i^{(2)}\partial\vartheta_j^{\prime (2)}} - \sum_{i=1}^N\sum_{j=0}^{N-i} \vartheta^{{\color{blue}(2)}}_{N-j} \vartheta^{{\color{blue}\prime (2)}}_{i+j}\frac{\partial}{\partial \vartheta^{{\color{Green} (4)}}_i} + \\
    &+\sum_{i,j=1}^{N}\left(\bar \vartheta^{{\color{blue} (2)}}_{i+1}\bar \vartheta^{{\color{blue}\prime (2)}}_j-\bar \vartheta^{{\color{blue} (2)}}_i\bar \vartheta^{{\color{blue}\prime (2)}}_{j+1}\right)\frac{\partial^2}{\partial\bar\vartheta_i^{(2)}\partial\bar \vartheta_j^{\prime (2)}} + \sum_{i=1}^N\sum_{j=0}^{N-i} \bar \vartheta^{{\color{blue}(2)}}_{N-j} \bar \vartheta^{{\color{blue}\prime (2)}}_{i+j}\frac{\partial}{\partial \vartheta^{{\color{Green} (4)}}_i}\,.
\end{aligned}
\end{equation}
Its kernel and image are
\ba \nn
\Ker(\hd_2) &= \Ker(\Sh)^{\scaleto{\{N+1\}}{5.5pt}} \oplus \Ker(\Sh)^{\scaleto{\{N+1\}}{5.5pt}} \oplus \langle \tilde{\vth}_2^{{\color{blue} (4)}},\, \tilde{\vth}_3^{{\color{blue} (4)}},\, \dots,\, \tilde{\vth}_N^{{\color{blue} (4)}}\rangle^{\scaleto{\{4\}}{5.5pt}}\,, \\
\Im(\hd_2) &= \Im(\Sh)^{\scaleto{\{N+3\}}{5.5pt}} \oplus \Im(\Sh)^{\scaleto{\{N+3\}}{5.5pt}} \oplus \langle \sum_{j=0}^{N-1} \vartheta^{{\color{blue}(2)}}_{N-j} \vartheta^{{\color{blue}\prime (2)}}_{1+j} - \sum_{j=0}^{N-1} \bar\vartheta^{{\color{blue}(2)}}_{N-j} \bar\vartheta^{{\color{blue}\prime (2)}}_{1+j} \rangle^{\scaleto{\{N+1\}}{5.5pt}}\,.
\ea 

\noindent $\bullet$ The $3$-rd differential is
\begin{equation}
    \hd_3 = m + m = \sum_{\substack{1\le i,j\le N\\i+j \le N+1}}\vartheta_{i+j-1}^{\prime (4)}\left(\frac{\partial^2}{\partial\vartheta_i^{{\color{blue}(2)}}\partial\vartheta_j^{{\color{blue}\prime (2)}}} + \frac{\partial^2}{\partial\bar\vartheta_i^{{\color{blue}(2)}}\partial\bar\vartheta_j^{{\color{blue}\prime (2)}}} \right)\,.
\end{equation}
Its kernel and image are
\ba \nn
\Ker(\hd_3) &= \Ker(m)^{\scaleto{\{N+3\}}{5.5pt}} \oplus (V\otimes V)^{\scaleto{\{N+3\}}{5.5pt}}\,, \\
\Im(\hd_3) &= V^{\scaleto{\{2(N+1)\}}{5.5pt}}\,.
\ea 
The cohomologies and their quantum dimensions are
{\small \ba \nn
\cH_0^{\bar 3_1} &= \Ker(\hd_0) \quad \Lra \quad \boxed{\dim_q\cH_0^{\bar 3_1} = [N]\,,} \\
\cH_1^{\bar 3_1} &= \Ker(\hd_1) \backslash \Im(\hd_0) = \varnothing \quad \Lra \quad \boxed{\dim_q\cH_1^{\bar 3_1} = 0\,,} \\
\cH_2^{\bar 3_1} &= \Ker(\hd_2) \backslash \Im(\hd_1) = \langle \tilde{\vth}_2^{{\color{blue} (4)}},\, \tilde{\vth}_3^{{\color{blue} (4)}},\, \dots,\, \tilde{\vth}_N^{{\color{blue} (4)}}\rangle^{\scaleto{\{4\}}{5.5pt}} \quad \Lra \quad \boxed{\dim_q\cH_2^{\bar 3_1} = q^4([N] - q^{N-1}) = q^3 [N-1]\,,} \\
\cH_3^{\bar 3_1} &= \Ker(\hd_3) \backslash \Im(\hd_2) = \langle \sum_{i+j=k+1} \vth_i^{-} \vth_j^{-} \rangle^{\scaleto{\{N+3\}}{5.5pt}}\,,\; k<N \quad \Lra \quad \boxed{\dim_q\cH_3^{\bar 3_1} = q^{N+3} \sum_{k=1}^{N-1} q^{N-2i+1}\cdot q^{N-2(k-i+1)+1}=q^{2N+3}[N-1]\,,} \\
\cH_4^{\bar 3_1} &= \Ker(0) \backslash \Im(\hd_3) = \varnothing \quad \Lra \quad \boxed{\dim_q\cH_4^{\bar 3_1} = 0\,,}
\ea} 

\noindent where $\vth_i^{-} \vth_j^{-} = \vth_i^{{\color{blue}(2)}} \vth_j^{{\color{blue}\prime (2)}} - \bar \vth_i^{{\color{blue}(2)}} \bar \vth_j^{{\color{blue}\prime (2)}}$. The Khovanov--Rozansky polynomial is
\begin{equation}
    \boxed{\boxed{P^{\bar 3_1}(A,q,T) = q^{2N-2} \left({\rm dim}_q {\cal H}_0^{3_1} + T^2 {\rm dim}_q {\cal H}_2^{3_1} + T^3 {\rm dim}_q {\cal H}_3^{3_1}\right) = q^{2N+1}T^{2}[N-1](1+q^{2N}T)  + q^{2N-2} [N]}}
\end{equation}
what reproduces the known answer in \cite{carqueville2014computing,anokhina2014towards-R}.

The graded Euler characteristic is
\begin{equation}
\begin{aligned}
    H^{\bar 3_1}(A,q) & \overset{{\rm Fig.\,}\ref{fig:mirr-tw-trefoil-complex}}{=} A^2 q^{-2} \left( D - 2 q^2 D + 2 q^{N+1} D^2 + q^4 D - 2 q^{N+3} D^2 + q^{2(N+1)} D  \right) = \\
    & \overset{{\rm Fig.\,}\ref{fig:mirr-tw-trefoil-2-complex}}{=} A^4 \cdot D \left( 1 + 2 \bphi D + \bphi^2 \right) = D\left(1-\left(1-A^{4}\right)\frac{\{Aq\}\{A/q\}}{\{A\}^2}\right).
\end{aligned}
\end{equation}

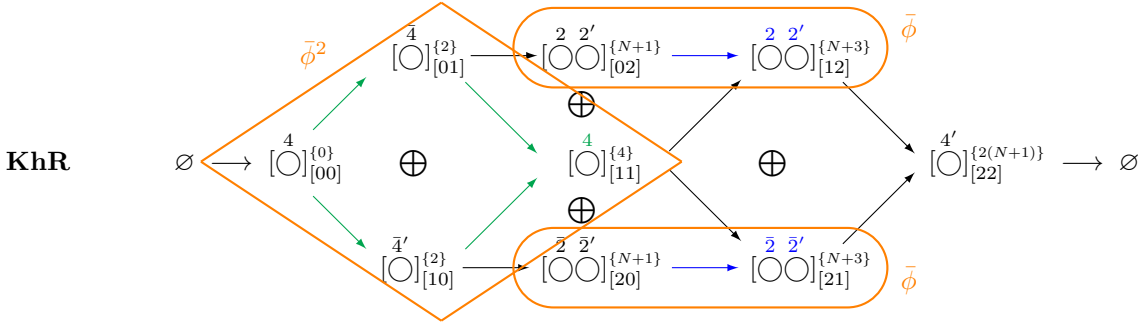
\begin{figure}[h!]
\centering
\begin{picture}(100,110)(40,-1)

\put(-150,50){\mbox{\bf KhR}}

\put(-87,50){\mbox{$\varnothing\;\longrightarrow \; [\overset{4}{\bigcirc}]_{[00]}^{\scaleto{\{0\}}{5.5pt}}$}}
{\color{Green}
\put(-35,65){\vector(1,1){20}}
\put(-35,40){\vector(1,-1){20}}}

\put(-6,90){\mbox{$[\overset{\bar 4}{\bigcirc}]_{[01]}^{\scaleto{\{2\}}{5.5pt}}$}}
\put(-10,10){\mbox{$[\overset{\bar 4'}{\bigcirc}]_{[10]}^{\scaleto{\{2\}}{5.5pt}}$}}
\put(-3,50){\mbox{$\bigoplus$}}

\put(-10,0){

\put(60,90){\mbox{$[\overset{2}{\bigcirc}\overset{2'}{\bigcirc}]_{[02]}^{\scaleto{\{N+1\}}{5.5pt}}$}}
\put(60,10){\mbox{$[\overset{\bar 2}{\bigcirc}\overset{\bar 2'}{\bigcirc}]_{[20]}^{\scaleto{\{N+1\}}{5.5pt}}$}}
\put(34,93){\vector(1,0){25}}
\put(31,13){\vector(1,0){25}}

\put(70,32){\mbox{$\bigoplus$}}
\put(70,72){\mbox{$\bigoplus$}}
\put(70,50){\mbox{$[\overset{{\color{Green} 4}}{\bigcirc}]_{[11]}^{\scaleto{\{4\}}{5.5pt}}$}}
{\color{Green}
\put(32,83){\vector(1,-1){27}}
\put(32,20){\vector(1,1){27}}}

\put(5,0){

{\color{blue}\put(101,93){\vector(1,0){25}}
\put(101,13){\vector(1,0){25}}}
\put(130,90){\mbox{$[\overset{{\color{blue} 2}}{\bigcirc}\overset{{\color{blue} 2'}}{\bigcirc}]_{[12]}^{\scaleto{\{N+3\}}{5.5pt}}$}}
\put(130,10){\mbox{$[\overset{{\color{blue}\bar 2}}{\bigcirc}\overset{{\color{blue}\bar 2'}}{\bigcirc}]_{[21]}^{\scaleto{\{N+3\}}{5.5pt}}$}}
\put(133,50){\mbox{$\bigoplus$}}
\put(100,57){\vector(1,1){27}}
\put(100,50){\vector(1,-1){27}}

\put(15,0){

\put(150,83){\vector(1,-1){27}}
\put(150,22){\vector(1,1){27}}
\put(182,50){\mbox{$[\overset{4'}{\bigcirc}]_{[22]}^{\scaleto{\{2(N+1)\}}{5.5pt}}\;\longrightarrow \; \varnothing$}}

}
}
}

\thicklines

{\color{orange}

\put(110,96){\oval(140,30)}

\put(110,13){\oval(140,30)}

\put(-215,0){

\put(138,53){\line(1.5,1){90}}
\put(138,53){\line(1.5,-1){90}}
\put(228,113){\line(1.5,-1){90}}
\put(228,-7){\line(1.5,1){90}}

\put(175,90){\mbox{$\bphi^2$}}
}

\put(185,100){\mbox{$\bphi$}}
\put(185,5){\mbox{$\bphi$}}

}
    
\end{picture}
\caption{\footnotesize The complex for the mirror trefoil knot in the bipartite presentation. We color in blue and green the arrows for the morphisms Sh and zero, respectively. The elements in rhomb and ovals connected by these colored arrows are shrunk when transferring to the precursor complex.}
\label{fig:mirr-tw-trefoil-complex-min}
\end{figure}

\paragraph{Precursor.} Let us rewrite the complex in Fig.\,\ref{fig:mirr-tw-trefoil-complex} in the minimalistic form, see Fig.\,\ref{fig:mirr-tw-trefoil-complex-min}.

\begin{figure}[h!]
    \centering
\begin{picture}(300,100)(-50,0)

\put(-125,45){\mbox{\bf Kh}}

\put(40,15){

\put(34,0){
\put(-100,25){\line(1,1){16}}
\put(-100,41){\line(1,-1){6}}
\put(-90,31){\line(1,-1){6}}
}

\put(-34,0){
\put(-50,25){\line(-1,1){16}}
\put(-50,41){\line(-1,-1){6}}
\put(-60,31){\line(-1,-1){6}}
}

\put(-75,25){\oval(18,10)[b]}
\put(-75,41){\oval(18,10)[t]}

\put(-75,25){\oval(50,20)[b]}
\put(-75,41){\oval(50,20)[t]}
}

\put(140,0){

\put(-87,50){\mbox{$\varnothing\;\overset{0}{\longrightarrow} \; [\overset{4}{\bigcirc}]_{[00]}^{\scaleto{\{-2\}}{5.5pt}}$}}
\put(-35,65){\vector(1,1){20}}
\put(-35,40){\vector(1,-1){20}}
\put(-45,75){\mbox{$-\Delta$}}
\put(-40,25){\mbox{$\Delta$}}

\put(-5,90){\mbox{$[\overset{2}{\bigcirc}\overset{2'}{\bigcirc}]_{[01]}^{\scaleto{\{-1\}}{5.5pt}}$}}
\put(-5,10){\mbox{$[\overset{\bar 2}{\bigcirc}\overset{\bar 2'}{\bigcirc}]_{[10]}^{\scaleto{\{-1\}}{5.5pt}}$}}
\put(-3,50){\mbox{$\bigoplus$}}

\put(10,0){

\put(65,50){\mbox{$[\overset{4'}{\bigcirc}]_{[11]}^{\scaleto{\{0\}}{5.5pt}}\; \overset{0}{\longrightarrow}\; \varnothing$}}
\put(32,83){\vector(1,-1){27}}
\put(32,20){\vector(1,1){27}}
\put(30,60){\mbox{$m$}}
\put(25,35){\mbox{$m$}}
}
}

\end{picture}
    \caption{\footnotesize The two unknots being the precursor diagram of the bipartite trefoil knot and its Khovanov complex.}
    \label{fig:Kh-prec-mirr-tw-tref}
\end{figure}
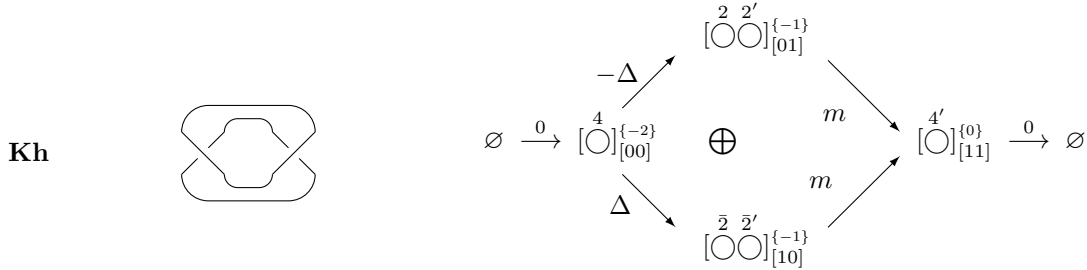

\noindent In order to transfer to the precursor Khovanov complex, we return to the HOMFLY 2-hypercube and arrange morphisms. In other words, in the complex in Fig.\,\ref{fig:mirr-tw-trefoil-complex-min} we shrink orange circled spaces into the unique space coming with the corresponding powers of $\bphi$ and forget the grading shifts in curly brackets (see Fig.\,\ref{fig:mirr-tw-trefoil-2-complex}). Then, we substitute $\bphi \rightarrow -q^{-1}$, and get the precursor complex in Fig.\,\ref{fig:Kh-prec-mirr-tw-tref}. The calculation of the Khovanov polynomial is the same as in Section~\ref{sec:tw-tref} up to the signs of morphisms and the grading shifts, i.e. another framing factor in the Khovanov polynomial. Thus, we do not repeat the calculation here.

\begin{figure}[h!]
\centering
\begin{picture}(100,100)(-15,5)

\put(-90,50){\mbox{$\varnothing\;\overset{0}{\longrightarrow} \; \bphi^2[\overset{4}{\bigcirc}]_{[00]}$}}
\put(-35,65){\vector(1,1){20}}
\put(-35,40){\vector(1,-1){20}}
\put(-45,75){\mbox{$-\Delta$}}
\put(-40,25){\mbox{$\Delta$}}

\put(-5,90){\mbox{$\bphi\,[\overset{2}{\bigcirc}\overset{2'}{\bigcirc}]_{[01]}$}}
\put(-5,10){\mbox{$\bphi\,[\overset{\bar 2}{\bigcirc}\overset{\bar 2'}{\bigcirc}]_{[10]}$}}
\put(0,50){\mbox{$\bigoplus$}}

\put(15,0){

\put(65,50){\mbox{$[\overset{4'}{\bigcirc}]_{[11]}\; \overset{0}{\longrightarrow}\; \varnothing$}}
\put(32,83){\vector(1,-1){27}}
\put(32,20){\vector(1,1){27}}
\put(30,60){\mbox{$m$}}
\put(25,35){\mbox{$m$}}
}
    
\end{picture}
\caption{\footnotesize The complex for the mirror trefoil knot in the bipartite presentation coming from the HOMFLY 2-hypercube. Note that it does not give the Khovanov--Rozansky polynomial but present here as an intermediate step towards the precursor Khovanov complex.}
\label{fig:mirr-tw-trefoil-2-complex}
\end{figure}
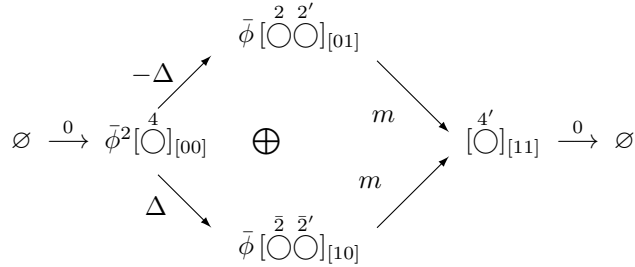

\newpage

\subsubsection{Twist figure-eight knot: Tw$_{-2}$}

This example is the first one being non-chiral, i.e. containing both the lock vertex and its mirror. The 3-hypercube has been already drawn in Fig.\,\ref{fig:tw-8-H-complex}. The only remaining thing is to place the morphisms, so that the complex is as follows
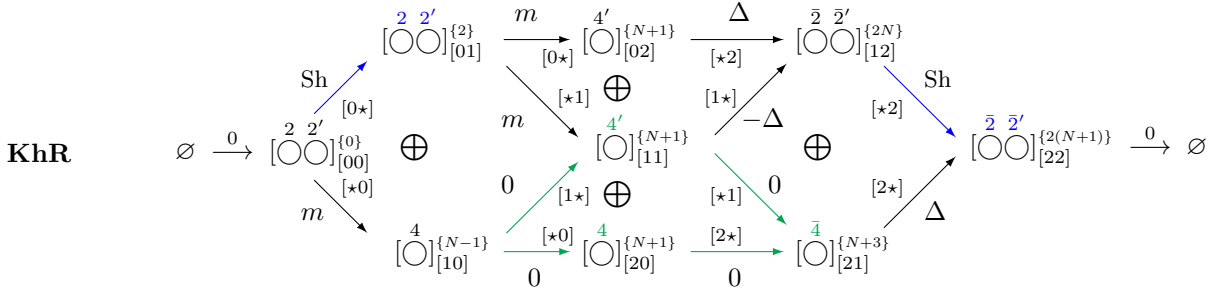
\begin{figure}[h!]
\centering
\begin{picture}(100,110)(40,-5)

\put(-150,47){\mbox{\bf KhR}}

\put(-87,48){\mbox{$\varnothing\;\overset{0}{\longrightarrow} \; [\overset{2}{\bigcirc}\overset{2'}{\bigcirc}]_{[00]}^{\scaleto{\{0\}}{5.5pt}}$}}
{\color{blue}\put(-35,65){\vector(1,1){20}}}
\put(-35,40){\vector(1,-1){20}}
\put(-40,75){\mbox{{\small Sh}}}
\put(-25,65){\mbox{\scriptsize $[0\star]$}}
\put(-40,25){\mbox{$m$}}
\put(-25,35){\mbox{\scriptsize $[\star 0]$}}

\put(-10,90){\mbox{$[\overset{{\color{blue} 2}}{\bigcirc}\overset{{\color{blue} 2'}}{\bigcirc}]_{[01]}^{\scaleto{\{2\}}{5.5pt}}$}}
\put(-5,10){\mbox{$[\overset{4}{\bigcirc}]_{[10]}^{\scaleto{\{N-1\}}{5.5pt}}$}}
\put(-3,50){\mbox{$\bigoplus$}}

\put(5,0){

\put(60,90){\mbox{$[\overset{4'}{\bigcirc}]_{[02]}^{\scaleto{\{N+1\}}{5.5pt}}$}}
\put(60,10){\mbox{$[\overset{{\color{Green} 4}}{\bigcirc}]_{[20]}^{\scaleto{\{N+1\}}{5.5pt}}$}}
\put(31,93){\vector(1,0){25}}
{\color{Green} 
\put(31,13){\vector(1,0){25}}}
\put(35,100){\mbox{$m$}}
\put(45,17){\mbox{\scriptsize $[\star 0]$}}
\put(45,84){\mbox{\scriptsize $[0\star]$}}
\put(40,0){\mbox{$0$}}

\put(68,32){\mbox{$\bigoplus$}}
\put(68,72){\mbox{$\bigoplus$}}
\put(65,50){\mbox{$[\overset{{\color{Green} 4'}}{\bigcirc}]_{[11]}^{\scaleto{\{N+1\}}{5.5pt}}$}}
\put(32,83){\vector(1,-1){27}}
\put(30,60){\mbox{$m$}}
\put(51,69){\mbox{\scriptsize $[\star 1]$}}
{\color{Green}
\put(32,20){\vector(1,1){27}}}
\put(30,35){\mbox{$0$}}
\put(51,32){\mbox{\scriptsize $[1\star]$}}

\put(10,0){

\put(105,0){\mbox{$0$}}
\put(105,100){\mbox{$\Delta$}}
\put(91,93){\vector(1,0){35}}
{\color{Green}
\put(91,13){\vector(1,0){35}}}
\put(130,90){\mbox{$[\overset{\bar 2}{\bigcirc}\overset{\bar 2'}{\bigcirc}]_{[12]}^{\scaleto{\{2N\}}{5.5pt}}$}}
\put(130,10){\mbox{$[\overset{{\color{Green} \bar 4}}{\bigcirc}]_{[21]}^{\scaleto{\{N+3\}}{5.5pt}}$}}
\put(133,50){\mbox{$\bigoplus$}}
\put(100,57){\vector(1,1){27}}
{\color{Green} 
\put(100,50){\vector(1,-1){27}}}
\put(98,17){\mbox{\scriptsize $[2\star]$}}
\put(98,84){\mbox{\scriptsize $[\star 2]$}}
\put(120,35){\mbox{$0$}}
\put(110,60){\mbox{$-\Delta$}}
\put(98,32){\mbox{\scriptsize $[\star 1]$}}
\put(96,69){\mbox{\scriptsize $[1 \star]$}}

\put(10,0){

{\color{blue} \put(150,83){\vector(1,-1){27}}
}
\put(150,22){\vector(1,1){27}}
\put(182,50){\mbox{$[\overset{{\color{blue} \bar 2}}{\bigcirc}\overset{{\color{blue} \bar 2'}}{\bigcirc}]_{[22]}^{\scaleto{\{2(N+1)\}}{5.5pt}}\;\overset{0}{\longrightarrow} \; \varnothing$}}

\put(165,25){\mbox{$\Delta$}}
\put(165,75){\mbox{{\small Sh}}}
\put(145,65){\mbox{\scriptsize $[\star 2]$}}
\put(145,35){\mbox{\scriptsize $[2 \star]$}}
}
}
}
    
\end{picture}
\caption{\footnotesize The complex for the figure-eight knot in the bipartite presentation. We color in blue and green the arrows for the morphisms Sh and zero, respectively, and the corresponding labellings of spaces. Spaces are enumerated by $[\alpha_1\, \alpha_2]$ with $\alpha_1=0,\,1,\,2$ corresponding to smoothings shown in Fig.\,\ref{fig:bip-vertex-smoothings}. Arrows are labelled by $[\alpha_1 \star]$ or $[\star \alpha_2]$ depending on the change of the enumerators of spaces.}
\label{fig:tw-8-complex}
\end{figure}

\noindent Here the subscripts of smoothings are their enumerators $[\alpha_1 \; \alpha_2]$, and curly brackets denote grading shifts. The numbers under the circles are numbers of edges and label the spaces. The full differentials act between direct sums of spaces sharing the same vertical.

\medskip

\noindent $\bullet$ The $0$-th differential is
\begin{equation}
    \hat{d}_0 = {\rm Sh} + m = \sum_{i,j=1}^{N}\left(\vartheta^{{\color{blue} (2)}}_{i+1}\vartheta^{{\color{blue}\prime (2)}}_j-\vartheta^{{\color{blue} (2)}}_i\vartheta^{{\color{blue}\prime (2)}}_{j+1}\right)\frac{\partial^2}{\partial\vartheta_i^{(2)}\partial\vartheta_j^{\prime (2)}} + \sum_{\substack{1\le i,j\le N\\i+j \le N+1}}\vartheta_{i+j-1}^{(4)}\frac{\partial^2}{\partial\vartheta_i^{(2)}\partial\vartheta_j^{\prime (2)}},\quad\vth_{N+1}\equiv 0\,.
\end{equation}
Its kernel and image are
\begin{equation} \nn
\begin{aligned}
    \Ker(\hat{d}_0) &= \Ker(\Sh)^{\scaleto{\{0\}}{5.5pt}} \cap \Ker(m)^{\scaleto{\{0\}}{5.5pt}} = \langle \vth_2^{(2)} \vth_N^{\prime (2)} +\dots + \vth_N^{(2)} \vth_2^{\prime (2)},\, \vth_N^{(2)} \vth_N^{\prime (2)} \rangle^{\scaleto{\{0\}}{5.5pt}}\,, 
    \\
    \Im(\hat{d}_0) &= \langle  \vartheta_{i+1}^{{\color{blue} (2)}}\vartheta_j^{{\color{blue}\prime (2)}} - \vartheta_i^{{\color{blue} (2)}}\vartheta_{j+1}^{{\color{blue} \prime (2)}}+\vartheta_{i+j-1}^{(4)},\, \vartheta^{(4)}_N \rangle^{\scaleto{\{N-1\}}{5.5pt}}\,.
\end{aligned}
\end{equation}

\noindent $\bullet$ The $1$-st differential is
\begin{equation}
    \hd_1 = m + m + 0 + 0 = \sum_{\substack{1\le i,j\le N\\i+j \le N+1}}\left(\vartheta_{i+j-1}^{\prime (4)}+\vartheta_{i+j-1}^{{\color{Green}\prime (4)}}\right)\frac{\partial^2}{\partial\vartheta_i^{{\color{blue}(2)}}\partial\vartheta_j^{{\color{blue}\prime (2)}}}\,.
\end{equation}
Its kernel and image are
\begin{equation} \nn
\begin{aligned}
    \Ker(\hd_1) &= \Ker(m)^{\scaleto{\{2\}}{5.5pt}} \cup V^{\scaleto{\{N-1\}}{5.5pt}}\,, \\
    \Im(\hd_1) &= V^{\scaleto{\{N+1\}}{5.5pt}}\,.
\end{aligned}
\end{equation}

\noindent $\bullet$ The $2$-nd differential is
\begin{equation}
    \hd_2 = \Delta + (-\Delta) + 0 + 0 = \sum_{i=1}^N\sum_{j=0}^{N-i} \bar\vartheta^{(2)}_{N-j}\bar \vartheta^{\prime (2)}_{i+j}\left(\frac{\partial}{\partial \vartheta^{\prime (4)}_i} - \frac{\partial}{\partial \vartheta^{{\color{Green}\prime (4)}}_i}\right)\,.
\end{equation}
Its kernel and image are
\ba \nn
    \Ker(\hd_2) &= V^{\scaleto{\{N+1\}}{5.5pt}}\oplus V^{\scaleto{\{N+1\}}{5.5pt}}\,, \\
    \Im(\hd_2) &= \Im(\Delta)^{\scaleto{\{2N\}}{5.5pt}}\,.
\ea 

\noindent $\bullet$ The $3$-rd differential is
\begin{equation}
\begin{aligned}
    \hd_3 = \Sh + \Delta = \sum_{i,j=1}^{N}\left(\bar\vartheta^{{\color{blue} (2)}}_{i+1}\bar\vartheta^{{\color{blue}\prime (2)}}_j-\bar\vartheta^{{\color{blue} (2)}}_i\bar\vartheta^{{\color{blue}\prime (2)}}_{j+1}\right)\frac{\partial^2}{\partial\bar\vartheta_i^{(2)}\partial\bar\vartheta_j^{\prime (2)}} + \sum_{i=1}^N\sum_{j=0}^{N-i} \bar\vartheta^{{\color{blue}(2)}}_{N-j}\bar \vartheta^{{\color{blue}\prime (2)}}_{i+j} \frac{\partial}{\partial \bar\vartheta^{{\color{Green} (4)}}_i}\,.
\end{aligned}
\end{equation}
Its kernel and image are
\ba \nn
\Ker(\hd_3) &= \Ker(\Sh)^{\scaleto{\{2N\}}{5.5pt}}\oplus \langle \tilde{\vth}_2^{{\color{Green} (4)}},\, \tilde{\vth}_3^{{\color{Green} (4)}},\, \dots,\, \tilde{\vth}_N^{{\color{Green} (4)}}\rangle^{\scaleto{\{N+3\}}{5.5pt}}\,, 
\\ 
\Im(\hd_3) &= (V\otimes V)^{\scaleto{\{2(N+1)\}}{5.5pt}} \backslash \langle \bvth_1^{{\color{blue} (2)}}\bvth_1^{{\color{blue}\prime (2)}},\,\bvth_1^{{\color{blue} (2)}}\bvth_2^{{\color{blue}\prime (2)}},\dots,\, \bvth_1^{{\color{blue} (2)}}\bvth_{N-1}^{{\color{blue}\prime (2)}} \rangle^{\scaleto{\{2(N+1)\}}{5.5pt}}\,.
\ea 
The cohomologies and their quantum dimensions are
{\small \ba \nn
{\cal H}_0^{4_1} &= \Ker(\hd_0) \quad \Lra \quad \boxed{\dim_q \cH_0^{4_1} = q^{-N+1}([N]-q^{N-1})=q^{-N}[N-1]\,,} \\
{\cal H}_1^{4_1} &= \Ker(\hd_1) \backslash \Im(\hd_0) = \langle \vth_i^{(4)} \rangle^{\scaleto{\{N-1\}}{5.5pt}},\, i < N \quad \Rightarrow \quad \boxed{\dim_q \cH_1^{4_1} = q^{N-1}([N]-q^{-N+1}) = q^N [N-1]} \\
{\cal H}_2^{4_1} &= \Ker(\hd_2) \backslash \Im(\hd_1) = V^{\scaleto{\{N+1\}}{5.5pt}} \quad \Rightarrow \quad \boxed{\dim_q \cH_2^{4_1} = q^{N+1}[N]\,,} \\
{\cal H}_3^{4_1} &= \Ker(\hd_3) \backslash \Im(\hd_2) = \langle \tilde{\vth}_2^{{\color{Green} (4)}},\, \tilde{\vth}_3^{{\color{Green} (4)}},\, \dots,\, \tilde{\vth}_N^{{\color{Green} (4)}}\rangle^{\scaleto{\{N+3\}}{5.5pt}} 
\quad \Rightarrow \quad \boxed{\dim_q \cH_3^{4_1} = q^{N+3} ([N] - q^{N-1}) = q^{N+2}[N-1]\,,} \\
{\cal H}_4^{4_1} &= \Ker(0) \backslash \Im(\hd_3) = \langle \bvth_1^{{\color{blue} (2)}}\bvth_1^{{\color{blue}\prime (2)}},\,\bvth_1^{{\color{blue} (2)}}\bvth_2^{{\color{blue}\prime (2)}},\dots,\, \bvth_1^{{\color{blue} (2)}}\bvth_{N-1}^{{\color{blue}\prime (2)}} \rangle^{\scaleto{\{2(N+1)\}}{5.5pt}} \; \Rightarrow \; \boxed{\dim_q \cH_4^{4_1} = q^{2(N+1)} \cdot q^{N-1}([N] - q^{-N+1}) = q^{3N+2}[N-1]\,.}
\ea} 

\noindent The Khovanov--Rozansky polynomial is
\begin{equation}\nn
\begin{aligned}
    P^{4_1}(A,q,T) &= q^{-N-1}T^{-2} \left({\rm dim}_q {\cal H}_0^{4_1} + T {\rm dim}_q {\cal H}_1^{4_1} + T^2 {\rm dim}_q {\cal H}_2^{3_1} + T^3 {\rm dim}_q {\cal H}_3^{3_1} + T^4 {\rm dim}_q {\cal H}_4^{3_1}\right) = \\
    &= q^{-N-1}T^{-2}\left( [N-1](q^{-N} + T q^{N}) + q^{N+1} T^2 [N] + q^{N+2} T^3 [N-1](1 + T q^{2N}) \right) \Longrightarrow 
\end{aligned}
\end{equation}
\begin{equation}
    \boxed{\boxed{P^{4_1}(A,q,T) = [N] + [N-1]\left( q^{-2N-1}T^{-2} + q^{-1} T^{-1} + q T + q^{2N+1} T^2 \right)}}
\end{equation}
what reproduces the known answer in \cite{carqueville2014computing,anokhina2014towards-R}.

The graded Euler characteristic is
\ba \nn
H^{4_1} & \overset{{\rm Fig.\,}\ref{fig:tw-8-complex}}{=} (Aq)^{-1} \left( D^2 - q^2 D^2 - q^{N-1} D + 3q^{N+1} D - q^{2N} D^2 - q^{N+3} D + q^{2(N+1)}D^2 \right) = \\ 
& \overset{{\rm Fig.\,}\ref{fig:tw-8-2-complex}}{=} (Aq)^{-1} \cdot Aq D \cdot \left( D \bphi + \phi\bphi + D \phi + 1 \right) = D \left( D \bphi + \phi\bphi + D \phi + 1 \right)\,.
\ea

\paragraph{Precursor.} Let us rewrite the complex in Fig.\,\ref{fig:tw-8-complex} in the minimalistic form:

\begin{figure}[h!]
\centering
\begin{picture}(100,110)(40,-5)

\put(-150,47){\mbox{\bf KhR}}

\put(-87,48){\mbox{$\varnothing\;\longrightarrow \; [\overset{2}{\bigcirc}\overset{2'}{\bigcirc}]_{[00]}^{\scaleto{\{0\}}{5.5pt}}$}}
{\color{blue}\put(-35,65){\vector(1,1){20}}}
\put(-35,40){\vector(1,-1){20}}

\put(-10,90){\mbox{$[\overset{{\color{blue} 2}}{\bigcirc}\overset{{\color{blue} 2'}}{\bigcirc}]_{[01]}^{\scaleto{\{2\}}{5.5pt}}$}}
\put(-5,10){\mbox{$[\overset{4}{\bigcirc}]_{[10]}^{\scaleto{\{N-1\}}{5.5pt}}$}}
\put(-3,50){\mbox{$\bigoplus$}}

\put(5,0){

\put(60,90){\mbox{$[\overset{4'}{\bigcirc}]_{[02]}^{\scaleto{\{N+1\}}{5.5pt}}$}}
\put(60,10){\mbox{$[\overset{{\color{Green} 4}}{\bigcirc}]_{[20]}^{\scaleto{\{N+1\}}{5.5pt}}$}}
\put(31,93){\vector(1,0){25}}
{\color{Green} 
\put(31,13){\vector(1,0){25}}}

\put(68,32){\mbox{$\bigoplus$}}
\put(68,74){\mbox{$\bigoplus$}}
\put(70,62){\mbox{\scriptsize {\color{Green} $4'$}}}
\put(65,50){\mbox{$[\bigcirc]_{[11]}^{\scaleto{\{N+1\}}{5.5pt}}$}}
\put(32,83){\vector(1,-1){27}}
{\color{Green}\put(32,20){\vector(1,1){27}}}

\put(10,0){

\put(91,93){\vector(1,0){35}}
{\color{Green}
\put(91,13){\vector(1,0){35}}}
\put(130,90){\mbox{$[\overset{\bar 2}{\bigcirc}\overset{\bar 2'}{\bigcirc}]_{[12]}^{\scaleto{\{2N\}}{5.5pt}}$}}
\put(130,10){\mbox{$[\overset{{\color{Green} \bar 4}}{\bigcirc}]_{[21]}^{\scaleto{\{N+3\}}{5.5pt}}$}}
\put(133,50){\mbox{$\bigoplus$}}
\put(100,57){\vector(1,1){27}}
{\color{Green} 
\put(100,50){\vector(1,-1){27}}}

\put(10,0){

{\color{blue} 
\put(150,83){\vector(1,-1){27}}
}
\put(150,22){\vector(1,1){27}}
\put(182,50){\mbox{$[\overset{{\color{blue} \bar 2}}{\bigcirc}\overset{{\color{blue} \bar 2'}}{\bigcirc}]_{[22]}^{\scaleto{\{2(N+1)\}}{5.5pt}}\;\longrightarrow \; \varnothing$}}

}
}
}

\thicklines

{\color{orange}

\put(-70,40){\line(1,1){65}}
\put(-15,40){\line(1,1){65}}
\put(-70,40){\line(1,0){55}}
\put(-5,105){\line(1,0){55}}

\put(192,40){\line(-1,1){65}}
\put(283,40){\line(-1,1){65}}
\put(192,40){\line(1,0){91}}
\put(127,105){\line(1,0){91}}

\put(37,60){\line(1,0){103}}
\put(-23,0){\line(1,0){223}}
\put(-23,0){\line(1,1){60}}
\put(200,0){\line(-1,1){60}}

\put(-28,95){\mbox{$\bphi$}}
\put(234,95){\mbox{$\phi$}}
\put(-37,5){\mbox{$\phi\bphi$}}

}
    
\end{picture}
\caption{\footnotesize The complex for the figure-eight knot in the bipartite presentation. We color in blue and green the arrows for the morphisms Sh and zero, respectively, and the corresponding labellings of spaces. The circled in orange elements connected by these blue arrows are shrunk when transferring to the precursor complex.}
\end{figure}
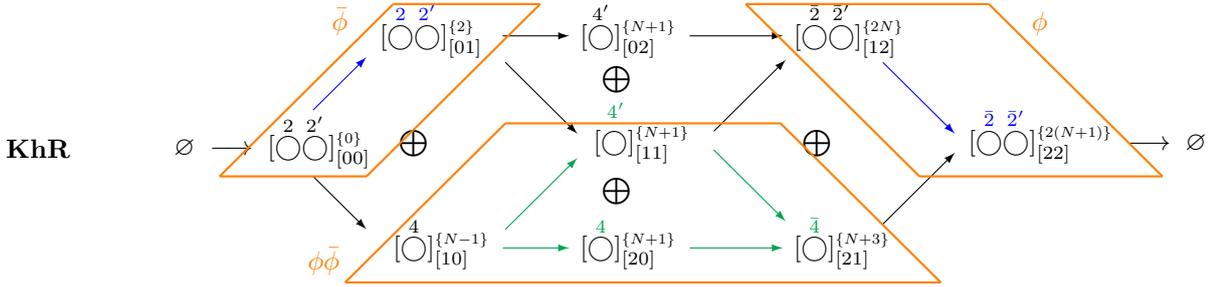

\noindent The circled spaces can be shrunk to single ones without changing the weighted sum, i.e. the HOMFLY polynomial. Instead of ordinary gradings, these spaces have powers of $\phi$ as multipliers:
\begin{figure}[h!]
\centering
\begin{picture}(100,100)(-15,5)

\put(-105,50){\mbox{$\varnothing\;\overset{0}{\longrightarrow} \; \bphi\,[\overset{2}{\bigcirc}\overset{2'}{\bigcirc}]_{[00]}$}}
\put(-35,65){\vector(1,1){20}}
\put(-35,40){\vector(1,-1){20}}
\put(-40,75){\mbox{$m$}}
\put(-40,25){\mbox{$m$}}

\put(-2,90){\mbox{$[\overset{4'}{\bigcirc}]_{[01]}$}}
\put(-8,10){\mbox{$\phi\bphi\,[\overset{4}{\bigcirc}]_{[10]}$}}
\put(-3,50){\mbox{$\bigoplus$}}

\put(5,0){

\put(65,50){\mbox{$\phi\,[\overset{\bar 2}{\bigcirc}\overset{\bar 2'}{\bigcirc}]_{[11]}\; \overset{0}{\longrightarrow}\; \varnothing$}}
\put(32,83){\vector(1,-1){27}}
\put(32,20){\vector(1,1){27}}
\put(30,60){\mbox{$\Delta$}}
\put(25,35){\mbox{$-\Delta$}}
}
    
\end{picture}
\caption{\footnotesize The HOMFLY 2-hypercube for the figure-eight knot in the bipartite presentation with morphisms coming in the precursor Khovanov complex.}
\label{fig:tw-8-2-complex}
\end{figure}
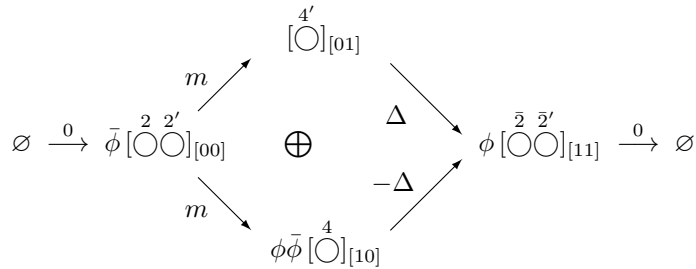

\begin{figure}[h!]
\centering
\begin{picture}(100,100)(-70,5)

\put(-240,45){\mbox{\bf Kh}}

\put(-80,20){

\put(-100,25){\line(1,1){16}}
\put(-100,41){\line(1,-1){6}}
\put(-90,31){\line(1,-1){6}}

\put(34,0){

\put(-100,25){\line(1,1){16}}
\put(-100,41){\line(1,-1){6}}
\put(-90,31){\line(1,-1){6}}
}


\put(-75,25){\oval(18,10)[b]}
\put(-75,41){\oval(18,10)[t]}

\put(-75,25){\oval(50,20)[b]}
\put(-75,41){\oval(50,20)[t]}
}

\put(-105,50){\mbox{$\varnothing\;\overset{0}{\longrightarrow} \; [\overset{2}{\bigcirc}\overset{2'}{\bigcirc}]_{[00]}^{\scaleto{\{-1\}}{5.5pt}}$}}
\put(-35,65){\vector(1,1){20}}
\put(-35,40){\vector(1,-1){20}}
\put(-40,75){\mbox{$m$}}
\put(-40,25){\mbox{$m$}}

\put(-8,90){\mbox{$[\overset{4'}{\bigcirc}]_{[01]}^{\scaleto{\{0\}}{5.5pt}}$}}
\put(-8,10){\mbox{$[\overset{4}{\bigcirc}]_{[10]}^{\scaleto{\{0\}}{5.5pt}}$}}
\put(-3,50){\mbox{$\bigoplus$}}

\put(-5,0){

\put(65,50){\mbox{$[\overset{\bar 2}{\bigcirc}\overset{\bar 2'}{\bigcirc}]_{[11]}^{\scaleto{\{1\}}{5.5pt}}\; \overset{0}{\longrightarrow}\; \varnothing$}}
\put(32,83){\vector(1,-1){27}}
\put(32,20){\vector(1,1){27}}
\put(30,60){\mbox{$\Delta$}}
\put(25,35){\mbox{$-\Delta$}}
}
    
\end{picture}
\caption{\footnotesize The Hopf link being the precursor diagram of the bipartite figure-eight knot and its Khovanov complex.}
\label{fig:Kh-prec-4-1}
\end{figure}
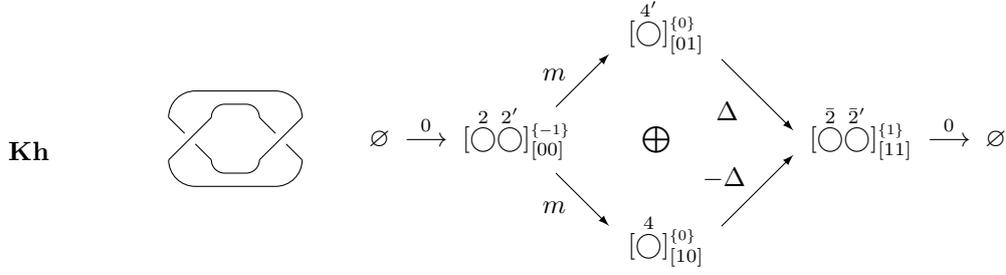

\noindent If we put $\phi \rightarrow -q$, $\bphi \rightarrow -q^{-1}$ and consider $N=2$, we get the precursor complex in Fig.\,\ref{fig:Kh-prec-4-1}.
\noindent The Khovanov polynomial for the Hopf link has been already calculated in Section~\ref{sec:Khovanov}. The answer for the Khovanov polynomial is:
\begin{equation}
    {\rm Kh}^{\rm Hopf}(q,T) = q^{-5}[2] \left( T^{-2} + q^{4}\right)\,, \quad [2] = q + q^{-1}\,. 
\end{equation}

\paragraph{Summary.} In this subsection, we have considered the example of non-chiral bipartite knot, i.e. containing both the positive and negative locks. We have again shown how to obtain the precursor Khovanov complex from the bipartite one.


\subsubsection{APT[2,4]}

The 2-strand antiparallel link $APT[2,4]$ has 2-dimensional 3-hypercube of resolutions. The grading shifts and morphisms are placed according to steps 3 and 4. The resulting complex is shown in Fig.\,\ref{fig:APT[2,4]-complex}. In this example, we see that there can be elementary 4-cycles (i.e. cycles of 4 vertices inside a complex) containing four Sh morphisms. We also see that Sh can be supplied with the minus sign.

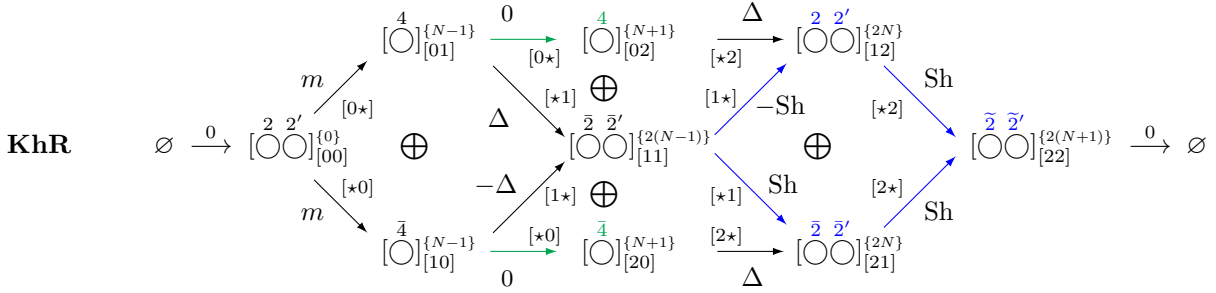
\begin{figure}[h!]
\centering
\begin{picture}(100,110)(40,-5)

\put(-150,50){\mbox{\bf KhR}}

\put(-95,50){\mbox{$\varnothing\;\overset{0}{\longrightarrow} \; [\overset{2}{\bigcirc}\overset{2'}{\bigcirc}]_{[00]}^{\scaleto{\{0\}}{5.5pt}}$}}
\put(-35,65){\vector(1,1){20}}
\put(-35,40){\vector(1,-1){20}}
\put(-40,75){\mbox{$m$}}
\put(-25,65){\mbox{\scriptsize $[0\star]$}}
\put(-40,25){\mbox{$m$}}
\put(-25,35){\mbox{\scriptsize $[\star 0]$}}

\put(-10,90){\mbox{$[\overset{4}{\bigcirc}]_{[01]}^{\scaleto{\{N-1\}}{5.5pt}}$}}
\put(-10,10){\mbox{$[\overset{\bar 4}{\bigcirc}]_{[10]}^{\scaleto{\{N-1\}}{5.5pt}}$}}
\put(-3,50){\mbox{$\bigoplus$}}

\put(0,0){

\put(65,90){\mbox{$[\overset{{\color{Green} 4}}{\bigcirc}]_{[02]}^{\scaleto{\{N+1\}}{5.5pt}}$}}
\put(65,10){\mbox{$[\overset{{\color{Green} \bar 4}}{\bigcirc}]_{[20]}^{\scaleto{\{N+1\}}{5.5pt}}$}}
{\color{Green} 
\put(31,93){\vector(1,0){25}}
\put(31,13){\vector(1,0){25}}}
\put(35,100){\mbox{\small 0}}
\put(45,17){\mbox{\scriptsize $[\star 0]$}}
\put(45,84){\mbox{\scriptsize $[0\star]$}}
\put(35,0){\mbox{\small 0}}

\put(68,32){\mbox{$\bigoplus$}}
\put(68,72){\mbox{$\bigoplus$}}
\put(60,50){\mbox{$[\overset{\bar 2}{\bigcirc}\overset{\bar 2'}{\bigcirc}]_{[11]}^{\scaleto{\{2(N-1)\}}{5.5pt}}$}}
\put(32,83){\vector(1,-1){27}}
\put(30,60){\mbox{$\Delta$}}
\put(51,69){\mbox{\scriptsize $[\star 1]$}}
\put(32,20){\vector(1,1){27}}
\put(25,35){\mbox{$-\Delta$}}
\put(51,32){\mbox{\scriptsize $[1\star]$}}

\put(15,0){

\put(110,0){\mbox{$\Delta$}}
\put(110,100){\mbox{$\Delta$}}
\put(101,93){\vector(1,0){25}}
\put(101,13){\vector(1,0){25}}
\put(130,90){\mbox{$[\overset{{\color{blue} 2}}{\bigcirc}\overset{{\color{blue} 2'}}{\bigcirc}]_{[12]}^{\scaleto{\{2N\}}{5.5pt}}$}}
\put(130,10){\mbox{$[\overset{{\color{blue} \bar 2}}{\bigcirc}\overset{{\color{blue} \bar 2'}}{\bigcirc}]_{[21]}^{\scaleto{\{2N\}}{5.5pt}}$}}
\put(133,50){\mbox{$\bigoplus$}}
{\color{blue} 
\put(100,57){\vector(1,1){27}}
\put(100,50){\vector(1,-1){27}}}
\put(98,17){\mbox{\scriptsize $[2\star]$}}
\put(98,84){\mbox{\scriptsize $[\star 2]$}}
\put(120,35){\mbox{$\Sh$}}
\put(115,65){\mbox{$-\Sh$}}
\put(98,32){\mbox{\scriptsize $[\star 1]$}}
\put(96,69){\mbox{\scriptsize $[1 \star]$}}

\put(10,0){

{\color{blue} 
\put(150,83){\vector(1,-1){27}}
\put(150,22){\vector(1,1){27}}}
\put(182,50){\mbox{$[\overset{{\color{blue} \widetilde{2}}}{\bigcirc}\overset{{\color{blue} \widetilde{2}'}}{\bigcirc}]_{[22]}^{\scaleto{\{2(N+1)\}}{5.5pt}}\;\overset{0}{\longrightarrow} \; \varnothing$}}

\put(165,25){\mbox{$\Sh$}}
\put(165,75){\mbox{$\Sh$}}
\put(145,65){\mbox{\scriptsize $[\star 2]$}}
\put(145,35){\mbox{\scriptsize $[2 \star]$}}
}
}
}
    
\end{picture}
\caption{\footnotesize The complex for the 2-strand aniparallel link $APT[2,4]$ from Fig.\,\ref{fig:two-bip-vert}. We color in blue and green the arrows for the morphisms Sh and zero, respectively, and the corresponding labellings of spaces. The Sh morphisms go after the $\Delta$ morphisms, while the zero morphisms go after the $m$ morphisms. 
Spaces are enumerated by $[\alpha_1\, \alpha_2]$ with $\alpha_1=0,\,1,\,2$ corresponding to smoothings shown in Fig.\,\ref{fig:bip-vertex-smoothings}. Arrows are labelled by $[\alpha_1 \star]$ or $[\star \alpha_2]$ depending on the change of the enumerators of spaces.}
\label{fig:APT[2,4]-complex}
\end{figure}

\noindent Write down differentials and their images and kernels.

\medskip

\noindent $\bullet$ The $0$-th differential is
\begin{equation}
    \hd_0 = m + m = \sum_{\substack{1\le i,j\le N\\i+j \le N+1}}\left(\vartheta_{i+j-1}^{ (4)}+\bar\vartheta_{i+j-1}^{ (4)}\right)\frac{\partial^2}{\partial\vartheta_i^{(2)}\partial\vartheta_j^{\prime (2)}}\,.
\end{equation}
Its kernel and image are
\ba \nn
\Ker(\hd_0) &= \Ker(m)^{\scaleto{\{0\}}{5.5pt}}\,, \\
\Im(\hd_0) &= \Im(m)^{\scaleto{\{N-1\}}{5.5pt}} = V^{\scaleto{\{N-1\}}{5.5pt}}\,.
\ea 

\noindent $\bullet$ The $1$-st differential is
\begin{equation}
    \hd_1 = 0 + \Delta + (-\Delta) + 0 = \sum_{i=1}^N\sum_{j=0}^{N-i} \bar\vartheta^{(2)}_{N-j}\bar \vartheta^{\prime (2)}_{i+j}\left(\frac{\partial}{\partial \vartheta^{(4)}_i} - \frac{\partial}{\partial \bar\vartheta^{(4)}_i}\right)\,.
\end{equation}
Its kernel and image are
\ba \nn
\Ker(\hd_1) &= V^{\scaleto{\{N-1\}}{5.5pt}}\,, \\
\Im(\hd_1) &= \Im(\Delta)^{\scaleto{\{2(N-1)\}}{5.5pt}}\,.
\ea 

\noindent $\bullet$ The $2$-nd differential is
\begin{equation}
\begin{aligned}
    \hd_2 &= \Delta + (-\Sh) + \Sh + \Delta = \sum_{i,j=1}^{N}\left(-\vartheta^{{\color{blue} (2)}}_{i+1}\vartheta^{{\color{blue}\prime (2)}}_j+\vartheta^{{\color{blue} (2)}}_i \vartheta^{{\color{blue}\prime (2)}}_{j+1}+\bar\vartheta^{{\color{blue} (2)}}_{i+1}\bar\vartheta^{{\color{blue}\prime (2)}}_j-\bar\vartheta^{{\color{blue} (2)}}_i\bar\vartheta^{{\color{blue}\prime (2)}}_{j+1}\right)\frac{\partial^2}{\partial\bar\vartheta_i^{(2)}\partial\bar\vartheta_j^{\prime (2)}} + \\
    & + \sum_{i=1}^N\sum_{j=0}^{N-i} \vartheta^{{\color{blue}(2)}}_{N-j} \vartheta^{{\color{blue}\prime (2)}}_{i+j} \frac{\partial}{\partial \vartheta^{{\color{Green} (4)}}_i} +  \sum_{i=1}^N\sum_{j=0}^{N-i} \bar\vartheta^{{\color{blue}(2)}}_{N-j}\bar \vartheta^{{\color{blue}\prime (2)}}_{i+j} \frac{\partial}{\partial \bar\vartheta^{{\color{Green} (4)}}_i}\,.
\end{aligned}
\end{equation}
Its kernel and image are
\ba \nn
\Ker(\hd_2) &= \Ker(\Sh)^{\scaleto{\{2(N-1)\}}{5.5pt}} \oplus \langle \tilde{\vth}_2^{-},\,\tilde{\vth}_3^{-},\,\dots,\,\tilde{\vth}_N^{-} \rangle^{\scaleto{\{N+1\}}{5.5pt}}\,, \\
\Im(\hd_2) &= \Im(\Sh)^{\scaleto{\{2N\}}{5.5pt}} \oplus \langle \sum_{j=0}^{N-1} \vartheta^{{\color{blue}(2)}}_{N-j} \vartheta^{{\color{blue}\prime (2)}}_{1+j},\, \sum_{j=0}^{N-1} \bar\vartheta^{{\color{blue}(2)}}_{N-j}\bar \vartheta^{{\color{blue}\prime (2)}}_{1+j},\, \sum_{j=0}^{N-i} \vartheta^{{\color{blue}(2)}}_{N-j} \vartheta^{{\color{blue}\prime (2)}}_{i+j} + \sum_{j=0}^{N-i} \bar\vartheta^{{\color{blue}(2)}}_{N-j}\bar \vartheta^{{\color{blue}\prime (2)}}_{i+j} \rangle^{\scaleto{\{2N\}}{5.5pt}},\, i>1\,.
\ea 

\noindent $\bullet$ The $3$-rd differential is
\begin{equation}
    \hd_3 = \Sh + \Sh = \sum_{i,j=1}^{N}\left(\tilde\vartheta^{{\color{blue} (2)}}_{i+1}\tilde\vartheta^{{\color{blue}\prime (2)}}_j-\tilde\vartheta^{{\color{blue} (2)}}_i\tilde\vartheta^{{\color{blue}\prime (2)}}_{j+1}\right)\left(\frac{\partial^2}{\partial\vartheta_i^{{\color{blue}(2)}}\partial\vartheta_j^{{\color{blue}\prime (2)}}}+\frac{\partial^2}{\partial\bar\vartheta_i^{{\color{blue}(2)}}\partial\bar\vartheta_j^{{\color{blue}\prime (2)}}}\right)\,.
\end{equation}
Its kernel and image are
\ba \nn
\Ker(\hd_3) &= \Ker(\Sh)^{\scaleto{\{2N\}}{5.5pt}} \oplus (V\otimes V)^{\scaleto{\{2N\}}{5.5pt}}\,, \\
\Im(\hd_3) &= \Im(\Sh)^{\scaleto{\{2(N+1)\}}{5.5pt}}\,.
\ea 
The cohomologies and their quantum dimensions are
\ba \nn
\cH_0 &= \Ker(\hd_0) \quad \Lra \quad \boxed{\dim_q \cH_0 = q^{-1}[N][N-1]\,,} \\
\cH_1 &= \Ker(\hd_1) \backslash \Im(\hd_0)=\varnothing \quad \Lra \quad \boxed{\dim_q \cH_1 = 0\,,} \\
\cH_2 &= \Ker(\hd_2) \backslash \Im(\hd_1) = \langle \tilde{\vth}_2^{-},\,\tilde{\vth}_3^{-},\,\dots,\,\tilde{\vth}_N^{-} \rangle^{\scaleto{\{N+1\}}{5.5pt}} \quad \Lra \quad \boxed{\dim_q \cH_2 = q^{N+1}([N]-q^{N-1})=q^N[N-1]\,,} \\
\cH_3 &= \Ker(\hd_3) \backslash \Im(\hd_2) \quad \Lra \quad \boxed{\dim_q \cH_3 = q^{2N}([N]^2 + q^{-N+1}[N] - q^{-1}[N][N-1] - q^{-N+1}[N] - 1) = q^{3N}[N-1]\,,} \\
\cH_4 &= \Ker(0) \backslash \Im(\hd_3) \quad \Lra \quad \boxed{\dim_q \cH_4 = q^{2(N+1)}([N]^2 - q^{-1}[N][N-1]) = q^{3N+1}[N]\,.}
\ea

\noindent The unreduced Khovanov--Rozansky polynomial is
\be
\boxed{\boxed{
\begin{array}{rl}
 P^{APT[2,4]}(A,q,T) &= (q^N T)^{-4} \left( \dim_q \cH_0 + T^2 \dim_q \cH_2 + T^3 \dim_q \cH_3 + T^4 \dim_q \cH_4 \right) =\\ \\
    &= q^{-N+1}([N] + (q^{-1}T^{-1} + q^{-2N-1}T^{-2})[N-1] + q^{-3N-2}[N][N-1]T^{-4})
\end{array}
}}    
\ee
what reproduces the known answer in \cite{carqueville2014computing,anokhina2014towards-R}.

The HOMFLY polynomial is the $q$-Euler characteristic of the complex:
\begin{equation}
    H^{APT[2,4]}(A,q,T) \overset{{\rm Fig.\,}\ref{fig:APT[2,4]-complex}}{=} A^{-4}(D^2 - 2 q^{N-1} D + 2 q^{N+1} D + q^{2(N-1)} D^2 - 2 q^{2N} D^2 + q^{2(N+1)}D^2) \overset{{\rm Fig.\,}\ref{fig:APT[2,4]-2-complex}}{=} A^{-4}(D^2 + 2 \phi D + \phi^2 D^2)\,.
\end{equation}

\paragraph{Precursor.} We again highlight parts of the complex in Fig.\,\ref{fig:APT[2,4]-complex} that can be compressed to one space with $\phi$ multipliers: 

\begin{figure}[h!]
\centering
\begin{picture}(100,120)(40,-5)

\put(-150,50){\mbox{\bf KhR}}

\put(-95,50){\mbox{$\varnothing\;\longrightarrow \; [\overset{2}{\bigcirc}\overset{2'}{\bigcirc}]_{[00]}^{\scaleto{\{0\}}{5.5pt}}$}}
\put(-35,65){\vector(1,1){20}}
\put(-35,40){\vector(1,-1){20}}

\put(0,90){\mbox{$[\overset{4}{\bigcirc}]_{[01]}^{\scaleto{\{N-1\}}{5.5pt}}$}}
\put(0,10){\mbox{$[\overset{\bar 4}{\bigcirc}]_{[10]}^{\scaleto{\{N-1\}}{5.5pt}}$}}
\put(5,50){\mbox{$\bigoplus$}}

\put(10,0){

\put(60,90){\mbox{$[\overset{{\color{Green} 4}}{\bigcirc}]_{[02]}^{\scaleto{\{N+1\}}{5.5pt}}$}}
\put(60,10){\mbox{$[\overset{{\color{Green} \bar 4}}{\bigcirc}]_{[20]}^{\scaleto{\{N+1\}}{5.5pt}}$}}
{\color{Green} 
\put(31,93){\vector(1,0){25}}
\put(31,13){\vector(1,0){25}}}

\put(68,32){\mbox{$\bigoplus$}}
\put(68,72){\mbox{$\bigoplus$}}
\put(60,50){\mbox{$[\overset{\bar 2}{\bigcirc}\overset{\bar 2'}{\bigcirc}]_{[11]}^{\scaleto{\{2(N-1)\}}{5.5pt}}$}}
\put(32,83){\vector(1,-1){27}}
\put(32,20){\vector(1,1){27}}

\put(10,0){

\put(101,93){\vector(1,0){25}}
\put(101,13){\vector(1,0){25}}
\put(130,90){\mbox{$[\overset{{\color{blue} 2}}{\bigcirc}\overset{{\color{blue} 2'}}{\bigcirc}]_{[12]}^{\scaleto{\{2N\}}{5.5pt}}$}}
\put(130,10){\mbox{$[\overset{{\color{blue} \bar 2}}{\bigcirc}\overset{{\color{blue} \bar 2'}}{\bigcirc}]_{[21]}^{\scaleto{\{2N\}}{5.5pt}}$}}
\put(133,50){\mbox{$\bigoplus$}}
{\color{blue} 
\put(105,57){\vector(1,1){27}}
\put(105,50){\vector(1,-1){27}}}

\put(0,0){

{\color{blue} 
\put(150,83){\vector(1,-1){27}}
\put(150,22){\vector(1,1){27}}}
\put(182,50){\mbox{$[\overset{{\color{blue} \widetilde{2}}}{\bigcirc}\overset{{\color{blue} \widetilde{2}'}}{\bigcirc}]_{[22]}^{\scaleto{\{2(N+1)\}}{5.5pt}}\;\longrightarrow \; \varnothing$}}

}
}
}

\thicklines

{\color{orange}

\put(52,96){\oval(140,30)}

\put(52,13){\oval(140,30)}

\put(-50,0){

\put(78,53){\line(2.5,1){150}}
\put(78,53){\line(2.5,-1){150}}
\put(228,113){\line(1.5,-1){90}}
\put(228,-7){\line(1.5,1){90}}

\put(275,90){\mbox{$\phi^2$}}
}

\put(-27,100){\mbox{$\phi$}}
\put(-27,5){\mbox{$\phi$}}

}
    
\end{picture}
\caption{\footnotesize The complex for the $APT[2,4]$ link. We color in blue the arrows for the morphisms Sh and in green the arrows for the zero morphisms. The elements in rhomb and ovals connected by these colored arrows are shrunk when transferring to the precursor complex.}
\label{fig:APT[2,4]-complex-min}
\end{figure}
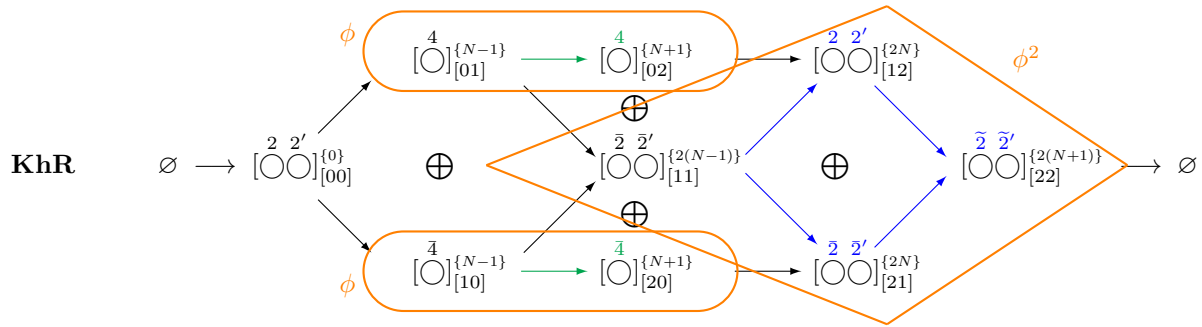

\noindent Now, we transfer to the HOMFLY 2-hypercube. Instead of grading shifts, we write the powers of $\phi$ as multipliers in Fig.\,\ref{fig:APT[2,4]-2-complex}. 
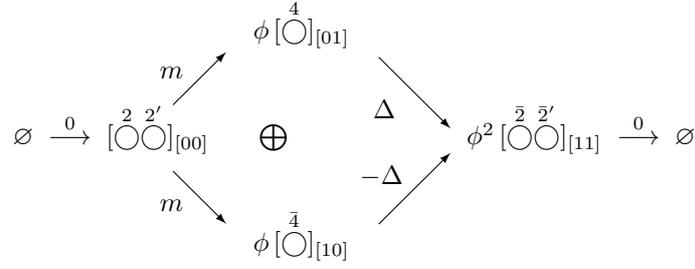
\begin{figure}[h!]
\centering
\begin{picture}(100,100)(-15,5)

\put(-95,50){\mbox{$\varnothing\;\overset{0}{\longrightarrow} \; [\overset{2}{\bigcirc}\overset{2'}{\bigcirc}]_{[00]}$}}
\put(-35,65){\vector(1,1){20}}
\put(-35,40){\vector(1,-1){20}}
\put(-40,75){\mbox{$m$}}
\put(-40,25){\mbox{$m$}}

\put(-5,90){\mbox{$\phi\,[\overset{4}{\bigcirc}]_{[01]}$}}
\put(-5,10){\mbox{$\phi\,[\overset{\bar 4}{\bigcirc}]_{[10]}$}}
\put(-3,50){\mbox{$\bigoplus$}}

\put(10,0){

\put(65,50){\mbox{$\phi^2\,[\overset{\bar 2}{\bigcirc}\overset{\bar 2'}{\bigcirc}]_{[11]}\; \overset{0}{\longrightarrow}\; \varnothing$}}
\put(32,83){\vector(1,-1){27}}
\put(32,20){\vector(1,1){27}}
\put(30,60){\mbox{$\Delta$}}
\put(25,35){\mbox{$-\Delta$}}
}
    
\end{picture}
\caption{\footnotesize The 2-hypercube for the $APT[2,4]$ link with shrunk colored edges. This complex cannot give the Khovanov--Rozansky polynomial but its weighted sum is the HOMFLY polynomial for the $APT[2,4]$ link.}
\label{fig:APT[2,4]-2-complex}
\end{figure}
\noindent The transition to $N=2$ and $\phi \rightarrow -q$ gives us exactly the complex for the Khovanov polynomial in Fig.\,\ref{fig:Kh-Hopf-complex}.

\paragraph{Summary.} By this example, we have demonstrated another combination of morphisms where four Sh are met in elementary 4-cycle inside the complex. We have also seen that Sh morphism can be supplied with the minus sign. 




\subsection{Tw$_{4}$}

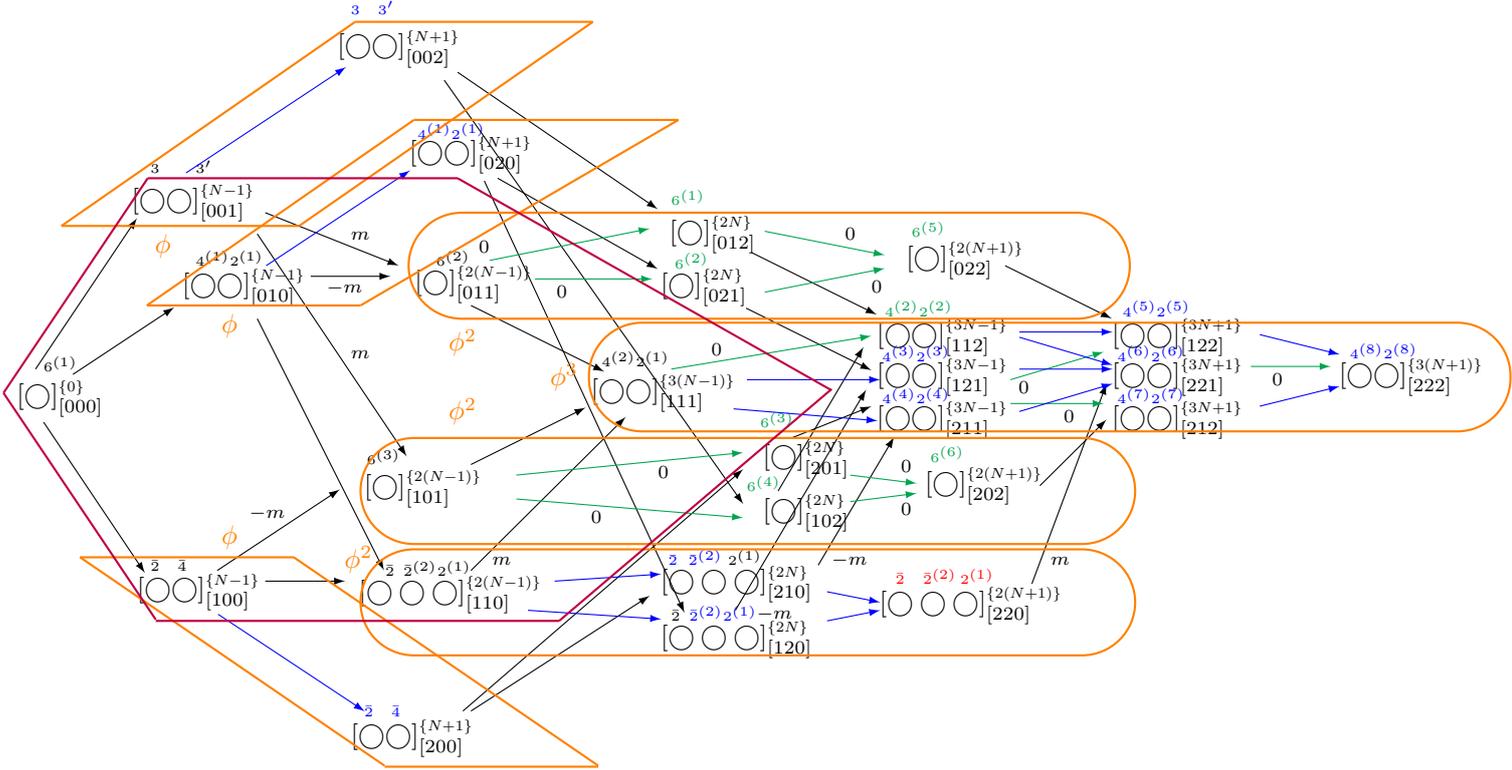
\begin{figure}[h!]
\begin{picture}(300,300)(-70,-50)

\put(-100,98){\mbox{$[\bigcirc]_{[000]}^{\scaleto{\{0\}}{5.5pt}}$}}

\put(-90,110){\mbox{\tiny $6^{(1)}$}}

\put(-5,0){

\put(-52,172){\mbox{$[\bigcirc\bigcirc]_{[001]}^{\scaleto{\{N-1\}}{5.5pt}}$}}

\put(-45,185){\mbox{\tiny $3 \qquad 3'$}}

{\color{blue} \put(30,245){\mbox{\tiny $3 \quad 3'$}}}

\put(-33,140){\mbox{$[\bigcirc\bigcirc]_{[010]}^{\scaleto{\{N-1\}}{5.5pt}}$}}

\put(-28,150){\mbox{\tiny $4^{(1)} 2^{(1)}$}}

\put(-50,25){\mbox{$[\bigcirc\bigcirc]_{[100]}^{\scaleto{\{N-1\}}{5.5pt}}$}}

\put(-45,35){\mbox{\tiny $\bar 2 \quad \bar 4$}}

\put(35,-20){\mbox{\tiny {\color{blue} $\bar 2 \quad \bar 4$}}}

}

\put(-25,0){

\put(50,-30){\mbox{$[\bigcirc\bigcirc]_{[200]}^{\scaleto{\{N+1\}}{5.5pt}}$}}

\put(45,230){\mbox{$[\bigcirc\bigcirc]_{[002]}^{\scaleto{\{N+1\}}{5.5pt}}$}}

\put(75,198){\mbox{\tiny {\color{blue} $4^{(1)} 2^{(1)}$}}}

\put(72,190){\mbox{$[\bigcirc\bigcirc]_{[020]}^{\scaleto{\{N+1\}}{5.5pt}}$}}

\put(82,150){\mbox{\tiny $6^{(2)}$}}

\put(74,141){\mbox{$[\bigcirc]_{[011]}^{\scaleto{\{2(N-1)\}}{5.5pt}}$}}

\put(0,-25){

\put(56,100){\mbox{\tiny $6^{(3)}$}}

\put(55,89){\mbox{$[\bigcirc]_{[101]}^{\scaleto{\{2(N-1)\}}{5.5pt}}$}}

\put(63,58){\mbox{\tiny $\bar 2 \,\; \bar 2^{(2)} 2^{(1)}$}}

\put(53,49){\mbox{$[\bigcirc\bigcirc\bigcirc]_{[110]}^{\scaleto{\{2(N-1)\}}{5.5pt}}$}}
}
}

\put(119,112){\mbox{\tiny $4^{(2)} 2^{(1)}$}}

\put(115,100){\mbox{$[\bigcirc\bigcirc]_{[111]}^{\scaleto{\{3(N-1)\}}{5.5pt}}$}}

\put(-20,0){

\put(161,140){\mbox{$[\bigcirc]_{[021]}^{\scaleto{\{2N\}}{5.5pt}}$}}

{\color{Green} \put(165,173){\mbox{\tiny $6^{(1)}$}}}

{\color{Green} \put(163,149){\mbox{\tiny $6^{(2)}$}}}

\put(161,160){\mbox{$[\bigcirc]_{[012]}^{\scaleto{\{2N\}}{5.5pt}}$}}
}

\put(15,-45){

{\color{Green} \put(155,110){\mbox{\tiny $6^{(4)}$}}}

\put(161,100){\mbox{$[\bigcirc]_{[102]}^{\scaleto{\{2N\}}{5.5pt}}$}}

{\color{Green} \put(160,134){\mbox{\tiny $6^{(3)}$}}}

\put(161,120){\mbox{$[\bigcirc]_{[201]}^{\scaleto{\{2N\}}{5.5pt}}$}}
}

\put(-20,-90){

 \put(165,106){\mbox{\tiny $\bar 2 \;\, {\color{blue} \bar 2^{(2)} 2^{(1)}}$}}

\put(161,97){\mbox{$[\bigcirc\bigcirc\bigcirc]_{[120]}^{\scaleto{\{2N\}}{5.5pt}}$}}

\put(164,127){\mbox{\tiny ${\color{blue} \bar 2 \;\; \bar 2^{(2)}} \; 2^{(1)}$}}

\put(161,118){\mbox{$[\bigcirc\bigcirc\bigcirc]_{[210]}^{\scaleto{\{2N\}}{5.5pt}}$}}
}

\put(233,150){\mbox{$[\bigcirc]_{[022]}^{\scaleto{\{2(N+1)\}}{5.5pt}}$}}

\put(235,161){\mbox{\tiny {\color{Green} $6^{(5)}$}}}

\put(-3,0){

\put(228,131){\mbox{\tiny {\color{Green} $4^{(2)} 2^{(2)}$}}}

\put(225,121){\mbox{$[\bigcirc\bigcirc]_{[112]}^{\scaleto{\{3N-1\}}{5.5pt}}$}}

\put(227,114){\mbox{\tiny {\color{blue} $4^{(3)} 2^{(3)}$}}}

\put(225,106){\mbox{$[\bigcirc\bigcirc]_{[121]}^{\scaleto{\{3N-1\}}{5.5pt}}$}}

\put(227,98){\mbox{\tiny {\color{blue} $4^{(4)} 2^{(4)}$}}}

\put(225,90){\mbox{$[\bigcirc\bigcirc]_{[211]}^{\scaleto{\{3N-1\}}{5.5pt}}$}}
}

\put(80,0){

\put(234,131){\mbox{\tiny {\color{blue} $4^{(5)} 2^{(5)}$}}}

\put(230,121){\mbox{$[\bigcirc\bigcirc]_{[122]}^{\scaleto{\{3N+1\}}{5.5pt}}$}}

\put(232,114){\mbox{\tiny {\color{blue} $4^{(6)}2^{(6)}$}}}

\put(230,106){\mbox{$[\bigcirc\bigcirc]_{[221]}^{\scaleto{\{3N+1\}}{5.5pt}}$}}

\put(232,98){\mbox{\tiny {\color{blue} $4^{(7)} 2^{(7)}$}}}

\put(230,90){\mbox{$[\bigcirc\bigcirc]_{[212]}^{\scaleto{\{3N+1\}}{5.5pt}}$}}

\put(319,115){\mbox{\tiny {\color{blue} $4^{(8)} 2^{(8)}$}}}

\put(315,106){\mbox{$[\bigcirc\bigcirc]_{[222]}^{\scaleto{\{3(N+1)\}}{5.5pt}}$}}
}

\put(242,76){\mbox{\tiny {\color{Green} $6^{(6)}$}}}

\put(240,65){\mbox{$[\bigcirc]_{[202]}^{\scaleto{\{2(N+1)\}}{5.5pt}}$}}

{\color{red} \put(229,30){\mbox{\tiny $\bar 2 \quad \bar 2^{(2)} \, 2^{(1)}$}}}

\put(223,20){\mbox{$[\bigcirc\bigcirc\bigcirc]_{[220]}^{\scaleto{\{2(N+1)\}}{5.5pt}}$}}

\put(70,135){\vector(2,-1){50}}

\put(-93,111){\vector(1,1.5){38}}
\put(-79,109){\vector(1.5,1){38}}
\put(-90,91){\vector(1,-1.5){38}}

\put(-25,35){\vector(1.5,1){46}}
\put(-7,31){\vector(1,0){30}}

\put(-7,170){\vector(2.5,-1){50}}
\put(10,146){\vector(1,0){30}}

\put(-10,130){\vector(1,-2){47.5}}
\put(-10,162){\vector(1,-1.5){56}}

\put(70,75){\vector(2,1){43}}

\put(70,35){\vector(1,1){58}}

\put(270,150){\vector(2,-1){40}}

\put(283,67){\vector(1,1){25}}

\put(280,30){\vector(1,2.7){28}}

\put(175,155){\vector(2,-1){47}}

\put(173,134){\vector(2,-1){47}}

\put(190,85){\vector(2.5,1){30}}

\put(185,65){\vector(1,1.7){32}}

\put(200,37){\vector(1,1.7){28.5}}

\put(169,20){\vector(1,1.7){49}}

\put(70,-18){\vector(2,1.3){67}}
\put(67,-18){\vector(1.15,1){105}}

\put(80,183){\vector(1.75,-1){60}}
\put(75,182){\vector(1.15,-2.5){75}}

\put(65,223){\vector(1.45,-1){75}}
\put(60,220){\vector(1.4,-2){112}}

\put(205,37){\mbox{\scriptsize $-m$}}
\put(177,17){\mbox{\scriptsize $-m$}}
\put(287,37){\mbox{\scriptsize $m$}}
\put(78,37){\mbox{\scriptsize $m$}}

\put(292,91){\mbox{\scriptsize $0$}}
\put(275,102){\mbox{\scriptsize $0$}}
\put(370,105){\mbox{\scriptsize $0$}}
\put(160,116){\mbox{\scriptsize $0$}}

\put(115,53){\mbox{\scriptsize $0$}}
\put(140,70){\mbox{\scriptsize $0$}}
\put(231,72){\mbox{\scriptsize $0$}}
\put(231,56){\mbox{\scriptsize $0$}}

\put(210,160){\mbox{\scriptsize $0$}}
\put(220,140){\mbox{\scriptsize $0$}}
\put(73,155){\mbox{\scriptsize $0$}}
\put(102,138){\mbox{\scriptsize $0$}}

\put(25,160){\mbox{\scriptsize $m$}}
\put(16,140){\mbox{\scriptsize $-m$}}
\put(25,115){\mbox{\scriptsize $m$}}
\put(-13,55){\mbox{\scriptsize $-m$}}

{\color{Green}

\put(212,61){\vector(8,1){25}}
\put(212,71){\vector(8,-1){25}}
\put(87,71){\vector(10,1){85}}
\put(87,62){\vector(12,-1){85}}

\put(180,163){\vector(5,-1){45}}
\put(180,140){\vector(5,1){45}}
\put(77,152){\vector(5,1){60}}
\put(94,145){\vector(1,0){44}}

\put(145,111){\vector(6,1){75}}

\put(272,107){\vector(5,1.5){35}}

\put(272,98){\vector(1,0){35}}

\put(362,112){\vector(1,0){30}}

}

{\color{blue} 

\put(-40,185){\vector(1.5,1){60}}
\put(-10,150){\vector(1.5,1){54}}

\put(-28,18.5){\vector(1.5,-1){55}}

\put(170,107){\vector(1,0){50}}
\put(165,96){\vector(12,-1){54}}

\put(272,125){\vector(1,0){35}}
\put(272,111){\vector(1,0){35}}

\put(272,123){\vector(5,-1.5){35}}
\put(272,95){\vector(5,1.5){35}}

\put(362,124){\vector(4,-1){30}}
\put(362,97){\vector(4,1){30}}

\put(98,31){\vector(15,1){40}}
\put(88,20){\vector(15,-1){50}}
\put(200,27){\vector(5,-1){20}}
\put(200,16){\vector(5,1){20}}
}

\thicklines

{\color{orange}
\put(175,150){\oval(270,40)}
\put(55,118){\mbox{$\phi^2$}}

\put(167,65){\oval(290,40)}
\put(55,92){\mbox{$\phi^2$}}

\put(167,23){\oval(290,40)}
\put(16,36){\mbox{$\phi^2$}}

\put(280,108){\oval(345,41)}
\put(93,105){\mbox{$\phi^3$}}

\put(-3,40){\line(1.45,-1){114}}
\put(-83,40){\line(1.45,-1){114}}
\put(-83,40){\line(1,0){80}}
\put(31,-39){\line(1,0){80}}
\put(-30,45){\mbox{$\phi$}}

\put(-58,135){\line(1.5,1.05){100}}
\put(22,135){\line(1.7,1){119}}
\put(-58,135){\line(1,0){80}}
\put(42,205){\line(1,0){99}}
\put(-30,125){\mbox{$\phi$}}

\put(-1,165){\line(1.5,1.05){110}}
\put(-90,165){\line(1.5,1.05){110}}
\put(-90,165){\line(1,0){89}}
\put(20,242){\line(1,0){89}}
\put(-55,155){\mbox{$\phi$}}
}

{\color{purple}
\put(-115,102){\line(1,1.5){54}}
\put(-115,102){\line(1,-1.5){57}}
\put(-61,183){\line(1,0){116}}
\put(-58,16){\line(1,0){151}}
\put(55,183){\line(7,-4){140}}
\put(93,16){\line(7,6){102}}
}
    
\end{picture}
\caption{\footnotesize The Khovanov--Rozansky complex for the twist knot Tw$_4$. Blue arrows indicate the morphisms $\pm\Sh$ and $0$, while black ones correspond to the morphisms $\pm\Delta$ and $\pm m$. Only operators $0$ and $\pm m$ are written explicitly. The part of the complex marked in red is left after shrinking parts circled in orange. In the shrinked complex, one formally replaces grading to multiplication by the corresponding power of $\phi$. The complex for the precursor diagram is then obtained by the substitution $\phi \rightarrow -q$. The following indices must be considered to be coincident: $2^{(1)}={\color{blue} 2^{(4)}}$, ${\color{blue} 2^{(1)}} = {\color{blue} 2^{( 3)}}$, ${\color{red} 2^{(1)}} = {\color{blue} 2^{(6)}}$. Note that the two Sh operators map to the same $[\overset{{\color{red} \bar 2}}{\bigcirc}\overset{{\color{red} \bar 2^{(2)}}}{\bigcirc}\overset{{\color{red} 2^{(1)}}}{\bigcirc}]_{[220]}$.}
\label{fig:Tw-4-complex}
\end{figure}

The complex in Fig.\,\ref{fig:Tw-4-complex} contains $3^3=27$ resolutions (vertices) and $2\cdot 3 \cdot 3^2=54$ edges\footnote{In generic 3-hypercube, there are $3^n$ vertices and $2n\cdot 3^{n-1}$ edges with $n$ being the number of lock tangles in a link.}. We explicitly indicate only the zero and $m$ morphisms. This is the first example with more than one cycle being mapped by the zero morphism. Emphasize that this happens after the action of the $m$ morphism. For example, this is the case in the sequence of morphisms $[\bigcirc \bigcirc \bigcirc]_{[110]}\overset{m}{\longrightarrow}[\bigcirc \bigcirc]_{[111]}\;\overset{0}{{\color{ Green} \longrightarrow}}\;[\bigcirc \bigcirc]_{[112]}$. On the contrary, if at first $\Delta$ acts, then the Sh morphism comes next. The example is $[\bigcirc]_{[101]}\overset{\Delta}{\longrightarrow}[\bigcirc \bigcirc]_{[111]}\;\overset{\Sh}{{\color{blue} \longrightarrow}}\;[\bigcirc \bigcirc]_{[121]}$. 

We circle in orange the parts of the complex being expanded after the transition from the HOMFLY 2-hypercube to the 3-hypercube. This 2-hypercube is circled in purple. 

\noindent The differentials in this complex are listed below.

\medskip

\noindent $\bullet$ The $0$-th differential is
\begin{equation}
    \hd_0 = \Delta + \Delta + \Delta = \sum_{i=1}^N\sum_{j=0}^{N-i} \left(\vartheta^{(3)}_{N-j} \vartheta^{\prime (3)}_{i+j}+\vartheta^{4^{(1)}}_{N-j} \vartheta^{2^{(1)}}_{i+j} + \bar\vartheta^{(2)}_{N-j} \bar\vartheta^{\prime (4)}_{i+j}\right)\frac{\partial}{\partial \vartheta^{6^{(1)}}_i}\,.
\end{equation}
Its kernel and image are
\ba \nn
\Ker(\hd_0) &= \varnothing\,, \\
\Im(\hd_0) &= \Im(\Delta)^{\scaleto{\{N-1\}}{5.5pt}}\,.
\ea 

\noindent $\bullet$ The $1$-st differential is
\begin{equation}
\begin{aligned}
    \hd_1 &= \Sh + m + \Sh + (-m) + m + \Delta + (-m) + (-\Delta) + \Sh = \\
    &= \sum_{i,j=1}^{N}\left(\vartheta^{{\color{blue} (3)}}_{i+1}\vartheta^{{\color{blue}\prime (3)}}_j-\vartheta^{{\color{blue} (3)}}_i\vartheta^{{\color{blue}\prime (3)}}_{j+1}\right)\frac{\partial^2}{\partial\vartheta_i^{(3)}\partial\vartheta_j^{\prime (3)}} + \sum_{\substack{1\le i,j\le N\\i+j \le N+1}}\left(\vartheta_{i+j-1}^{6^{(2)}} + \vartheta_{i+j-1}^{6^{(3)}} \right) \frac{\partial^2}{\partial\vartheta_i^{(3)}\partial\vartheta_j^{\prime (3)}} + \\
    &+ \sum_{i,j=1}^{N}\left(\vartheta^{{\color{blue} 4^{(1)}}}_{i+1}\vartheta^{{\color{blue} 2^{(1)}}}_j-\vartheta^{{\color{blue} 4^{(1)}}}_i\vartheta^{{\color{blue} 2^{(1)}}}_{j+1}\right)\frac{\partial^2}{\partial\vartheta_i^{4^{(1)}}\partial\vartheta_j^{2^{(1)}}} - \sum_{\substack{1\le i,j\le N\\i+j \le N+1}}\vartheta_{i+j-1}^{6^{(2)}}\frac{\partial^2}{\partial\vartheta_i^{4^{(1)}}\partial\vartheta_j^{2^{(1)}}} + \sum_{i=1}^N\sum_{j=0}^{N-i} \bar\vartheta^{(2)}_{N-j} \bar\vartheta^{2^{(2)}}_{i+j} \frac{\partial}{\partial \vartheta^{4^{(1)}}_i} + \\
    &+ \sum_{i,j=1}^{N}\left(\bar\vartheta^{{\color{blue} (2)}}_{i+1}\bar\vartheta^{{\color{blue} (4)}}_j-\bar\vartheta^{{\color{blue} (2)}}_i\bar\vartheta^{{\color{blue} (4)}}_{j+1}\right)\frac{\partial^2}{\partial\bar\vartheta_i^{(2)}\partial\bar\vartheta_j^{(4)}} - \sum_{\substack{1\le i,j\le N\\i+j \le N+1}}\vartheta_{i+j-1}^{6^{(3)}}\frac{\partial^2}{\partial\bar\vartheta_i^{(2)}\partial\bar\vartheta_j^{(4)}} + \sum_{i=1}^N\sum_{j=0}^{N-i} \bar\vartheta^{2^{(2)}}_{N-j} \vartheta^{2^{(1)}}_{i+j} \frac{\partial}{\partial \bar\vartheta^{(4)}_i}\,.
\end{aligned}
\end{equation}
Its kernel and image are
\ba \nn
\Ker(\hd_1) &= ({\rm Ker}({\rm Sh}) \cap {\rm Ker}(m))^{\scaleto{\{N-1\}}{5.5pt}} \oplus {\rm Ker}({\rm Sh})^{\scaleto{\{N-1\}}{5.5pt}}\,, \\
\Im(\hd_1) &= 3\, \Im(\Sh)^{\scaleto{\{N+1\}}{5.5pt}} \oplus \langle \vth_N^{6^{(2)}} + \vth_N^{6^{(3)}} \rangle^{\scaleto{\{2(N-1)\}}{5.5pt}} \oplus (\langle \tilde{\vth}_N \rangle \otimes \Im(\Delta))^{\scaleto{\{2(N-1)\}}{5.5pt}}\,.
\ea 

\noindent $\bullet$ The $2$-nd differential is
\begin{equation}
\begin{aligned}
    \hd_2 &= m + m + m + \Delta + 0 + 0 + \Delta + (-\Delta) + 0 + 0 + m + (-\Sh) + \Sh + m + \Delta = \\
    &= \sum_{\substack{1\le i,j\le N\\i+j \le N+1}}\left(\vartheta_{i+j-1}^{{\color{Green} 6^{(1)}}} + \vartheta_{i+j-1}^{{\color{Green} 6^{(3)}}} \right) \frac{\partial^2}{\partial\vartheta_i^{{\color{blue}(3)}}\partial\vartheta_j^{{\color{blue}\prime (3)}}} + \sum_{\substack{1\le i,j\le N\\i+j \le N+1}}\vartheta_{i+j-1}^{{\color{Green} 6^{(2)}}} \frac{\partial^2}{\partial\vartheta_i^{{\color{blue}4^{(1)}}}\partial\vartheta_j^{{\color{blue} 2^{(1)}}}} + \sum_{i=1}^N\sum_{j=0}^{N-i} \bar\vartheta^{(2)}_{N-j} \bar\vartheta^{{\color{blue} 2^{(2)}}}_{i+j} \frac{\partial}{\partial \vartheta^{{\color{blue} 4^{(1)}}}_i} + \\
    &+ \sum_{i=1}^N\sum_{j=0}^{N-i} \vartheta^{4^{(2)}}_{N-j} \vartheta^{2^{(1)}}_{i+j} \left(\frac{\partial}{\partial \vartheta^{6^{(2)}}_i} - \frac{\partial}{\partial \vartheta^{6^{(3)}}_i}\right) + \sum_{\substack{1\le i,j\le N\\i+j \le N+1}}\vartheta_{i+j-1}^{4^{(2)}} \frac{\partial^2}{\partial \bar\vartheta_i^{(2)}\partial \bar\vartheta_j^{2^{(2)}}} + \sum_{i,j=1}^{N}\left(\bar \vartheta^{{\color{blue} (2)}}_{i+1}\bar \vartheta^{{\color{blue}2^{(2)}}}_j-\bar \vartheta^{{\color{blue} (2)}}_i\bar \vartheta^{{\color{blue} 2^{(2)}}}_{j+1}\right)\frac{\partial^2}{\partial \bar \vartheta_i^{(2)}\partial \bar \vartheta_j^{2^{(2)}}} - \\
    &- \sum_{i,j=1}^{N}\left(\bar \vartheta^{{\color{blue} 2^{(2)}}}_{i+1} \vartheta^{{\color{blue}2^{(1)}}}_j-\bar \vartheta^{{\color{blue} 2^{(2)}}}_i \vartheta^{{\color{blue} 2^{(1)}}}_{j+1}\right)\frac{\partial^2}{\partial \bar \vartheta_i^{2^{(2)}}\partial \vartheta_j^{2^{(1)}}} + \sum_{\substack{1\le i,j\le N\\i+j \le N+1}}\vartheta_{i+j-1}^{{\color{Green} 6^{(4)}}} \frac{\partial^2}{\partial\vartheta_i^{{\color{blue}(2)}}\partial\vartheta_j^{{\color{blue} (4)}}} + \sum_{i=1}^N\sum_{j=0}^{N-i} \bar\vartheta^{{\color{blue} 2^{(2)}}}_{N-j} \vartheta^{2^{(1)}}_{i+j} \frac{\partial}{\partial \vartheta^{{\color{blue} (4)}}_i}\,.
\end{aligned}
\end{equation}
Its kernel and image are
\ba \nn
\Ker(\hd_2) &= 3 \, \Ker(m)^{\scaleto{\{N+1\}}{5.5pt}} \oplus V^{\scaleto{\{2(N-1)\}}{5.5pt}} \oplus (\langle \tilde{\vth}_N \rangle \otimes \Ker(\Sh))^{\scaleto{\{2(N-1)\}}{5.5pt}}\,, \\
\Im(\hd_2) &= 3\,V^{\scaleto{\{2N\}}{5.5pt}} \oplus \Im(\Delta)^{\scaleto{\{3(N-1)\}}{5.5pt}} \oplus  (V\otimes \Im(\Sh))^{\scaleto{\{2N\}}{5.5pt}} \oplus (V\otimes(\Im(\Delta)\cap \Im(\Sh)))^{\scaleto{\{2N\}}{5.5pt}}\,.
\ea 

\noindent $\bullet$ The $3$-rd differential is
\begin{equation}
\begin{aligned}
    \hd_3 &= 0 + 0 + \Delta + \Delta + 0 + (-\Sh) + \Sh + (-\Delta) + 0 + 0 + (-m) + \Delta + (-m) + \Sh + \Sh = \\
    &= \sum_{i=1}^N\sum_{j=0}^{N-i} \vartheta^{{\color{Green} 4^{(2)}}}_{N-j} \vartheta^{{\color{Green} 2^{(2)}}}_{i+j} \frac{\partial}{\partial \vartheta^{{\color{Green} 6^{(1)}}}_i} + \sum_{i=1}^N\sum_{j=0}^{N-i} \vartheta^{{\color{blue} 4^{(3)}}}_{N-j} \vartheta^{{\color{blue} 2^{(3)}}}_{i+j} \frac{\partial}{\partial \vartheta^{{\color{Green} 6^{(2)}}}_i} - \sum_{i,j=1}^{N}\left( \vartheta^{{\color{blue} 4^{(3)}}}_{i+1} \vartheta^{{\color{blue}2^{(3)}}}_j-\vartheta^{{\color{blue} 4^{(3)}}}_i \vartheta^{{\color{blue} 2^{(3)}}}_{j+1}\right)\frac{\partial^2}{\partial \vartheta_i^{4^{(2)}}\partial \vartheta_j^{2^{(1)}}} + \\
    &+ \sum_{i,j=1}^{N}\left( \vartheta^{{\color{blue} 4^{(4)}}}_{i+1} \vartheta^{{\color{blue}2^{(4)}}}_j-\vartheta^{{\color{blue} 4^{(4)}}}_i \vartheta^{{\color{blue} 2^{(4)}}}_{j+1}\right)\frac{\partial^2}{\partial \vartheta_i^{4^{(2)}}\partial \vartheta_j^{2^{(1)}}} - \sum_{i=1}^N\sum_{j=0}^{N-i} \vartheta^{{\color{Green} 4^{(2)}}}_{N-j} \vartheta^{{\color{Green} 2^{(2)}}}_{i+j} \frac{\partial}{\partial \vartheta^{{\color{Green} 6^{(3)}}}_i} + \sum_{i=1}^N\sum_{j=0}^{N-i} \vartheta^{{\color{blue} 4^{(4)}}}_{N-j} \vartheta^{{\color{blue} 2^{(4)}}}_{i+j} \frac{\partial}{\partial \vartheta^{{\color{Green} 6^{(4)}}}_i} - \\
    &- \sum_{\substack{1\le i,j\le N\\i+j \le N+1}}\vartheta_{i+j-1}^{{\color{blue} 4^{(3)}}} \frac{\partial^2}{\partial\bar \vartheta_i^{(2)}\partial \bar \vartheta_j^{{\color{blue} 2^{(2)}}}} - \sum_{\substack{1\le i,j\le N\\i+j \le N+1}}\vartheta_{i+j-1}^{{\color{blue} 4^{(4)}}} \frac{\partial^2}{\partial\bar \vartheta_i^{{\color{blue}(2)}}\partial \bar \vartheta_j^{{\color{blue} 2^{(2)}}}} + \sum_{i,j=1}^{N}\left( \bar\vartheta^{{\color{red} (2)}}_{i+1} \vartheta^{{\color{red} 2^{(1)}}}_j-\bar\vartheta^{{\color{red} (2)}}_i \vartheta^{{\color{red} 2^{(1)}}}_{j+1}\right)\frac{\partial^2}{\partial \bar\vartheta_i^{(2)}\partial \vartheta_j^{{\color{blue} 2^{(1)}}}} + \\
    &+ \sum_{i,j=1}^{N}\left( \bar\vartheta^{{\color{red} 2^{(2)}}}_{i+1} \vartheta^{{\color{red} 2^{(1)}}}_j-\bar\vartheta^{{\color{red} 2^{(2)}}}_i \vartheta^{{\color{red} 2^{(1)}}}_{j+1}\right)\frac{\partial^2}{\partial \bar\vartheta_i^{{\color{blue} 2^{(2)}}}\partial \vartheta_j^{2^{(1)}}}\,.
\end{aligned}
\end{equation}
Its kernel and image are
\ba \nn
\Ker(\hd_3) &= 3\,V^{\scaleto{\{2N\}}{5.5pt}} \oplus \Ker(\Sh)^{\scaleto{\{3(N-1)\}}{5.5pt}} \oplus (V\otimes \Ker(m))^{\scaleto{\{2N\}}{5.5pt}} \oplus (V\otimes (\Ker(\Sh)\cap\Ker(m)))^{\scaleto{\{2N\}}{5.5pt}} \oplus \langle \tilde{\vth}_N \tilde{\vth}_1 \tilde{\vth}_j \rangle^{\scaleto{\{2N\}}{5.5pt}},\, j > 1\,, \\
\Im(\hd_3) &= 3\, \Im(\Delta)^{\scaleto{\{3N-1\}}{5.5pt}} \oplus (\Im(\Sh) \otimes V)^{\scaleto{\{2(N+1)\}}{5.5pt}} \oplus \langle \tilde{\vth}_{i+1} \tilde{\vth}_j - \tilde{\vth}_{i} \tilde{\vth}_{j+1},\, \tilde{\vth}_N \tilde{\vth}_k \rangle^{\scaleto{\{3N-1\}}{5.5pt}},\, i,\,j < N - 1,\, k = 2,\dots,N-1\,,
\ea 
where in the last line $\tilde{\vth}_{i} \tilde{\vth}_j = \vth_i^{{\color{blue} 4^{(3)}}}\vth_j^{{\color{blue} 2^{(3)}}}-\vth_i^{{\color{blue} 4^{(4)}}}\vth_j^{{\color{blue} 2^{(4)}}}$.

\medskip

\noindent $\bullet$ The $4$-th differential is
\begin{equation}
\begin{aligned}
    \hd_4 &= \Delta + (-\Sh) + \Sh + 0 + \Sh + \Sh + 0 + \Delta + m = \\
    &= \sum_{i=1}^N\sum_{j=0}^{N-i} \vartheta^{{\color{blue} 4^{(5)}}}_{N-j} \vartheta^{{\color{blue} 2^{(5)}}}_{i+j} \frac{\partial}{\partial \vartheta^{{\color{Green} 6^{(5)}}}_i} + \sum_{i=1}^N\sum_{j=0}^{N-i} \vartheta^{{\color{blue} 4^{(7)}}}_{N-j} \vartheta^{{\color{blue} 2^{(7)}}}_{i+j} \frac{\partial}{\partial \vartheta^{{\color{Green} 6^{(6)}}}_i} + \sum_{\substack{1\le i,j\le N\\i+j \le N+1}}\vartheta_{i+j-1}^{{\color{blue} 4^{(6)}}} \frac{\partial^2}{\partial\bar \vartheta_i^{{\color{red} (2)}}\partial \bar \vartheta_j^{{\color{red} 2^{(2)}}}} - \\
    &- \sum_{i,j=1}^{N}\left( \vartheta^{{\color{blue} 4^{(5)}}}_{i+1} \vartheta^{{\color{blue} 2^{(5)}}}_j-\vartheta^{{\color{blue} 4^{(5)}}}_i \vartheta^{{\color{blue} 2^{(5)}}}_{j+1} -  \vartheta^{{\color{blue} 4^{(7)}}}_{i+1} \vartheta^{{\color{blue} 2^{(7)}}}_j + \vartheta^{{\color{blue} 4^{(7)}}}_i \vartheta^{{\color{blue} 2^{(7)}}}_{j+1}\right)\frac{\partial^2}{\partial \vartheta_i^{{\color{Green} 4^{(2)}}}\partial \vartheta_j^{{\color{Green} 2^{(2)}}}} + \\
    &+ \sum_{i,j=1}^{N}\left( \vartheta^{{\color{blue} 4^{(6)}}}_{i+1} \vartheta^{{\color{blue} 2^{(6)}}}_j-\vartheta^{{\color{blue} 4^{(6)}}}_i \vartheta^{{\color{blue} 2^{(6)}}}_{j+1} \right)\left(\frac{\partial^2}{\partial \vartheta_i^{{\color{blue} 4^{(3)}}}\partial \vartheta_j^{{\color{blue} 2^{(3)}}}} + \frac{\partial^2}{\partial \vartheta_i^{{\color{blue} 4^{(4)}}}\partial \vartheta_j^{{\color{blue} 2^{(4)}}}}\right)\,. 
\end{aligned}
\end{equation}
Its kernel and image are
\ba \nn
    \Ker(\hd_4) &= 2\,\Ker(\Sh)^{\scaleto{\{3N-1\}}{5.5pt}} \oplus 2\,(V \otimes V)^{\scaleto{\{3N-1\}}{5.5pt}} \oplus (V \otimes \Ker(m))^{\scaleto{\{2(N+1)\}}{5.5pt}} \oplus \langle \tilde{\vth}_2,\, \dots,\, \tilde{\vth}_N \rangle^{\scaleto{\{2(N+1)\}}{5.5pt}}\,, \\
    \Im(\hd_4) &= (V \otimes V)^{\scaleto{\{3N+1\}}{5.5pt}} \oplus 2\, \Im(\Delta)^{\scaleto{\{3N+1\}}{5.5pt}} \oplus \langle \tilde{\vth}_{i+1} \tilde{\vth}_j - \tilde{\vth}_{i} \tilde{\vth}_{j+1},\, \tilde{\vth}_N \tilde{\vth}_k \rangle^{\scaleto{\{3N+1\}}{5.5pt}},\, i,\,j < N - 1,\, k = 2,\dots,N-1\,.
\ea 
where in the last line $\tilde{\vth}_{i} \tilde{\vth}_j = \vth_i^{{\color{blue} 4^{(5)}}}\vth_j^{{\color{blue} 2^{(5)}}}-\vth_i^{{\color{blue} 4^{(7)}}}\vth_j^{{\color{blue} 2^{(7)}}}$.

\medskip

\noindent $\bullet$ The $5$-th differential is
\begin{equation}
    \hd_5 = \Sh + 0 + \Sh = \sum_{i,j=1}^{N}\left( \vartheta^{{\color{blue} 4^{(8)}}}_{i+1} \vartheta^{{\color{blue} 2^{(8)}}}_j-\vartheta^{{\color{blue} 4^{(8)}}}_i \vartheta^{{\color{blue} 2^{(8)}}}_{j+1}\right)\left(\frac{\partial^2}{\partial \vartheta_i^{4^{(5)}}\partial \vartheta_j^{2^{(5)}}} + \frac{\partial^2}{\partial \vartheta_i^{4^{(7)}}\partial \vartheta_j^{2^{(7)}}}\right)\,.
\end{equation}
Its kernel and image are
\ba \nn
\Ker(\hd_5) &= 2\,(V \otimes V)^{\scaleto{\{3N+1\}}{5.5pt}} \oplus \Ker(\Sh)^{\scaleto{\{3N+1\}}{5.5pt}}\,, \\
\Im(\hd_5) &= \Im(\Sh)^{\scaleto{\{3(N+1)\}}{5.5pt}}\,.
\ea 
The cohomologies and their quantum dimensions are
{\footnotesize\begin{equation} \nn
\begin{aligned}
    \cH_0 &= \Ker(\hd_0) = \varnothing \quad \Lra \quad \boxed{\dim_q \cH_0 = 0\,,} \\
    \cH_1 &= \Ker(\hd_1) \backslash \Im(\hd_0) = ({\rm Ker}({\rm Sh}) \cap {\rm Ker}(m))^{\scaleto{\{N-1\}}{5.5pt}} \quad \Lra \quad \boxed{\dim_q \cH_1 = q^{-1}[N-1]\,,} \\
    \cH_2 &= \Ker(\hd_2) \backslash \Im(\hd_1) = V^{\scaleto{\{2(N-1)\}}{5.5pt}} \backslash \langle \vth_N^{6^{(2)}} + \vth_N^{6^{(3)}} \rangle^{\scaleto{\{2(N-1)\}}{5.5pt}} \quad \Lra \quad \boxed{\dim_q \cH_2 = q^{2(N-1)}([N]-q^{-N+1}) = q^{2N-1}[N-1]\,,} \\
    \cH_3 &= \Ker(\hd_3) \backslash \Im(\hd_2) = \langle \tilde{\vth}_N \tilde{\vth}_1 \tilde{\vth}_j \rangle^{\scaleto{\{2N\}}{5.5pt}} \quad \Lra \quad \boxed{\dim_q \cH_3 = q^{2N}([N]-q^{N-1})=q^{2N-1}[N-1]\,,} \\
    \cH_4 &= \Ker(\hd_4) \backslash \Im(\hd_3) = \langle \tilde{\vth}_2,\, \dots,\, \tilde{\vth}_N \rangle^{\scaleto{\{2(N+1)\}}{5.5pt}} \oplus ((V \otimes V)\backslash (\Im(\Delta) \oplus \langle \tilde{\vth}_{i+1} \tilde{\vth}_j - \tilde{\vth}_{i} \tilde{\vth}_{j+1},\, \tilde{\vth}_N \tilde{\vth}_k \rangle))^{\scaleto{\{3N-1\}}{5.5pt}},\, i,\,j < N - 1,\, k = 2,\dots,N-1 \\
    &\Lra \quad \boxed{\dim_q \cH_4 = q^{2(N+1)}([N] - q^{N-1}) + q^{3N-1}([N]^2-q^{-N+1}[N]-q^{-2}[N]^2) = (q^{2N+1} + q^{4N-1})[N-1]\,,} \\
    \cH_5 &= \Ker(\hd_5) \backslash \Im(\hd_4) = ((V \otimes V)\backslash (\Im(\Delta) \oplus \langle \tilde{\vth}_{i+1} \tilde{\vth}_j - \tilde{\vth}_{i} \tilde{\vth}_{j+1},\, \tilde{\vth}_N \tilde{\vth}_k \rangle))^{\scaleto{\{3N+1\}}{5.5pt}},\, i,\,j < N - 1,\, k = 2,\dots,N-1 \\
    &\Lra \quad \boxed{\dim_q \cH_5 = q^{3N+1} \cdot q^N[N-1] = q^{4N+1}[N-1]\,,} \\
    \cH_6 &= \Ker(0) \backslash \Im(\hd_5) = ((V \otimes V)\backslash \Im(\Sh))^{\scaleto{\{3(N+1)\}}{5.5pt}} \quad \Lra \quad \boxed{\dim_q \cH_6 = q^{3(N+1)}([N]^2 - q^{-1}[N][N-1]) = q^{4N+2}[N]\,.}
\end{aligned}
\end{equation}}

\noindent The Khovanov--Rozansky polynomial is
\be
\boxed{\boxed{
\begin{array}{c}
P^{{\rm Tw}_4}(A,q,T) = (q^N T)^{-6}(T \dim_q \cH_1 + T^2 \dim_q \cH_2 + T^3 \dim_q \cH_3 + T^4 \dim_q \cH_4 + T^5 \dim_q \cH_5 + T^6 \dim_q \cH_6) = \\ \\
= q^{-2N+2}[N] + \left( q^{-2N+1} T^{-1} + (q^{-2N-1} + q^{-4N+1})T^{-2} + q^{-4N-1}(T^{-3}+T^{-4}) + q^{-6N-1}T^{-5} \right)[N-1]
\end{array}
}}
\nn
\ee
what reproduces the known answer in \cite{carqueville2014computing,anokhina2014towards-R}.

The HOMFLY polynomial is the $q$-Euler characteristic of the complex in Fig.\,\ref{fig:Tw-4-complex} and also can be computed via the 2-hypercube (as described in Section~\ref{sec:H-hypercubes}) in Fig.\,\ref{fig:Tw-4-2-hypercube}:

\begin{equation}
\begin{aligned}
    H^{{\rm Tw}_4}(A,q) & \overset{{\rm Fig.}\,\ref{fig:Tw-4-complex}}{=} A^{-6}\Big( D - 3 q^{N-1} D^2 + 3 q^{N+1} D^2 + 2 q^{2(N-1)} D + q^{2(N-1)} D^3 - 4 q^{2N} D - q^{3(N-1)} D^2 -\\
    &- 2 q^{2N} D^3 + 2 q^{2(N+1)} D + 3 q^{3N-1} D^2 + q^{2(N+1)} D^3 - 3 q^{3N+1} D^2 + q^{3(N+1)} D^2 \Big) = \\ 
    &\overset{{\rm Fig.}\,\ref{fig:Tw-4-2-hypercube}}{=} A^{-6}\left( D + 3 \phi D^2 + 2\phi^2 D + \phi^2 D^3 + \phi^3 D^2 \right)\,.
\end{aligned}
\end{equation}


\paragraph{Precursor.} We first shrink orange circled spaces in Fig.\,\ref{fig:Tw-4-complex} into single ones and obtain the circled in purple complex with powers of $\phi$ as multipliers forming the HOMFLY 2-hypercube, see Section~\ref{sec:H-hypercubes}:

\begin{figure}[h!]
    \centering
\begin{picture}(300,120)(90,-10)

\put(80,45){\mbox{$\varnothing\;\longrightarrow\;[\overset{6^{(1)}}{\bigcirc}]_{[000]}$}}

\put(19,0){

\put(160,93){\mbox{$\phi\,[\overset{3}{\bigcirc}\overset{3'}{\bigcirc}]_{[001]}$}}

\put(163,45){\mbox{$\phi\,[\overset{4^{(1)}}{\bigcirc}\overset{2^{(1)}}{\bigcirc}]_{[010]}$}}

\put(160,0){\mbox{$\phi\,[\overset{\bar 2}{\bigcirc}\overset{\bar 4}{\bigcirc}]_{[100]}$}}

\put(130,57){\vector(1,1){30}}

\put(133,47){\vector(1,0){25}}

\put(130,38){\vector(1,-1){30}}

\put(17,0){

\put(225,93){\mbox{$\phi^2\,[\overset{6^{(2)}}{\bigcirc}]_{[011]}$}}

\put(215,45){\mbox{$\phi^2\,[\overset{6^{(3)}}{\bigcirc}]_{[101]}$}}

\put(220,-5){\mbox{$\phi^2\,[\overset{\bar 2}{\bigcirc}\overset{\bar 2^{(2)}}{\bigcirc}\overset{2^{(1)}}{\bigcirc}]_{[110]}$}}

\put(192,57){\vector(1,1){30}}

\put(192,10){\vector(1,1){30}}

\put(192,38){\vector(1,-1){30}}

\put(192,85){\vector(1,-1){30}}

\put(194,95){\vector(1,0){25}}

\put(194,2){\vector(1,0){25}}

\put(255,10){\vector(1,1){30}}

\put(255,85){\vector(1,-1){30}}

\put(290,45){\mbox{$\phi^3\,[\overset{4^{(2)}}{\bigcirc}\overset{2^{(1)}}{\bigcirc}]_{[111]}\;\longrightarrow\;\varnothing$}}

\put(260,47){\vector(1,0){25}}

}

}

\end{picture}
    \caption{\footnotesize The HOMFLY 2-hypercube for the Tw$_4$ knot. The morphisms and their signs can be restored uniquely, but are needed only in the Khovanov precursor complex. The Khovanov--Rozansky polynomial cannot be calculated with the use of this 2-hypercube, as we have discussed in Section~\ref{sec:H-hypercubes}.}
    \label{fig:Tw-4-2-hypercube}
\end{figure}
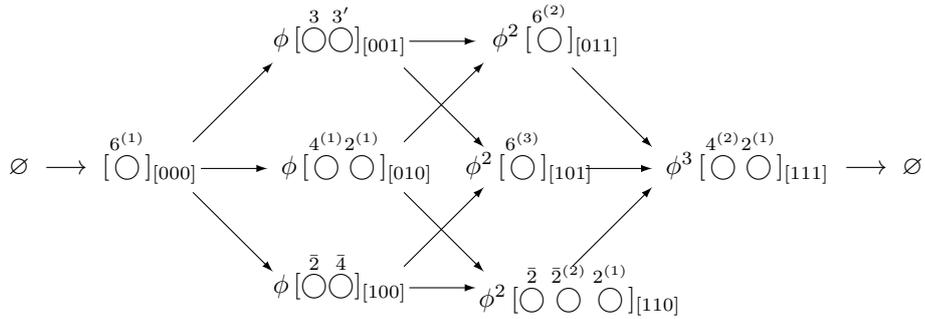

\begin{figure}[h!]
    \centering
\begin{picture}(300,120)(90,-10)

\put(80,45){\mbox{$\varnothing\;\overset{0}{\longrightarrow}\;[\overset{6^{(1)}}{\bigcirc}]_{[000]}^{\scaleto{\{0\}}{5.5pt}}$}}

\put(19,0){

\put(160,93){\mbox{$[\overset{3}{\bigcirc}\overset{3'}{\bigcirc}]_{[001]}^{\scaleto{\{1\}}{5.5pt}}$}}

\put(163,45){\mbox{$[\overset{4^{(1)}}{\bigcirc}\overset{2^{(1)}}{\bigcirc}]_{[010]}^{\scaleto{\{1\}}{5.5pt}}$}}

\put(160,0){\mbox{$[\overset{\bar 2}{\bigcirc}\overset{\bar 4}{\bigcirc}]_{[100]}^{\scaleto{\{1\}}{5.5pt}}$}}

\put(130,57){\vector(1,1){30}}

\put(133,47){\vector(1,0){25}}

\put(130,38){\vector(1,-1){30}}

\put(17,0){

\put(225,93){\mbox{$[\overset{6^{(2)}}{\bigcirc}]_{[011]}^{\scaleto{\{2\}}{5.5pt}}$}}

\put(223,45){\mbox{$[\overset{6^{(3)}}{\bigcirc}]_{[101]}^{\scaleto{\{2\}}{5.5pt}}$}}

\put(220,-5){\mbox{$[\overset{\bar 2}{\bigcirc}\overset{\bar 2^{(2)}}{\bigcirc}\overset{2^{(1)}}{\bigcirc}]_{[110]}^{\scaleto{\{2\}}{5.5pt}}$}}

\put(110,68){\mbox{$\Delta$}}

\put(120,50){\mbox{$\Delta$}}

\put(110,25){\mbox{$\Delta$}}

\put(200,100){\mbox{$m$}}

\put(185,75){\mbox{$m$}}

\put(175,60){\mbox{$-m$}}

\put(195,-10){\mbox{$-\Delta$}}

\put(175,13){\mbox{$-m$}}

\put(180,27){\mbox{$\Delta$}}

\put(280,25){\mbox{$m$}}

\put(280,65){\mbox{$\Delta$}}

\put(260,50){\mbox{$-\Delta$}}

\put(192,57){\vector(1,1){30}}

\put(192,10){\vector(1,1){30}}

\put(192,38){\vector(1,-1){30}}

\put(192,85){\vector(1,-1){30}}

\put(194,95){\vector(1,0){25}}

\put(194,2){\vector(1,0){25}}

\put(255,10){\vector(1,1){30}}

\put(255,85){\vector(1,-1){30}}

\put(290,45){\mbox{$[\overset{4^{(2)}}{\bigcirc}\overset{2^{(1)}}{\bigcirc}]_{[111]}^{\scaleto{\{3\}}{5.5pt}}\;\overset{0}{\longrightarrow}\;\varnothing$}}

\put(260,47){\vector(1,0){25}}

}

}

\end{picture}
    \caption{\footnotesize The precursor knot for the Tw$_4$ knot is the tied unknot. Its Knovanov complex is shown in this picture.}
    \label{fig:prec-Tw-4-2-hypercube}
\end{figure}
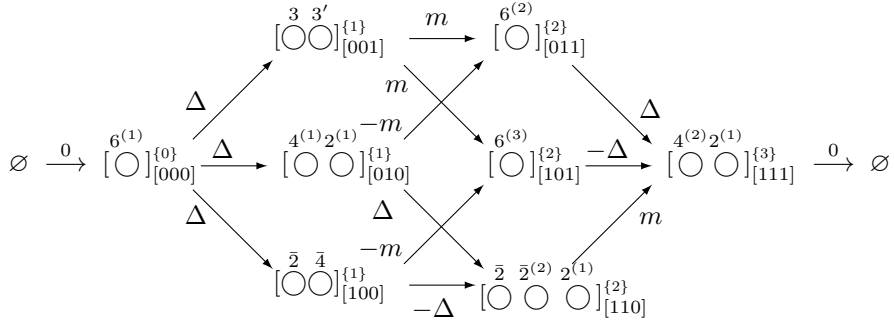

\noindent Now, we make the substitution $\phi\rightarrow -q$ and $N=2$ and transfer to the Khovanov complex in Fig.\,\ref{fig:prec-Tw-4-2-hypercube}. 
\noindent To calculate the Khovanov polynomial, one writes down the differentials and finds their kernels and images.

\medskip

\noindent $\bullet$ The $0$-th differential is
{\small \begin{equation} \nn
    \hd_0^{\,\rm unknot} = \underbrace{\left( \vth_2^{(3)} \vth_2^{\prime (3)} + \vth_2^{4^{(1)}} \vth_2^{2^{(1)}} + \bar \vth_2^{(2)} \bar \vth_2^{(4)} \right)}_{\vth_2^{+}\vth_2^{\prime +}}\frac{\partial}{\partial \vth_2^{6^{(1)}}} + \Big(\underbrace{ \vth_1^{(3)} \vth_2^{\prime (3)} + \vth_1^{4^{(1)}} \vth_2^{2^{(1)}} + \bar \vth_1^{(2)} \bar \vth_2^{(4)}}_{\vth_1^{+}\vth_2^{\prime +}} + \underbrace{ \vth_2^{(3)} \vth_1^{\prime (3)} + \vth_2^{4^{(1)}} \vth_1^{2^{(1)}} + \bar \vth_2^{(2)} \bar \vth_1^{(4)}}_{\vth_2^{+}\vth_1^{\prime +}} \Big)\frac{\partial}{\partial \vth_1^{6^{(1)}}}\,.
\end{equation}}
Its image and kernel are
\ba \nn
    \Im(\hd_0^{\,\rm unknot}) &= \langle \vth_2^{+}\vth_2^{\prime +},\,\vth_2^{+}\vth_1^{\prime +} + \vth_1^{+}\vth_2^{\prime +} \rangle\,, \\
    \Ker(\hd_0^{\,\rm unknot}) &= \varnothing\,.
\ea 

\noindent $\bullet$ The $1$-st differential is
\ba 
\hd_1^{\,\rm unknot} &= \vth^{6^{(2)}}_2 \Bigg( \underbrace{\frac{\partial^2}{\partial \vth^{(3)}_1 \partial \vth_2^{\prime (3)}} - \frac{\partial^2}{\partial \vth^{4^{(1)}}_1 \partial \vth_2^{2^{(1)}}}}_{\frac{\partial^2}{\partial \vth^{(3)-4^{(1)}}_1 \partial \vth_2^{\prime (3)-2^{(1)}}}} + \underbrace{\frac{\partial^2}{\partial \vth^{(3)}_2 \partial \vth_1^{\prime (3)}} - \frac{\partial^2}{\partial \vth^{4^{(1)}}_2 \partial \vth_1^{2^{(1)}}}}_{\frac{\partial^2}{\partial \vth^{(3)-4^{(1)}}_2 \partial \vth_1^{\prime (3)-2^{(1)}}}} \Bigg) + \vth^{6^{(2)}}_1 \underbrace{\left( \frac{\partial^2}{\partial \vth^{(3)}_1 \partial \vth_1^{\prime (3)}} - \frac{\partial^2}{\partial \vth^{4^{(1)}}_1 \partial \vth_1^{2^{(1)}}} \right)}_{\frac{\partial^2}{\partial \vth^{(3)-4^{(1)}}_1 \partial \vth_1^{\prime (3)-2^{(1)}}}}+ \\ 
&+ \vth^{6^{(3)}}_2 \Bigg( \underbrace{\frac{\partial^2}{\partial \vth^{(3)}_1 \partial \vth_2^{\prime (3)}} - \frac{\partial^2}{\partial \bar\vth^{(2)}_1 \partial \bar\vth_2^{(4)}}}_{\frac{\partial^2}{\partial \vth^{(3)-(2)}_1 \partial \vth_2^{\prime (3)-(4)}}} + \underbrace{\frac{\partial^2}{\partial \vth^{(3)}_2 \partial \vth_1^{\prime (3)}} - \frac{\partial^2}{\partial \bar\vth^{(2)}_2 \partial \bar\vth_1^{(4)}}}_{\frac{\partial^2}{\partial \vth^{(3)-(2)}_2 \partial \vth_1^{\prime (3)-(4)}}} \Bigg) + \vth^{6^{(3)}}_1 \underbrace{\left( \frac{\partial^2}{\partial \vth^{(3)}_1 \partial \vth_1^{\prime (3)}} - \frac{\partial^2}{\partial \bar \vth^{(2)}_1 \partial \bar \vth_1^{(4)}} \right)}_{\frac{\partial^2}{\partial \vth^{(3)-(2)}_1 \partial \vth_1^{\prime (3)-(4)}}} + \\
&+ \bar \vth_2^{(2)} \bar \vth_2^{2^{(2)}} \frac{\partial}{\partial \vth_2^{4^{(1)}}} + \left( \bar \vth_1^{(2)} \bar \vth_2^{2^{(2)}} + \bar \vth_2^{(2)} \bar \vth_1^{2^{(2)}} \right) \frac{\partial}{\partial \vth_1^{4^{(1)}}} - \bar \vth_2^{2^{(2)}} \vth_2^{2^{(1)}} \frac{\partial}{\partial \bar \vth_2^{(4)}} - \left( \bar \vth_1^{2^{(2)}} \vth_2^{2^{(1)}} + \bar \vth_2^{2^{(2)}} \vth_1^{2^{(1)}} \right) \frac{\partial}{\partial \bar \vth_1^{(4)}}\,.
\ea 
Its image and kernel are
\ba \nn
\Im(\hd_1^{\,\rm unknot}) &= \langle \vth_1^{6^{(2)}} + \vth_1^{6^{(3)}},\, \vth_2^{6^{(2)}} + \vth_2^{6^{(3)}} \rangle \oplus \langle \bar \vth_2^{2^{(2)}} \vth_2^{2^{(1)}} \bar \vth_2^{(2)},\,\bar \vth_1^{2^{(2)}} \vth_2^{2^{(1)}} \bar \vth_2^{(2)},\, \bar \vth_2^{2^{(2)}} \vth_1^{2^{(1)}} \bar \vth_2^{(2)} + \vth_2^{6^{(2)}},\, \bar \vth_2^{2^{(2)}} \vth_2^{2^{(1)}} \bar \vth_1^{(2)} + \vth_2^{6^{(3)}},\\
&\bar \vth_2^{2^{(2)}} \vth_1^{2^{(1)}} \bar \vth_1^{(2)} + \bar \vth_1^{2^{(2)}} \vth_1^{2^{(1)}} \bar \vth_2^{(2)} + \vth_1^{6^{(2)}},\, \bar \vth_1^{2^{(2)}} \vth_2^{2^{(1)}} \bar \vth_1^{(2)} + \bar \vth_2^{2^{(2)}} \vth_1^{2^{(1)}} \bar \vth_1^{(2)} + \vth_1^{6^{(3)}} \rangle \cong \\
&\cong V \oplus V \oplus V \oplus \langle \bar \vth_2^{2^{(2)}} \vth_2^{2^{(1)}} \bar \vth_2^{(2)},\, \bar \vth_1^{2^{(2)}} \vth_2^{2^{(1)}} \bar \vth_2^{(2)} \rangle\,, \\
\Ker(\hd_1^{\,\rm unknot}) &= \langle \vth_2^{(3)}\vth_2^{\prime (3)},\, \vth_1^{(3)}\vth_2^{\prime (3)} - \vth_2^{(3)}\vth_1^{\prime (3)} \rangle \oplus \\
&\oplus \langle \vth_2^{4^{(1)}}\vth_2^{2^{(1)}} - \bar \vth_2^{(2)} \bar \vth_2^{(4)},\, \frac{1}{2}\left( \vth_1^{4^{(1)}}\vth_2^{2^{(1)}} - \bar \vth_1^{(2)} \bar \vth_2^{(4)} - \left( \vth_2^{4^{(1)}}\vth_1^{2^{(1)}} - \bar \vth_2^{(2)} \bar \vth_1^{(4)} \right) \right) + \vth_1^{(3)}\vth_2^{\prime (3)} \rangle\,.
\ea

\noindent $\bullet$ The $2$-nd differential is
\begin{equation}
\begin{aligned}
    \hd_2^{\,\rm unknot} &= \vth_2^{4^{(2)}} \vth_2^{2^{(1)}}\left( \frac{\partial}{\partial \vth_2^{6^{(2)}}} - \frac{\partial}{\partial \vth_2^{6^{(3)}}} \right) + \left( \vth_1^{4^{(2)}} \vth_2^{2^{(1)}} + \vth_2^{4^{(2)}} \vth_1^{2^{(1)}} \right)\left( \frac{\partial}{\partial \vth_1^{6^{(2)}}} - \frac{\partial}{\partial \vth_1^{6^{(3)}}} \right) + \\
    &+ \vth_2^{4^{(2)}}\left( \frac{\partial^2}{\partial \bar \vth_1^{(2)} \bar \vth_2^{2^{(2)}}} + \frac{\partial^2}{\partial \bar \vth_2^{(2)} \bar \vth_1^{2^{(2)}}} \right) + \vth_1^{4^{(2)}} \frac{\partial^2}{\partial \bar \vth_1^{(2)} \bar \vth_1^{2^{(2)}}}\,.
\end{aligned}
\end{equation}
Its image and kernel are
{\small \begin{equation} \nn
\begin{aligned}
    \Im(\hd_2^{\,\rm unknot}) &= V \otimes V\,,\\
    \Ker(\hd_2^{\,\rm unknot}) &= \langle \vth_1^{6^{(2)}} + \vth_1^{6^{(3)}},\, \vth_2^{6^{(2)}} + \vth_2^{6^{(3)}} \rangle \oplus \langle (\bar \vth_1^{(2)} \bar \vth_2^{2^{(2)}} - \bar \vth_2^{(2)} \bar \vth_1^{2^{(2)}})\vth_1^{2^{(1)}},\, (\bar \vth_1^{(2)} \bar \vth_2^{2^{(2)}} - \bar \vth_2^{(2)} \bar \vth_1^{2^{(2)}})\vth_2^{2^{(1)}},\, \bar \vth_2^{(2)} \bar \vth_2^{2^{(2)}}\vth_1^{2^{(1)}},\,\bar \vth_2^{(2)} \bar \vth_2^{2^{(2)}}\vth_2^{2^{(1)}}, \\
    &\bar \vth_1^{(2)} \bar \vth_2^{2^{(2)}}\vth_2^{2^{(1)}} + \vth_2^{6^{(2)}},\, \bar \vth_1^{(2)} \bar \vth_1^{2^{(2)}}\vth_2^{2^{(1)}} + \vth_1^{(2)} \bar \vth_2^{2^{(2)}}\vth_1^{2^{(1)}} + \vth_1^{6^{(2)}} \rangle \,.
\end{aligned}
\end{equation}}
The cohomologies and their quantum dimensions are
\ba \nn
\cH_0^{\,\rm unknot} &= \Ker(\hd_0^{\,\rm unknot}) = \varnothing \quad \Lra \quad \boxed{\dim_q\cH_0^{\,\rm unknot} = 0\,,} \\
\cH_1^{\,\rm unknot} &= \Ker(\hd_1^{\,\rm unknot}) \backslash \Im(\hd_0^{\,\rm unknot}) \cong \langle \vth_2^{(3)}\vth_2^{\prime (3)},\, \vth_1^{(3)}\vth_2^{\prime (3)} - \vth_2^{(3)}\vth_1^{\prime (3)} \rangle \quad \Lra \quad \boxed{\dim_q\cH_1^{\,\rm unknot} = q^{-2}+1\,,} \\
\cH_2^{\,\rm unknot} &= \Ker(\hd_2^{\,\rm unknot}) \backslash \Im(\hd_1^{\,\rm unknot}) = \varnothing \quad \Lra \quad \boxed{\dim_q\cH_2^{\,\rm unknot} = 0\,,} \\
\cH_3^{\,\rm unknot} &= \Ker(0) \backslash \Im(\hd_2^{\,\rm unknot}) = \varnothing \quad \Lra \quad \boxed{\dim_q\cH_3^{\,\rm unknot} = 0\,.}
\ea  
The Khovanov polynomial is
\begin{equation}
    \Kh^{\,\rm unknot} = T^{-1}\cdot (qT) \dim_q \cH_1^{\,\rm unknot} = D_2\,.
\end{equation}

\paragraph{Summary.} In this example, we have seen that not only one cycle can be mapped by the zero morphism. There also appear zero morphism following $m$ operators. When the $\Delta$ morphism comes, the Sh operator acts next. 

\setcounter{equation}{0}



\setcounter{equation}{0}

\section{Khovanov precursor complex}
\label{sec:precursors}

By examples in the previous section, we have demonstrated how to get the Khovanov complex for a precursor diagram. A precursor diagram is obtained from a bipartite diagram\footnote{There is also an open problem how to construct CAJ-like operators (see Section~\ref{sec:CAJ-ext-Jones}) acting on extended HOMFLY polynomials by reversing one lock vertex. Such operators should be well defined when transferring to a precursor diagram.} by replacing each lock tangle with a single vertex as shown in Fig.\,\ref{fig:precursor-bip}, see also the whole Section~\ref{sec:H-J-duality}. There are two equivalent methods to get the Khovanov precursor complex.

\medskip

\noindent {\bf 1.} The first variant is to draw a 2-hypercube for the HOMFLY polynomial for a bipartite link, see examples in Figs.\,\ref{fig:H-Hopf-complex}, \ref{fig:H-tw-trefoil-complex}, \ref{fig:H-tw-4-1-complex}. Change $\phi \rightarrow -q$, $\bphi \rightarrow -q^{-1}$ and consider $N=2$. Shift gradings of all resolutions so that the initial point of a hypercube starts with the grading shift $=\{0\}$. Then, we place morphisms as described in Section~\ref{sec:Khovanov} and get the Khovanov complex that is calculated by the same formula~\eqref{Kh-def}:
\begin{equation}\label{Kh-def-2}
    {\rm Kh}^{\cal L}(q,T) = q^{n_\bullet}\cdot (T q^2)^{-n_\circ} \sum_{i=0}^n (qT)^i \dim_q \cH_i^{\cal L}
\end{equation}
where $\cH_i$ are cohomologies of a resulting complex.

\medskip

\noindent {\bf 2.} Having a $3$-complex for the Khovanov--Rozansky polynomial, we shrink resolutions connected by colored (in blue or green) arrows to a single one. Then, forget about the grading shifts and put instead corresponding powers of $\phi$, $\bphi$ by the shrunk resolutions. The weighted with these multipliers sum is the HOMFLY polynomial expressed through $\phi$, $\bphi$, $D_N$. Morphisms and enumeration of resolutions are inherited from an initial Khovanov--Rozansky complex. Next, as in the first method, we change $\phi \rightarrow -q$, $\bphi \rightarrow -q^{-1}$ and consider $N=2$. Shift gradings of all resolutions so that the initial point of a hypercube starts with the grading shift $=\{0\}$. The Khovanov polynomial is then given by~\eqref{Kh-def-2}.

\medskip
 
\noindent An interesting question is the possibility of the inverse process:
to {\it lift} the data from a precursor diagram to bipartite ones --
i.e. to extract the Khovanov--Rozansky answer for a bipartite link
from the Khovanov one for a corresponding precursor diagram.
This is hardly possible in full; still, the exact estimate of the difference  
between information in the bipartite and precursor complexes 
is not yet available. 




\section{Conclusion}
\label{sec:concl}

In this paper, we have proposed an efficient algorithm of {\it cycle calculus} for the Khovanov–Rozansky polynomials
for a special family of knots, known as bipartite.
The method is a direct analogue to the planar Khovanov calculus, see Section~\ref{sec:Khovanov}. 
All the morphisms are provided by cut-and-join operators, only now they act on $N$-dimensional vector spaces.
And one needs to add one more type of morphism, mapping a pair of cycles to another one. 
As a result, the full set of morphisms act along the edges of the 3-hypercube substituting the ordinary Kauffman--Khovanov
2-hypercube.
 
We have started in Section~\ref{sec:bip-exp} from planar decomposition of the antiparallel lock tangle (antiparallel
composition of two single crossings), see Fig.\,\ref{fig:pladeco}. 
We have noted that this fact allows us to organize the expansion
of the HOMFLY polynomial in $\phi$, $\bphi$, $D$ into a 2-hypercube (similar to the Jones case in Section~\ref{sec:Jones-2-hyp}).
However, $\phi$ and $\bar\phi$ are not monomials. 
In order to get a lift to the Khovanov--Rozansky complex, this 2-hypercube must be blown up to a 3-hypercube by expanding $\phi = q^{N+1} - q^{N-1}$ and $\bphi = q^{-N-1} - q^{-N+1}$, see Fig.\,\ref{fig:pladeco-3-hyp}. This allows for the $T$-deformation described in Section~\ref{sec:method}. In Section~\ref{sec:examples}, we provide examples of the proposed Khovanov--Rozansky cycle calculus. 

The planar decomposition of the lock element (in Fig.\,\ref{fig:pladeco}) being compared with the Kauffman bracket for a single vertex for $N=2$ (see Fig.\,\ref{fig:Kauff}) provides the duality between the HOMFLY polynomials for bipartite diagrams and Jones polynomial for a corresponding precursor diagram (obtained from a bipartite one by shrinking lock tangles via the rule in Fig.\,\ref{fig:precursor-bip}). This correspondence is actually lifted to the $T$-deformed case. We describe the algorithms to obtain the Khovanov precursor complex in Section~\ref{sec:precursors} and analyze the examples in Section~\ref{sec:examples}.

We do not prove the topological invariance of our answers --
which is not so simple for bipartite knots \cite{ALM},
because bipartite {\it realization} itself is not invariant under Redemeister moves. Moreover, the full set of equality relations transforming a bipartite diagram into another bipartite one is unknown.
Instead, we demonstrate that the answers coincide with the known Khovanov--Rozansky results,
which are usually deduced in a far more complicated way
but instead are proved to be topological invariants.
More details about 
equivalence of bipartite and the Khovanov--Rozansky (matrix factorization) calculations
and the bipartite reduction of the Khovanov--Rozansky calculus will be presented in our forthcoming paper. This will be an evidence for topological invariance of our Khovanov--Rozansky cycle calculus.

Another important comment is that the cohomological calculations involve quantum dimensions of subspaces
of spaces with $N$-dependent dimensions, which vanish at small values of $N$. This fact affects the Khovanov--Rozanky homology so that their dimensions are not necessarily continuous functions of $N$. A conventional way to handle this problem is in terms of superpolynomials, which are smooth in $A = q^N$ and reproduce the Khovanov–Rozansky formulas for $N \geq N_0^{\cal L}$ with $N_0^{\cal L}$ being some integer number and dependent on a choice of a link $\cal L$.
Although the break of continuity does not happen at our simple examples, it is known to take place for more complicated knots, such as $8_{19} = T[3,4] = {\rm pretzel}(3,3,-2)$. The latter one is bipartite, and it is an interesting question whether our bipartite algorithm reproduces the discontinuity of the Khovanov--Rozansky polynomial at the known point $N=N_0^{8_{19}}=3$. In other words, the question is whether the discontinuity of the Khovanov--Rozansky polynomial in $N$ can be understood already in terms of the simple cycle calculus, not only in complicated terms of matrix factorizations. Thus, the bipartite Khovanov--Rozansky calculus may spread more light on the origin of these discontinuities. In particular, it is an interesting question, if bipartite links can demonstrate simpler and more understandable behavior in this respect.
For example, maybe for bipartite links this $N_0^{\cal L}$ is bounded from above.

Even more intriguing are discontinuities of Khovanov--Rozansky polynomials
in {\it evolution} parameters  within {\it families} of knots \cite{dunin2013superpolynomials,dunin2022evolution,anokhina2019nimble,willis2021khovanov,nakagane2019action,anokhina2018khovanov}.
Our examples in this paper are too few to reveal the peculiarities of evolution within bipartite families --
this is one more question for future analysis.


A separate interesting subject is the case of $N=2$, where Khovanov 2-hypercube complexes
are reduced to the 3-hypercube ones -- it would be interesting to describe the reduction. We will also do it in another text.

There are four more open problems to be solved. First, it is tempting to write morphisms not for links, which resolutions are formed by closed cycles, but for tangles with free ends. This would allow us to first reduce (simplify) complexes constructed from combinations of tangles, and then to glue from these tangles different links having already simplified complexes. Moreover, such a tangle calculus would allow us to find out how many types of morphisms can appear. In our examples, there have appeared only four types of morphisms, but in general, we have not proved that it is always the case. 

The second question is: What happens for the reduced Khovanov--Rozansky polynomial? Can cycle technique be applied to compute the reduced Khovanov--Rozansky polynomial? How does this reduction relate to the reduction of complexes in the Khovanov case? Do they commute, or does the reduction of complexes completely differ for the reduced case?

The third problem is to find operators acting on the extended bipartite HOMFLY polynomial\footnote{The extended bipartite HOMFLY polynomial should be constructed the same way as the extended Jones polynomial in Section~\ref{sec:CAJ-ext-Jones}. Namely, one distinguishes quantum dimensions and makes the substitutions $-q^{N-1}\rightarrow t_1$, $-q^2 \rightarrow t_2$.} by changing one lock vertex to the mirror one. One of the operators should be the ordinary cut-and-join operator from Section~\ref{sec:CAJ-ext-Jones}, and it is the only one remaining in the downgrade to the precursor case. However, another operator should be introduced. This operator supersymmetrization must give the Sh morphism. 

The fourth question concerns not-fully-bipartite links being non-bipartite links but still having lock tangles inside a corresponding diagram. In the HOMFLY case, one can apply the planar decomposition for the lock tangle (Fig.\,\ref{fig:pladeco}) only to these lock tangles. The resulting diagrams correspond to simpler links, and if these links are bipartite then the planar technique allows one to calculate the HOMFLY polynomial even for not-fully-bipartite knots. A question is whether this fact can be somehow lifted to the Khovanov--Rozansky polynomials.

\bigskip
 
\noindent To summarize, we have extended the bipartite calculus of~\cite{ALM} to the field of superpolynomials,
and again it appeared much simpler and handled by the Khovanov-like 
rather than the generic Khovanov--Rozansky  formalism.
This adds more arguments to the suggestion that
{\bf the theory of bipartite knots can turn to be a closed subset of knot theory
which is much simpler and more ``planar’’} than its other parts.  
As usual in particle and string physics, the planar ``limit’’ is not fully independent
(in this particular case the independent treatment of topological invariance is not so simple),
but in many respects it can be considered as a self-contained piece of science -- the
{\bf bipartite knot theory}. 

It is a separate challenging question, whether this has a clear QFT counterpart —
the {\bf bipartite Chern--Simons theory}.

\section*{Acknowledgments}

We are grateful for enlightening discussions to D. Galakhov, A. Popolitov, A. Sleptsov and R. Stepanov. 

Our work is partly
funded within the state assignment of the Institute for Information Transmission Problems of RAS. 
It is partly supported by the grants of the Foundation for the Advancement of Theoretical Physics and Mathematics “BASIS”. The work of E.L. is partly supported by the Ministry of Science and Higher Education of the Russian Federation (agreement no. 075–15–2022–287).

\printbibliography

\end{document}